%% file: sample.tex
\documentclass[fleqn,usenatbib]{mnras}

\usepackage{newtxtext,newtxmath}

\usepackage[T1]{fontenc}

\DeclareRobustCommand{\VAN}[3]{#2}
\let\VANthebibliography\thebibliography
\def\thebibliography{\DeclareRobustCommand{\VAN}[3]{##3}\VANthebibliography}

\usepackage{graphics,graphicx}	
\usepackage{amsmath}	
\input{macros.tex}

\usepackage{tikz}
   \usetikzlibrary{positioning,spy}
\usepackage{graphicx}
\usepackage{subfig}
\usepackage{tikz,xcolor,hyperref}

\definecolor{lime}{HTML}{A6CE39}
\DeclareRobustCommand{\orcidicon}{%
	\begin{tikzpicture}
	\draw[lime, fill=lime] (0,0) 
	circle [radius=0.16] 
	node[white] {{\fontfamily{qag}\selectfont \tiny ID}};
	\draw[white, fill=white] (-0.0625,0.095) 
	circle [radius=0.007];
	\end{tikzpicture}
	\hspace{-2mm}
}

\foreach \x in {A, ..., Z}{%
	\expandafter\xdef\csname orcid\x\endcsname{\noexpand\href{https://orcid.org/\csname orcidauthor\x\endcsname}{\noexpand\orcidicon}}
}

\title[$\suzaku$ and $\chandra$ observations of MKW4]{Joint $\suzaku$ 
and $\chandra$ observations of the MKW4 galaxy group out to the virial radius}

\author[Sarkar et al.]{
Arnab Sarkar,$^{1}$\thanks{E-mail: \href{mailto:arnab.sarkar@uky.edu}{arnab.sarkar@uky.edu}}\orcidA{}
Yuanyuan Su,$^{1}$\orcidB{}
Scott Randall,$^{2}$\orcidC{}
Fabio Gastaldello,$^{3}$\orcidD{}
Isabella Trierweiler,$^{4}$
\newauthor
Raymond White,$^{5}$
Ralph Kraft,$^{2}$\orcidF{}
Eric Miller$^{6}$\orcidG{}
\\
$^{1}$Physics and Astronomy, University of Kentucky, 505 Rose street, Lexington, KY 40506, USA\\
$^{2}$Harvard-Smithsonian Center for Astrophysics, 60 Garden Street, Cambridge, MA 02138, USA\\
$^{3}$Instituto di Astrofisica Spaziale e Fisica Cosmica (INAF-IASF), Milano, via A. Corti 12, I-20133 Milano, Italy\\
$^{4}$Division of Astronomy and Astrophysics, University of California Los Angeles, Los Angeles, CA\\
$^{5}$Department of Physics and Astronomy, University of Alabama, Box 870324, Tuscaloosa, AL 35487, USA\\
$^{6}$Kavli Institute for Astrophysics \& Space Research, Massachusetts Institute of Technology, 77 Massachusetts Ave, Cambridge, MA 02139, USA
}

\date{Accepted for publication in MNRAS \\ Accepted 2020 December 8. Received 2020 December 7; in original form 2020 November 4}

\begin{document}
\label{firstpage}
\pagerange{\pageref{firstpage}--\pageref{lastpage}}
\maketitle

\begin{abstract}
We present joint $\suzaku$ and $\chandra$ 
observations 
of MKW4. With a global temperature of 1.6 keV, 
MKW4 is one of the smallest galaxy groups 
that have been mapped in X-rays out to the 
virial radius. We measure its gas properties 
from its center to the virial radius in the 
north, east, and northeast directions. Its 
entropy profile follows a power-law of 
$\propto r^{1.1}$ between 
R$_{500}$ and $R_{200}$ in all directions, 
as expected from the purely gravitational 
structure formation model. 
The well-behaved entropy profiles at the 
outskirts of MKW4 disfavor the presence of 
gas clumping or thermal non-equilibrium
between ions and electrons in this system. 
We measure an enclosed baryon fraction of 11\% 
at R$_{200}$, remarkably smaller
than the cosmic baryon fraction of 15\%.
{  {We note that the enclosed gas fractions
at R$_{200}$ are systematically smaller
for groups than for clusters from existing
studies in the literature. The low baryon fraction
of galaxy groups, such as MKW4, suggests that
their shallower gravitational potential well
may make them more vulnerable to 
baryon losses due to AGN 
feedback or galactic winds.}}
We find that 
the azimuthal scatter of various gas 
properties at the outskirts of MKW4 is 
significantly lower than in other 
systems, suggesting that MKW4 is a spherically 
symmetric and highly relaxed system.

\end{abstract}




\section{Introduction}\label{sec:intro}

A significant fraction of all the 
baryons of the Universe, 
including more than half of 
the galaxies, reside in groups and 
low mass clusters 
\citep[e.g.,][]{10.1111/j.1365-2966.2004.07408.x,
10.1046/j.1365-8711.2003.06207.x}.
In the standard CDM structure formation
model, 
clusters continue to grow and evolve through 
mergers and accretions, largely along  
cosmic filaments, in their outer regions 
\citep{2019SSRv..215....7W}. These 
processes may leave distinctive marks in the gas 
properties at cluster outskirts, including 
inhomogeneous gas density distributions, turbulent 
gas motions, and electrons that are not in 
thermodynamic equilibrium with the ions in the 
intra-cluster medium (ICM). 
Groups are expected to be more evolved 
than massive clusters, as their sound 
crossing times are short compared to the 
Hubble time \citep[0.1--0.5 H$_{0}^{-1}$, ][]{2017MNRAS.471....2P}. 
However, due to their shallower gravitational 
potential wells, groups are more sensitive to 
non-gravitational processes, such as 
galactic winds, star formation, and feedback 
from active galactic nuclei 
\citep[AGN, e.g.,][]{2015A&A...573A.118L,2010A&A...511A..85P,Mathews_2011,2012ApJ...748...11H,2016A&A...592A..37T}. 
Probing the gas properties of galaxy 
groups out to their virial radii thus
provides a powerful approach to investigating their 
growth and evolution.

In galaxy cluster studies, 
entropy, as a function of radius,  
records the thermal history of the ICM 
\citep[e.g.,][]{10.1111/j.1365-2966.2005.09158.x,2010MNRAS.406..822M}. 
{  {Entropy is empirically defined as 
K(r)=T/n$_{\rm e}^{2/3}$, where n$_{\rm e}$
and T are the electron density and gas temperature,
respectively. 
}}
A growing number of X-ray observations have been
made for
cluster outskirts 
\citep[e.g.,][]{2017MNRAS.469.1476S,2018A&A...614A...7G}. 
The majority 
of these works study the 
properties of massive galaxy clusters 
($\textrm{T}_\textrm{X}$ $>$ 3 keV), 
while there is a lack of 
detailed studies of the outskirts
of galaxy groups. { The 
studies of 
massive clusters have brought up unexpected 
results 
\citep[e.g.,][]{2014MNRAS.437.3939U,2017MNRAS.469.1476S,2018A&A...614A...7G}
such as the flattening or 
even a drop of entropy 
profiles 
between R$_{500}$\footnote[2]{$r_{\Delta}\ =$ radius from cluster core where matter density is $\Delta$ times the critical density of the Universe.} 
and R$_{200}$
relative to the expectation from
numerical simulations of the 
gravitational collapse
model \citep{2005MNRAS.364..909V}. 
Several explanations 
have been proposed 
to explain the deviation
of entropy profiles from the self-similar value, 
e.g., the breakdown of thermal
equilibrium between electrons and protons 
\citep{10.1093/pasj/63.sp3.S1019} or 
inhomogeneous gas density distribution (gas clumping)
at cluster outskirts \citep{Nagai_2011}}. 
Accretion shocks at the 
cluster outskirts tend to heat the heavier ions 
faster than 
electrons, causing thermal non-equilibrium between 
electrons and ions and leading to a lower gas entropy
\citep[e.g.,][]{10.1093/pasj/63.sp3.S1019,10.1093/pasj/62.2.371}. 
{ Unresolved 
cool gas clumps were invoked by \citet{Simionescu1576} and 
\citet{Nagai_2011} }to explain 
the observed entropy flattening, since the denser, 
cooler clumps have a higher emissivity than the 
local ICM. 
The clumping factor is defined as
\begin{equation}\label{eqn:clump}
   \ \ \ \ \ \ \ \ \ \ \ \ \ \ \ \ \ \ \ \ \ \ \ \ \ \ \ \ \ \ \ \ \ \ \ \textrm{C} = \frac{\langle \textrm{n}_\textrm{e}^2 \rangle}{\langle \textrm{n}_\textrm{e} \rangle^2}, 
\end{equation}
where $\textrm{n}_\textrm{e}$ is the electron 
density. 
\citet{Simionescu1576} estimate a 
clumping factor of C $\approx$ 16 for 
the Perseus cluster. \citet{2013MNRAS.428.2812B} 
and \citet{10.1111/j.1365-2966.2012.21282.x} 
report $\textrm{C}$ $\approx$ 7 for Abell 1835 
and $\textrm{C}$ $\approx$ 9 for PKS 0745-191, 
respectively.
In contrast, a handful of observations 
have indicated 
that low mass clusters 
(T$_\textrm{X}$ $<$ 3 keV) show little 
to no flattening in { their} entropy profiles 
(e.g., RXJ1159, \citealt{2015ApJ...805..104S}; 
A1750, 
\citealt{2016ApJ...831...55B}; UGC 03957, 
\citealt{2016A&A...592A..37T}), presumably because 
groups have lower clumping factors at their 
outskirts. { Galaxy groups may provide essential 
constraints on whether
their entropy profiles behave in a 
self-similar way compared to galaxy
clusters.}

MKW4 is a cool core cluster at $z \sim 0.02$ 
with a global temperature of 1.6 keV 
\citep{Sun_2009}.  
It contains nearly 50 member galaxies, 
including NGC 4073, 
the brightest group galaxy (BGG). The 
NGC 4073 is about 1.5 times brighter than the 
second-brightest galaxy in the group 
\citep{2003MNRAS.346..525O}. 
Unlike massive clusters, the outskirts of 
galaxy groups 
are relatively unexplored due to their low surface 
brightness. { The typical ICM surface 
brightness in 
group outskirts falls below 20\% of the  total
emission.} 
The measurement of their gas 
properties is therefore extremely 
challenging.
The now-defunct 
$\suzaku$ X-ray telescope with its low particle 
background helped unravel this new frontier.
However, $\suzaku$ can only resolve point sources 
down to a flux level of $10^{-13}$ erg cm$^{-2}$ 
s$^{-1}$ due to its 
modest PSF ($\sim 3'$), which causes significant 
statistical and systematic 
uncertainties in the measurement of gas properties 
at the group outskirts. 
{ Thanks to the superb angular 
resolution of 
$\chandra$, even with the modest exposures
it increases the number of detected
point sources by 1 dex, which allows
tighter constrains on the cosmic X-ray
background (CXB) uncertainties} \citep{2012AIPC.1427...13M}.

{ We utilize deep $\suzaku$ 
and snapshot $\chandra$ observations to 
probe the thermal properties of MKW4 out 
to its virial radius in multiple directions, 
which we present in this paper. {The metallicity distribution of MKW4 will be presented in
a following paper.}}
Using NASA/IPAC Extragalactic 
Database,\footnote[3]{\url{http://ned.ipac.caltech.edu}}
we calculate a luminosity distance of 83 Mpc 
(1$\arcsec$ = 0.443 kpc) for $z$ = 0.02, adopting a 
cosmology of H$_0$ = 70 km s$^{-1}$ Mpc$^{-1}$, 
$\Omega_{\Lambda}$ = 0.7, 
and $\Omega_\textrm{m}$ = 0.3. 
All reported uncertainties in this 
paper are at 1$\sigma$ confidence level 
unless mentioned otherwise.

\section{Observations and data reduction} \label{sec:data}
\begin{figure*}

\begin{tabular}{ccc}
\includegraphics[width=0.083\textwidth]{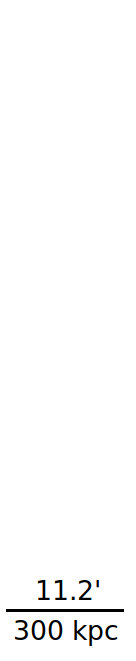}    &  \hspace{-20pt}
\includegraphics[width=0.4\textwidth]{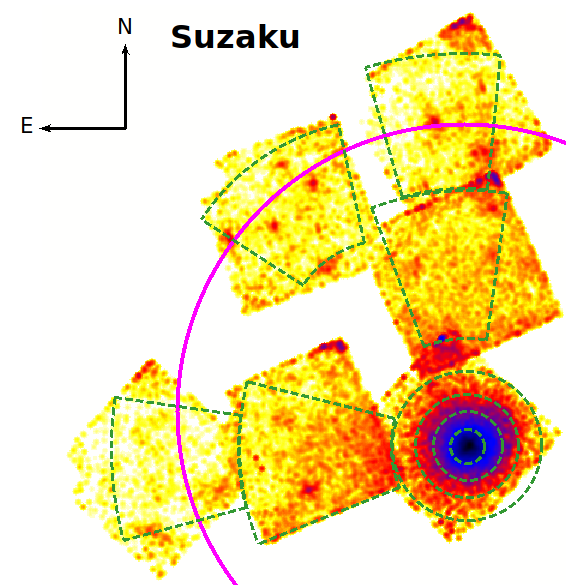}    &  \hspace{10pt} 
\includegraphics[width=0.36\textwidth]{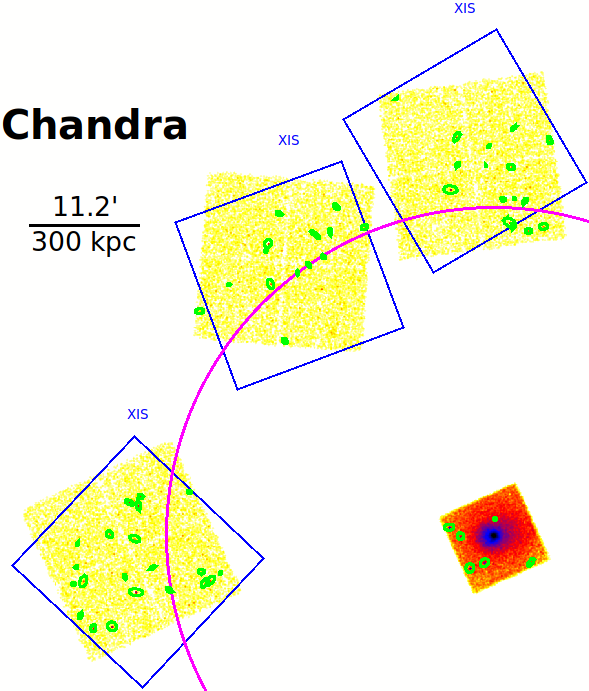}
\end{tabular}
    \caption{Left: $\suzaku$ mosaic 
image of MWK4 in the 0.5-2 keV energy band 
with NXB subtracted and exposure corrected. 
Green annuli represent  
the regions used for spectra extraction. 
Magenta circle indicates R$_{200}$ = 
33.2$\arcmin$ 
(884 kpc). Right: {$\chandra$ mosaic image of 
MKW4 center (ACIS-S3) and  
three outer regions (ACIS-I) in the 0.5-2.0 keV 
energy band. The image of MKW4 is smoothed with a Gaussian 
kernel and a number of radii of 30. Blue boxes represent the 
$\suzaku$ FOV. Green elliptical regions are the resolved point 
sources. Magenta circle indicates R$_{200}$.
}}
\label{fig:suzaku_mo}
\end{figure*}

MKW4 has been mapped with 6 $\suzaku$ 
pointings from its center out to the virial 
radii in the north, east, and north-east 
directions, as shown in Figure \ref{fig:suzaku_mo}. 
It has also been observed with a deep $\chandra$ 
ACIS-S observation at its 
center and three snapshot $\chandra$ ACIS-I 
observations overlapping with the three outer 
$\suzaku$ pointings 
in three directions. Below we describe the data 
reduction processes and spectral fitting
for each of the observations.

\subsection{Suzaku}
MKW4 has been mapped with three 
X-ray Imaging Spectrometer (XIS) 
instruments onboard $\suzaku$ -- 
two front-illuminated CCDs (XIS0, XIS3), 
and one back-illuminated CCD (XIS1). 
The observation logs are summarized 
in Table \ref{tab:obs_log}. 
All three XIS instruments were in 
the normal clocking 
mode without window and burst options 
during the observations.
\begin{table*}
\centering
\caption{Observational log}
\label{tab:obs_log}
\begin{tabular}{llllclll}
\hline
Name & Obs Id & Instrument & Obs Date & Exposure time (ks) & RA ($^\circ$) & DEC ($^\circ$) & P.I.\\
\hline
$$\suzaku$$ central & 808066010 & XIS0, XIS1, 
XIS3 & 2013 Dec 30 & 34.6 & 181.1346 & 1.9097 & 
F. Gastaldello \\
$$\suzaku$$ Offset 1 & 805081010 & XIS0, XIS1, 
XIS3 & 2010 Nov 30 & 77.23 & 181.1270 & 2.2181 &
F. Gastaldello\\
$$\suzaku$$ N2 & 808067010 & XIS0, XIS1, XIS3 & 
2010 Nov 30 & 97 & 181.1504 & 2.5206 & Y. Su\\
$$\suzaku$$ Offset 2 & 805082010 & XIS0, XIS1, 
XIS3 & 2010 Nov 30 & 80 & 181.4270 & 1.8972 & Y.
Su\\
$$\suzaku$$ E1 & 808065010 & XIS0, XIS1, XIS3 & 
2013 Dec 29 & 100 & 181.7142 & 1.8506 & F. 
Gastaldello\\
$$\suzaku$$ NE & 809062010 & XIS0, XIS1, XIS3 & 
2013 Dec 29 & 87.5 & 181.4583 & 2.3361 & Y. Su\\
$$\chandra$$ central & 3234 & ACIS-S & 2002 Nov 
24 & 30 & 181.1283 & 1.9286 & Y. Fukazawa\\
$$\chandra$$ N & 20593 & ACIS-I & 2019 Feb 25 & 
14 & 181.1369 & 2.5209 & Y. Su\\
$$\chandra$$ E & 20592 & ACIS-I & 2018 Nov 17 & 
15 & 181.7146 & 1.8715 & Y. Su\\
$$\chandra$$ NE & 20591 & ACIS-I & 2019 Mar 08 &
14 & 181.4548 & 2.3585 & Y. Su\\
\hline
\end{tabular}
\end{table*}

\subsubsection{Data reduction}
The $\suzaku$ data was 
reduced using HEAsoft 6.25, 
CIAO\ 4.11, and the XIS 
calibration database (CALDB) version 20181010. 
We followed a standard data reduction 
thread
\footnote[4]{\url{https://heasarc.gsfc.nasa.gov/docs/suzaku/analysis/abc/}} to process all the event files.
The 5$\times$5 mode event files were converted to 
the 3$\times$3 mode event files and combined with
the other 3$\times$3 mode  
event files. The resulting event files 
were filtered for calibration source 
regions and bad pixels with {\tt cleansis}. 
We selected events with GRADE 0, 2, 3, 4, and 6. 
Light curves were filtered for 
flares using the {\tt\string lc$\_$clean} 
task of CIAO\ 4.11. 
The resolved point sources were identified 
visually. 
We did not exclude 
those sources while extracting spectra, 
because exclusion would result in a very 
low photon count available for the spectral 
analysis. We instead took the 
advantage of $\chandra$ observations to 
estimate their position, flux and incorporated 
them in the background fit (discussed in Section 
\ref{sec:back_sub}).

\begin{figure*}
\includegraphics[width=0.8\textwidth]{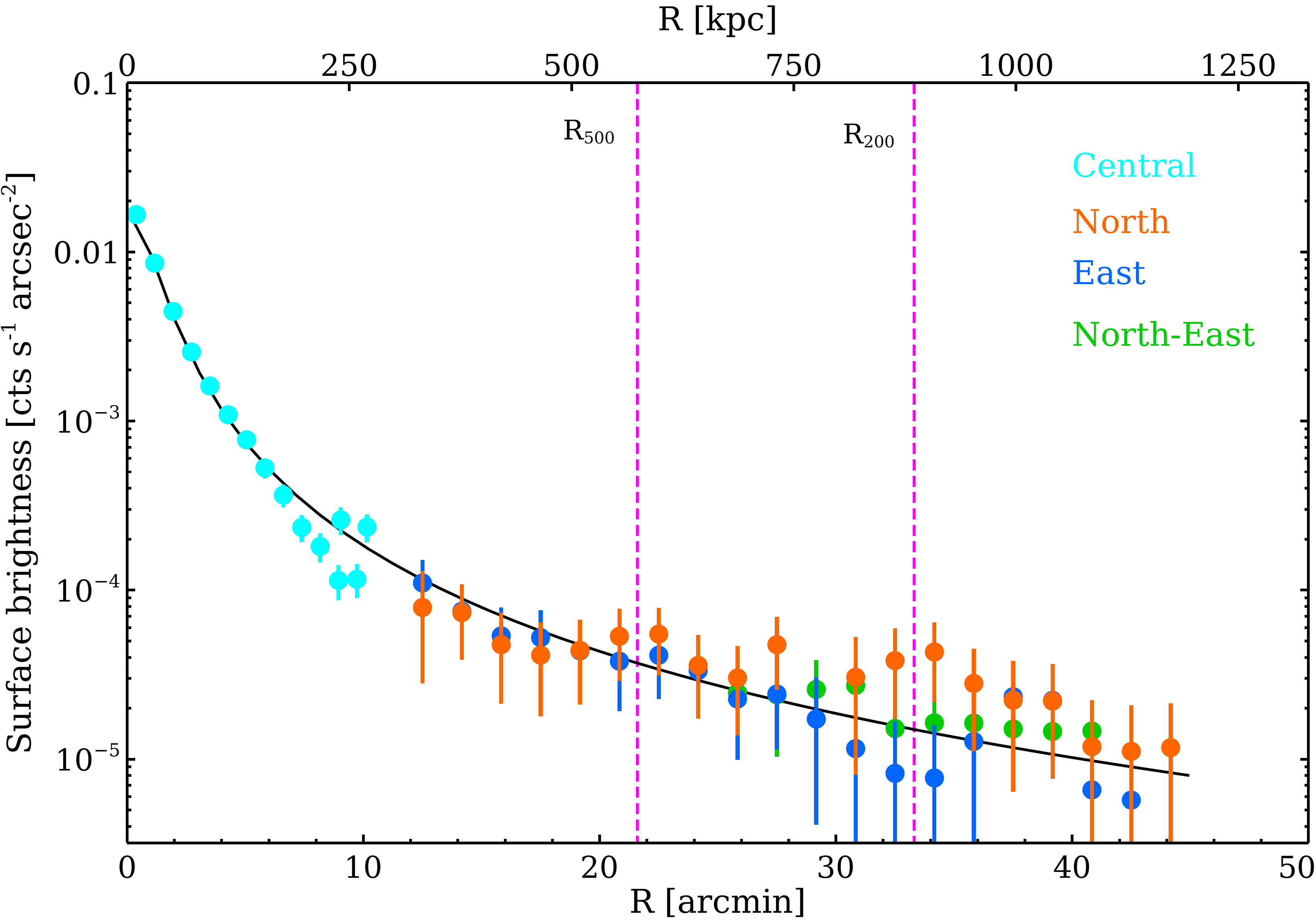}
\caption{Surface brightness profiles of MKW4  
derived from the $\suzaku$ 
XIS image in the 0.5 - 2.0 keV energy band
in three different directions. 
The profiles have been corrected for 
exposure, instrumental background (NXB), 
and the point sources detected with $\chandra$.
Black solid line is the best-fit $\beta$-profile.}
\label{fig:sur_bri}
\end{figure*}

We extracted spectra from four concentric 
annuli, $0\arcmin-2\arcmin$, $2\arcmin-4\arcmin$, 
$4\arcmin-6\arcmin$, and $6\arcmin-8.6\arcmin$ for  
the central pointing. We extracted
spectra from two partial annuli 
$12.4\arcmin-29\arcmin$, $29\arcmin-48.3\arcmin$
for the pointings in the north direction and 
from $8.6\arcmin-26\arcmin$, 
$26\arcmin-45\arcmin$ for 
the pointings in the east direction. 
For the north-east direction, 
we extracted spectra from a partial annulus of 
$27\arcmin-42\arcmin$. All annuli had
widths ranging from 8 kpc 
at the central region to 223 kpc at 
the outermost bin.
We generated 
redistribution matrix files (RMF) for all regions 
and detectors using FTOOL 
{\tt\string xisrmfgen} and instrumental 
background files (NXB) using FTOOL {\tt\string 
xisnxbgen}. The ancillary response 
files (ARF) were generated using {\tt\string 
xissimarfgen} by providing an appropriate 
$\beta$-image derived from the $\chandra$
surface brightness profile of MKW4 at the 
central region. 
Another ARF was produced to model the X-ray 
background by considering uniform sky emission 
in a circular region of 20$\arcmin$ radius.
An exposure corrected and NXB subtracted 
mosaic image of MKW4 in the 0.5-2 keV energy 
band is shown in Figure \ref{fig:suzaku_mo}. 
{ We obtained $\suzaku$ surface brightness profiles 
of MKW4 in three different 
directions, as shown in
Figure \ref{fig:sur_bri}. 
We fitted the profiles with a single 
$\beta$-model \citep{2009A&A...500..103A}:
\begin{equation}
    {\rm S(r)} = {\rm S_{0}} \left [ 1 + \left(\frac{\rm r}{\rm r_{c}}\right)^{2}\right]^{-3\beta+1/2},
\end{equation}
which yielded the 
best-fit ($\beta$, r$_\textrm{c}$) =
(0.515 $\pm$ 0.001, 30 $\pm$ 2.5 kpc).}

\subsubsection{Spectral analysis} \label{sec:back_sub}
\begin{figure*}
\begin{tabular}{cc}
\vspace{-40pt}\\
\includegraphics[width=0.54\textwidth]{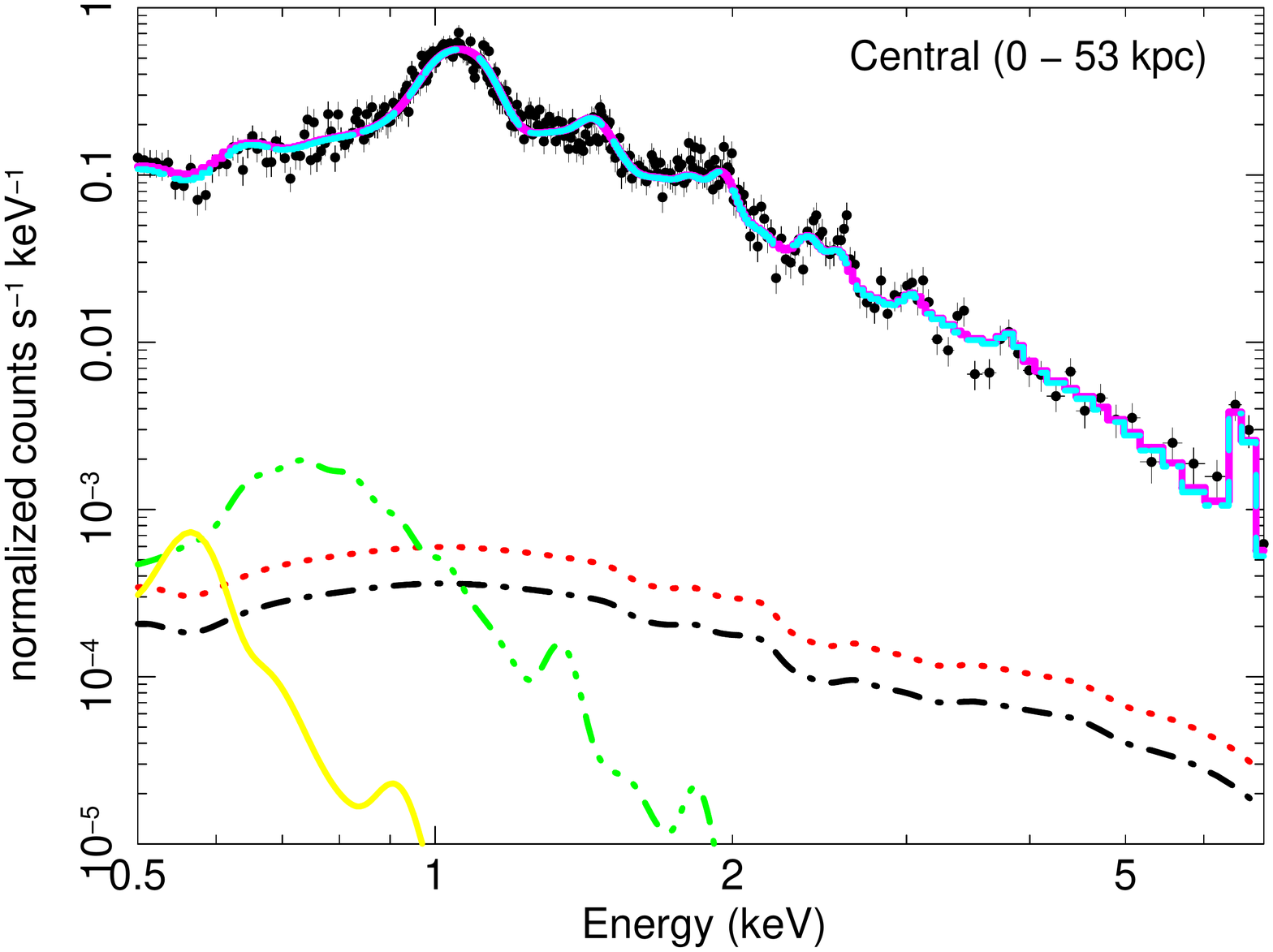}    &  \hspace{-40pt} \includegraphics[width=0.54\textwidth]{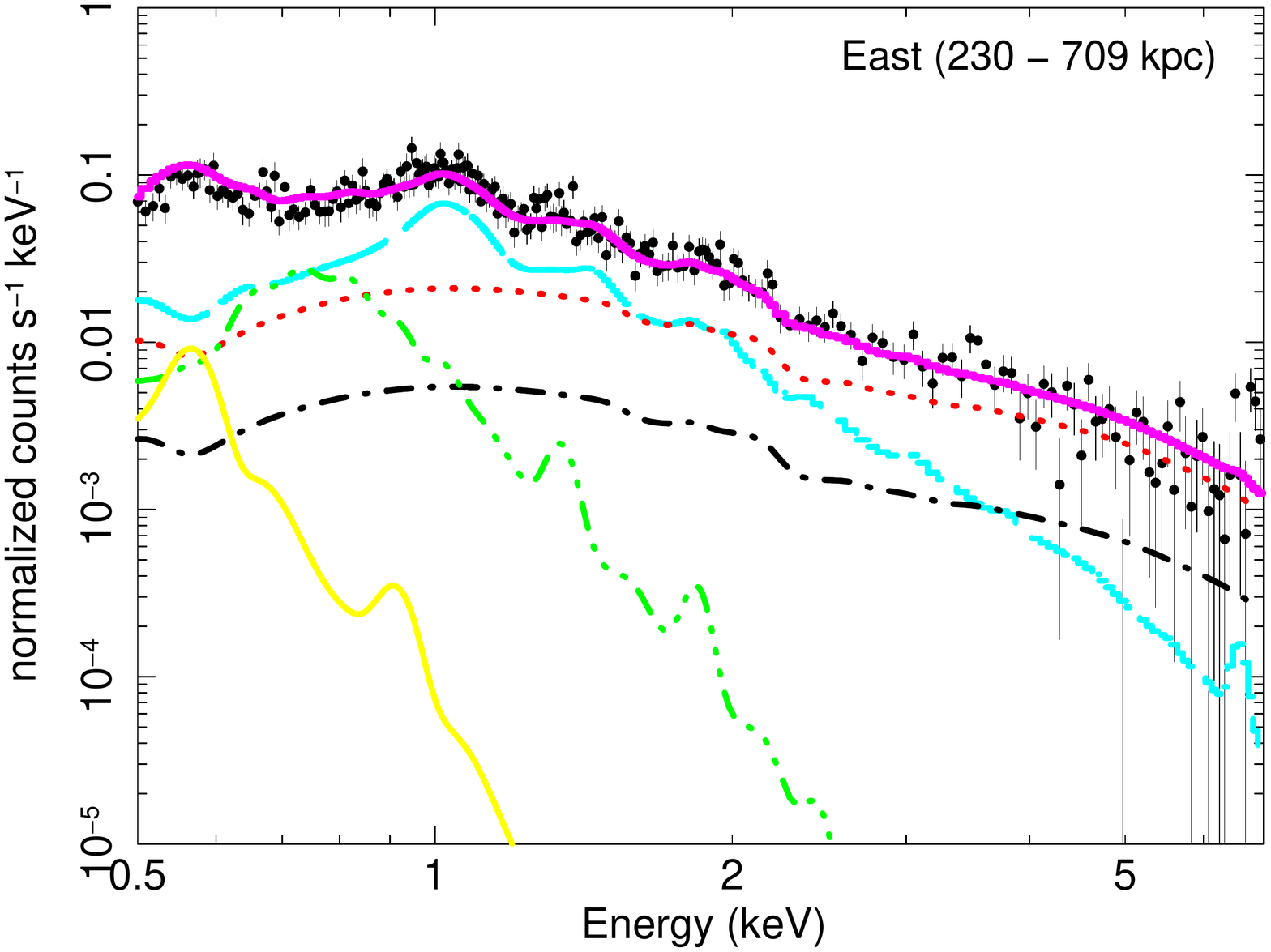} \vspace{-30pt}\\
\includegraphics[width=0.54\textwidth]{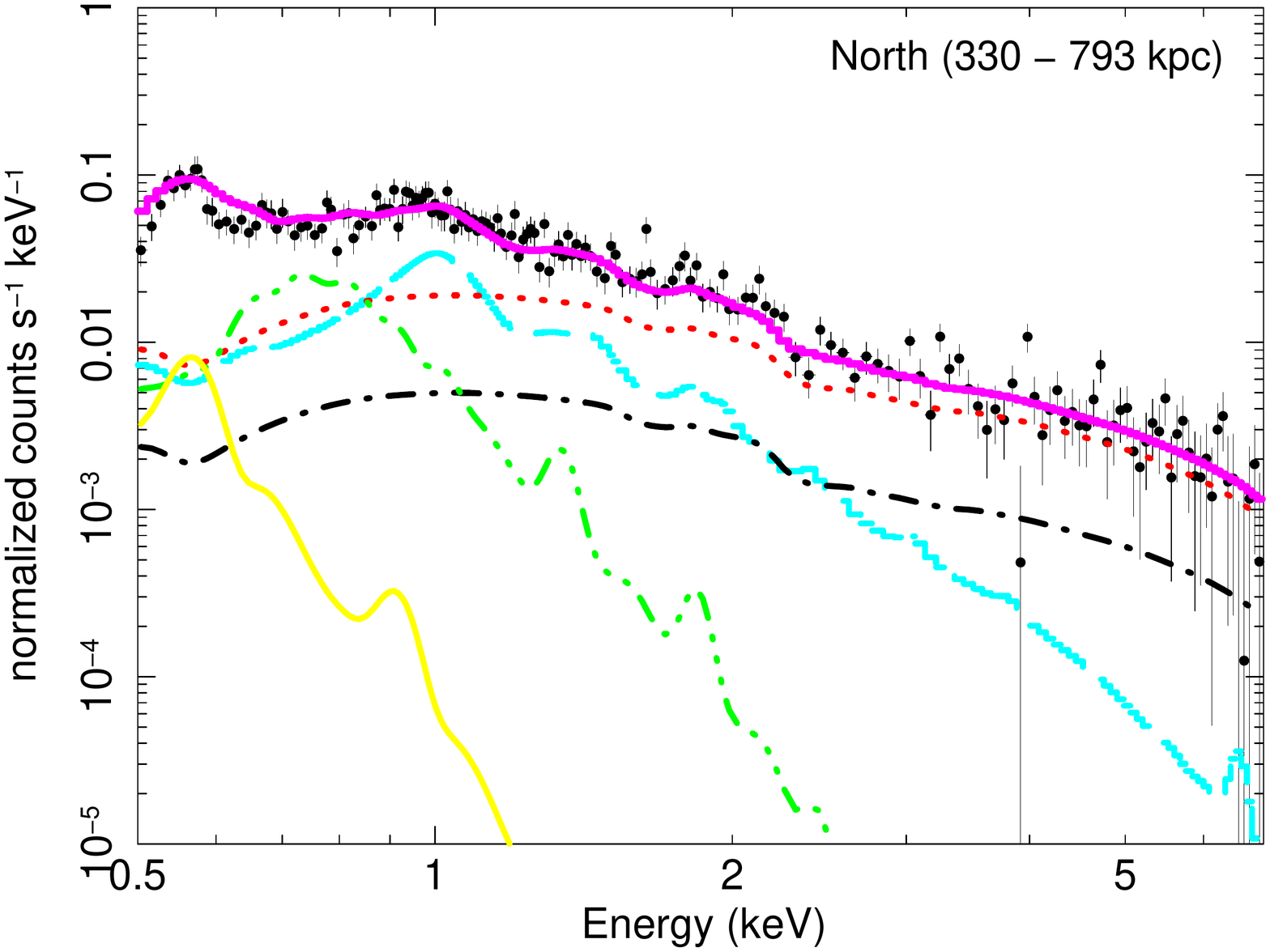}    &  \hspace{-40pt} \includegraphics[width=0.54\textwidth]{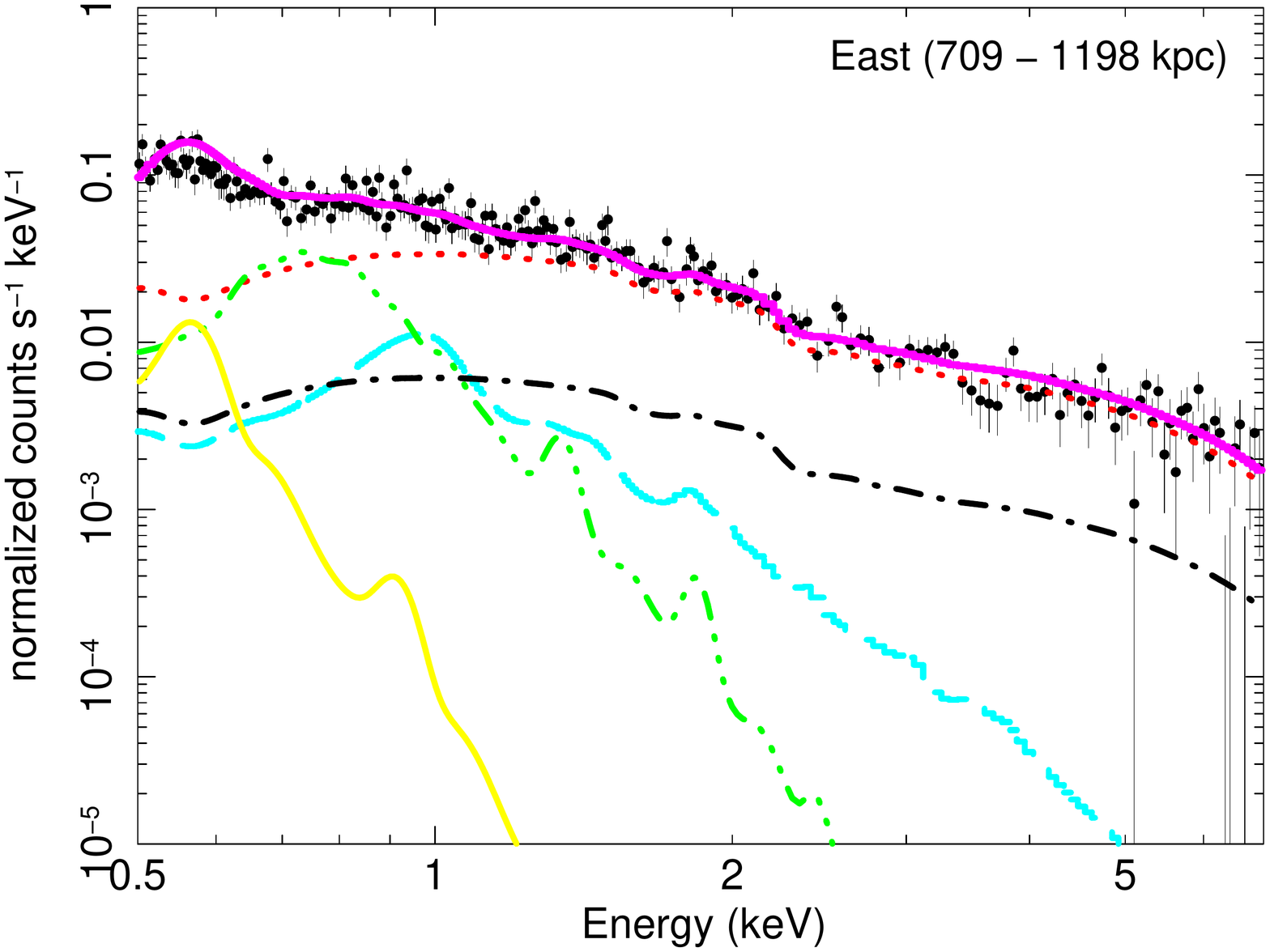}\vspace{-30pt}\\
\includegraphics[width=0.54\textwidth]{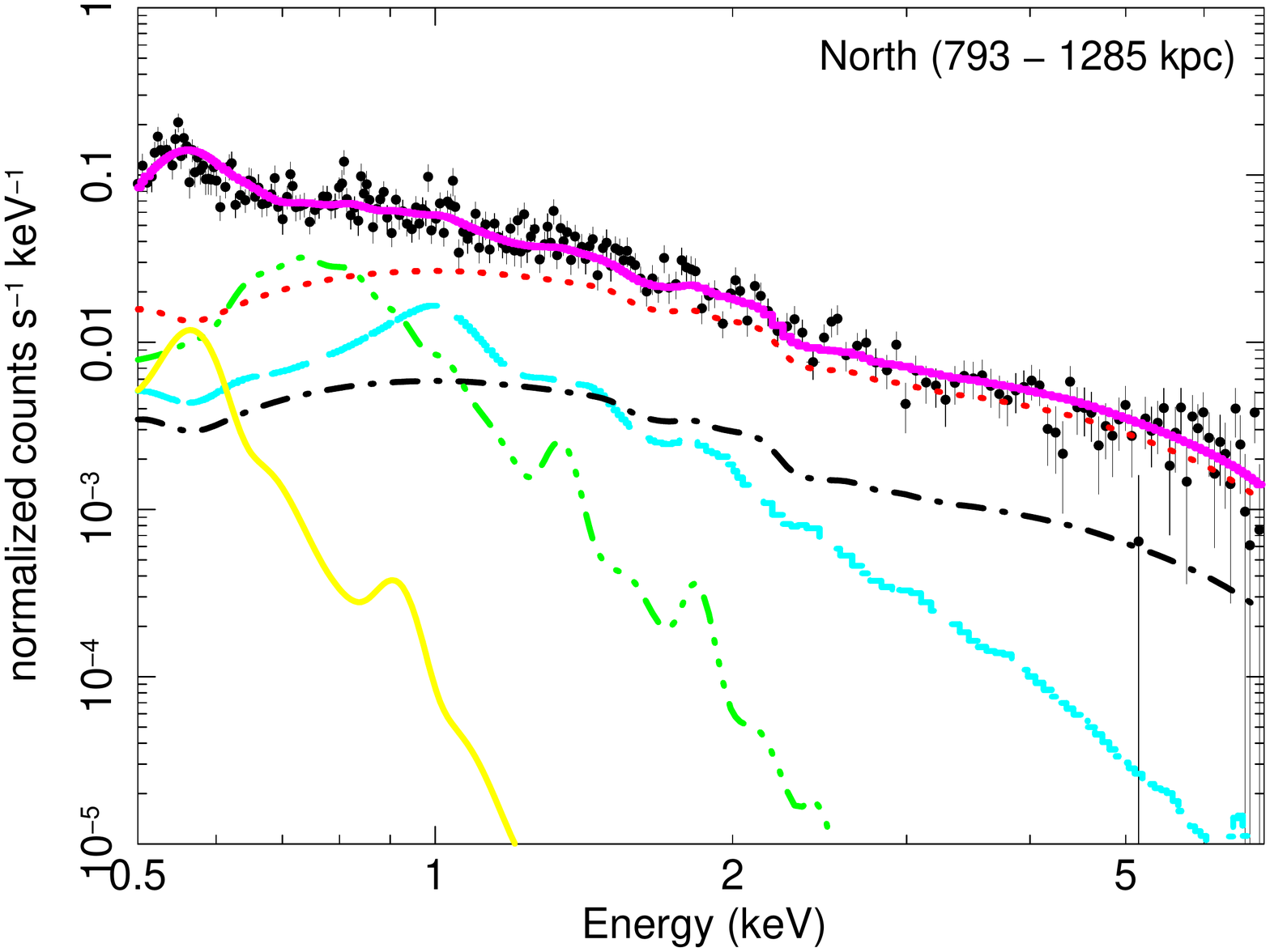}    &  \hspace{-40pt} \includegraphics[width=0.54\textwidth]{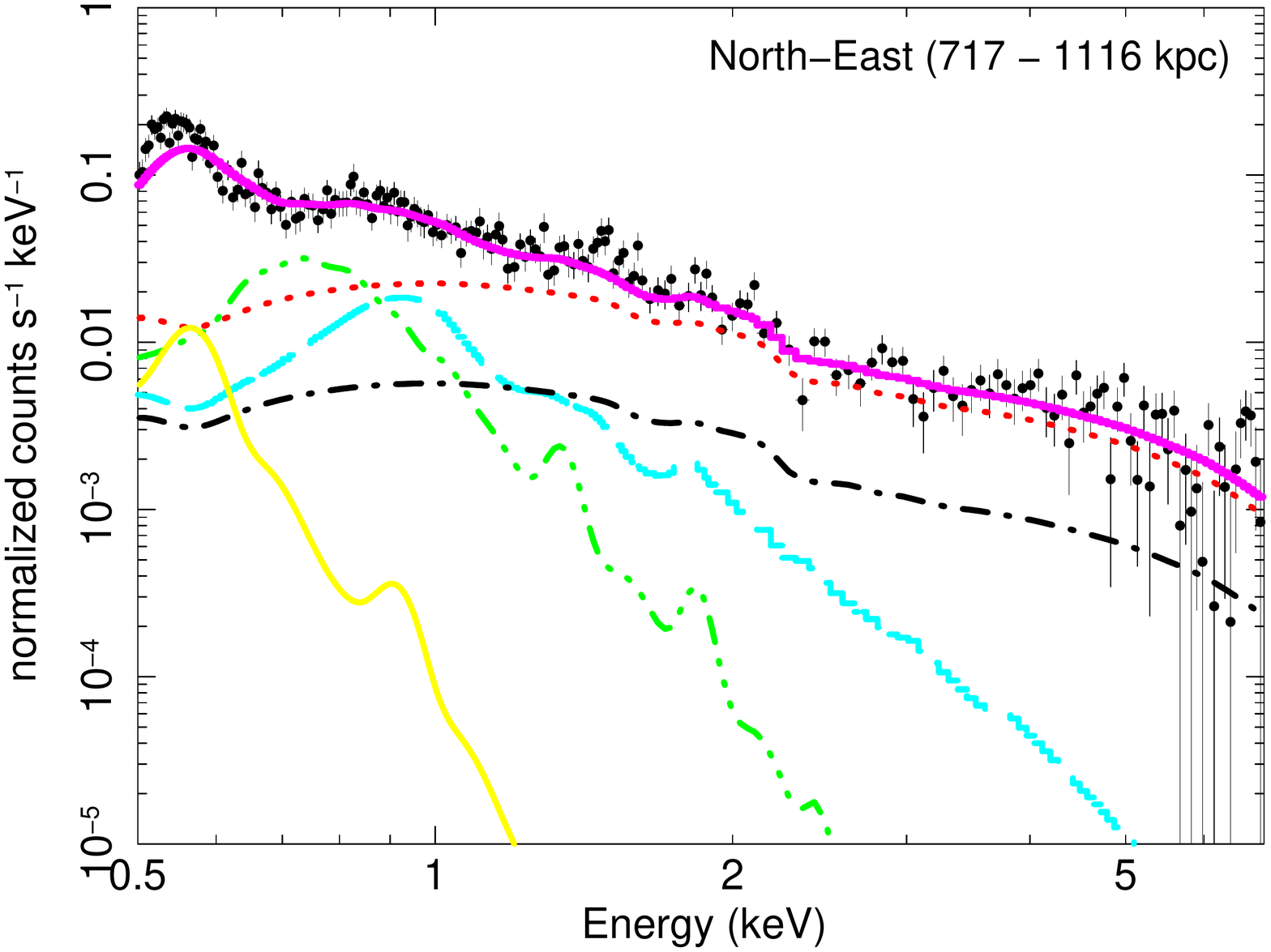}\\
\end{tabular}
\caption{$\suzaku$ XIS 1 spectra of MKW4 
for representative regions of central and two outer 
bins in north and east directions. 
Black dots are the data points.
Cyan, red, black, green, and yellow lines
are the best-fit ICM emission, 
resolved CXB, unresolved CXB,
MW,  
and LHB components, respectively.
Magenta line is the model of the best-fit ICM emission
and X-ray background components together. 
}
\label{fig:xspec}
\end{figure*}

Spectral analysis was performed using {\tt\string 
XSPEC--12.10.1} and C-statistics. 
Each spectrum was rebinned to guarantee 
20 photons
per bin to aid the convergence and 
the computational speed. 
Spectra extracted from the XIS0, XIS1, and XIS3 
were simultaneously fitted. The spectral 
fitting was restricted to the 0.4 - 7.0 keV 
energy band for XIS1, and to the 0.6 - 7.0 keV 
energy band for XIS0 and XIS3 
\citep[e.g.,][]{10.1093/pasj/59.sp1.S1}. 
We fitted each spectrum with an ICM emission 
model plus a multi-component X-ray 
background model. The ICM emission model
contains a thermal {\tt\string apec} 
component associated with a photoelectric absorption 
({\tt\string 
phabs}) component - {\tt phabs $\times$ 
apec}, as shown in Figure \ref{fig:xspec}. 
The temperature, 
abundance, and normalization of the 
{\tt\string apec} component were allowed to 
vary independently for regions at the group center
and at the intermediate radii. 
For those three regions at R$_{200}$, 
we find it necessary to fix their 
abundances at 0.2 $Z_{\odot}$. { {Similar
metallicity was observed 
at R$_{200}$ for the
RX J1159+5531 group \citep{2015ApJ...805..104S}.} 
We discuss the systematic 
uncertainties caused by this choice of abundance
in Section \ref{sec:sys}.}

\begin{figure*}
\centering
\begin{tabular}{cc}
 \includegraphics[width=0.47\textwidth]{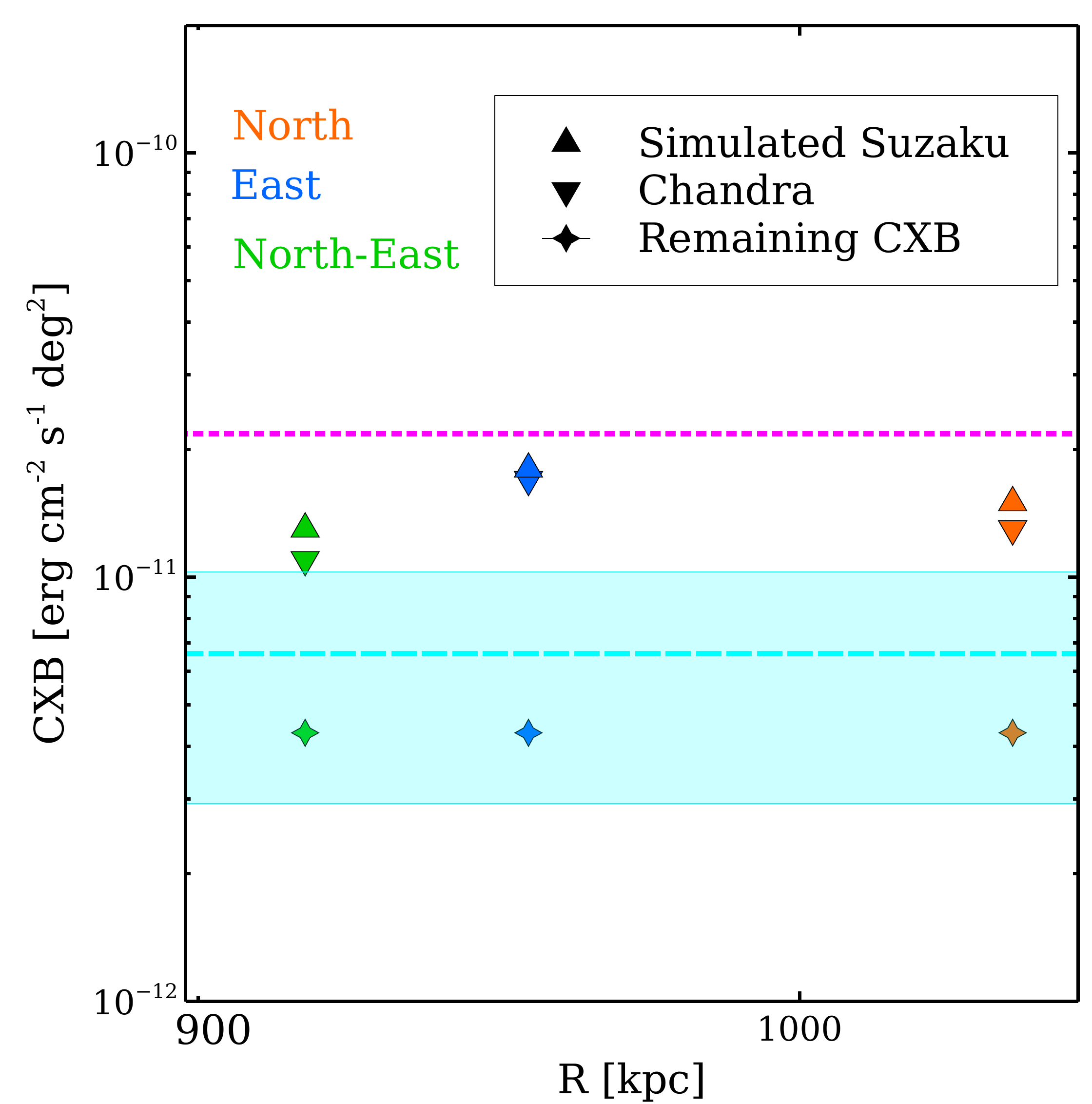}    &   
 \includegraphics[width=0.47\textwidth]{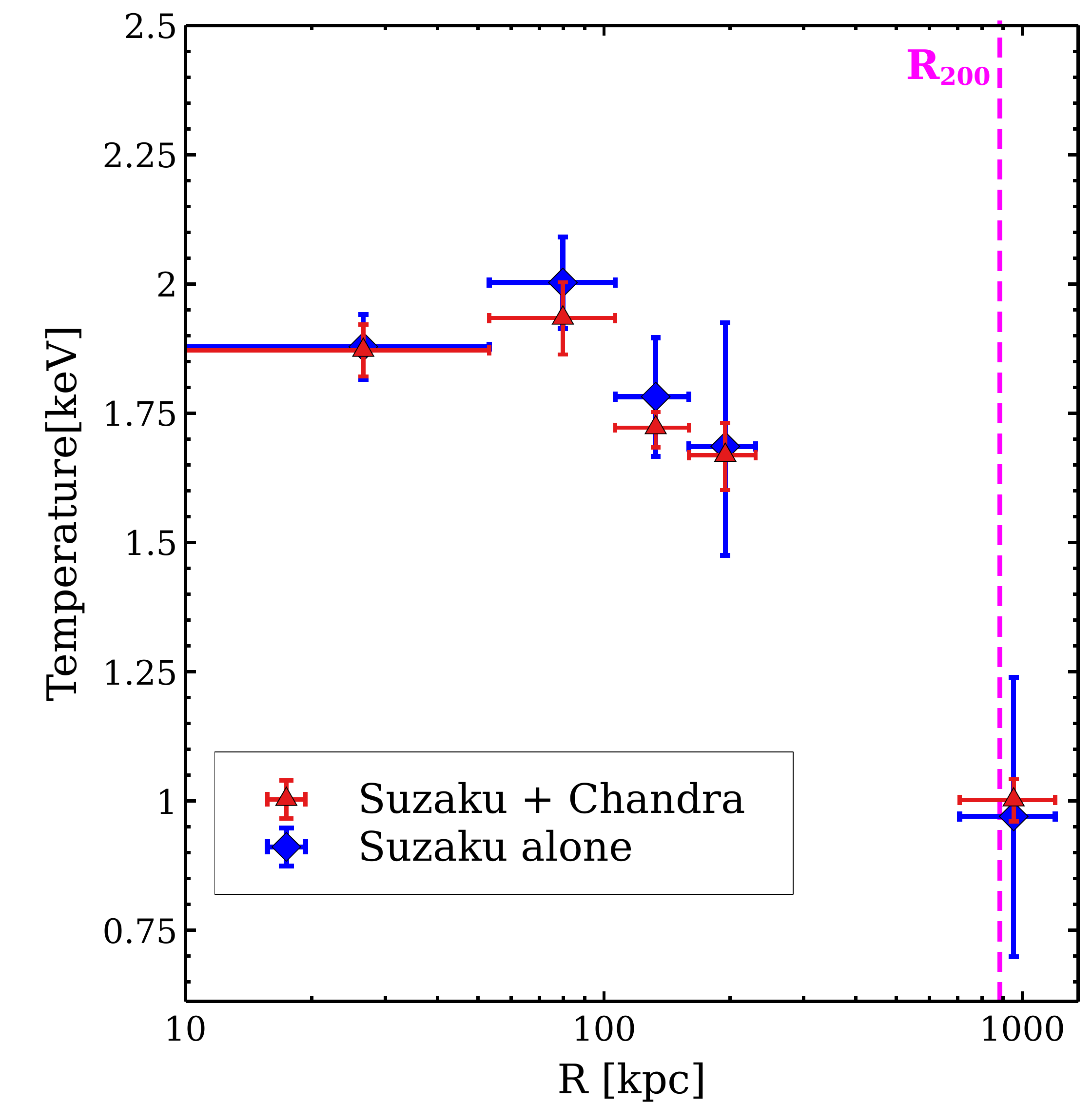}\\
\end{tabular}
\caption{{\sl Left}: Triangle down: 
surface brightness of point sources 
resolved by $\chandra$. Triangle up: surface brightness of 
point sources obtained from
mock $\suzaku$ 
sky. Magenta dotted: total surface 
brightness limit of resolved+unresolved 
point sources taken from 
\citet{2009A&A...493..501M}. 
{ Cyan dashed: expected surface brightness
limit of unresolved point sources with 1$\sigma$ 
error (shaded) calculated using Equations \ref{eqn:cxb}
and \ref{eqn:NS}.}
Star: estimated surface brightness 
of unresolved point sources from $\suzaku$ fitting. 
{\sl Right}: 
Blue: spectroscopic temperature profile of MKW4 in 
the north-east direction
obtained from spectral fitting of the $\suzaku$ data only. 
Red: same profile obtained 
from the joint fitting of 
$\suzaku$ and $\chandra$ data. 
Magenta dashed line indicates R$_{200}$ of MKW4.}
\vspace{5pt}
\label{fig:CXB}
\end{figure*}

To model the X-ray background, we adopted
{\tt\string 
phabs}$\times$({\tt\string pow$_\textrm{resolved\ 
CXB}$} + {\tt\string pow$_\textrm{unresolved\ CXB}$}
+ {\tt\string apec}$_\textrm{MW}$) + 
{\tt\string apec}$_{\textrm{LHB}}$, 
where {\tt\string pow$_\textrm{resolved\ CXB}$} 
and {\tt\string pow$_\textrm{unresolved\ CXB}$} 
are the power-law components to model the 
resolved CXB and unresolved CXB, 
respectively. The thermal {\tt\string apec}$_\textrm{MW}$
and {\tt\string apec}$_\textrm{LHB}$ represent two
foreground components to account 
for emissions from the 
Milky Way (MW) and Local Hot Bubble (LHB), 
respectively. We made use of the $\chandra$ 
observations that cover the $\suzaku$ pointings of 
the outskirts of MKW4 to mitigate 
much of the CXB contribution. 
We detected a total of 78 point 
sources with $\chandra$ { (see Figure \ref{fig:suzaku_mo})}. 
{Point sources that  
fall onto the ACIS-S1, S2, and S4 chips
are not included in our analysis.} 
The faintest point 
source was detected at a flux 
of $8.1\times 10^{-15}$ erg cm$^{-2}$ s$^{-1}$.
We converted the count rates of resolved point 
sources to the fluxes assuming a 
power-law model with a photon index of 
1.41 
\citep{2004A&A...419..837D}.
We produced mock $\suzaku$ observations for the 
resolved point sources using
{\tt\string xissim}, based on their 
positions and fluxes determined with $\chandra$. 
An exposure time of 100 ks was 
set to ensure good photon statistics. 
We extracted spectra from the mock 
$\suzaku$ observations 
using the same extraction regions used for 
the actual $\suzaku$ observations.
Figure \ref{fig:CXB} ({\sl left}) 
compares the surface 
brightnesses of the actual point sources 
resolved by $\chandra$
and that of the simulated $\suzaku$ observations,
{ {which are in good agreement
with each other.}}
We fitted the mock spectra with an absorbed 
power-law model ({\tt phabs} $\times$ {\tt pow}) 
with a photon index of 1.41. The 
resulting best-fit normalizations were used as 
the normalizations of the resolved component 
({\tt\string pow$_\textrm{resolved\ CXB}$}) 
in the background model and kept frozen. 
Figure \ref{fig:CXB} ({\sl right}) compares 
the spectral temperature profiles
(discussed in Section 
\ref{sec:result})
obtained with $\suzaku$ 
observations alone and with joint $\suzaku$ and 
$\chandra$ 
observations of MKW4 in the 
north-east direction. The significant improvement
of uncertainties demonstrates that the addition of 
$\chandra$ observations can help
to constrain the CXB and 
increase the accuracy of the measurement of ICM 
properties at the outskirts 
\citep[e.g.,][]{2012AIPC.1427...13M}.
For the regions with no $\chandra$ 
coverages (two regions of intermediate 
radii in north and 
east directions), we let the normalizations of 
{\tt\string pow$_\textrm{resolved\ CXB}$}
component to vary independently.

The normalizations of {\tt\string 
pow$_\textrm{unresolved\ 
CXB}$}
component { {were allowed to vary
collectively for 
different regions}},
assuming little 
fluctuations in the surface brightness of the
remaining unresolved point sources. We obtained a 
best-fit normalization 
of 4.2 $\times$ 10$^{-12}$ erg cm$^{-2}$ s$^{-1}$ 
deg$^{-2}$ for the unresolved
point sources in the 2.0 - 10.0 keV energy band,
as shown in Figure \ref{fig:CXB}.
The normalization of unresolved CXB can be estimated as 
\citep{2009A&A...493..501M} -
\begin{equation}\label{eqn:cxb}
    \textrm{F}_\textrm{CXB} = 2.18 \pm 0.13\times 10^{-11} - 
    \int_{\textrm{s}_\textrm{min}}^{\textrm{s}_\textrm{max}}{\left(\frac{\textrm{dN}}{\textrm{dS}}
    \right)\times \textrm{SdS}},
\end{equation}
in units of erg cm$^{-2}$ s$^{-1}$ deg$^{-2}$, 
assuming a total CXB surface brightness of 
$2.18\pm0.13\times 10^{-11}$ erg cm$^{-2}$ 
s$^{-1}$ deg$^{-2}$ in the 2.0 - 10.0 keV energy 
band. 
We adopted an analytical form of the integral 
source flux distribution from \citet{2003ApJ...588..696M} -
\begin{equation}\label{eqn:NS}
    \textrm{N}(> \textrm{S}) = 
    \textrm{N}_\textrm{S(H)}\left[\frac{(2\times10^{-15})^{\alpha_{1,\textrm{S(H)}}}}{\textrm{S}^{\alpha_{1,\textrm{S(H)}}}+\textrm{S}_\textrm{0,H}^{\alpha_{1,\textrm{S(H)}}-\alpha_{2,\textrm{S(H)}}}\textrm{S}^{\alpha_{2,\textrm{S(H)}}}} \right],
\end{equation}
where $\alpha_{1,\textrm{S(H)}}=1.57^{+0.10}_{-0.08}
$, 
$\alpha_{2,\textrm{S(H)}}=0.44^{+0.12}_{-0.13}$, 
$\textrm{S}_\textrm{{0,H}}=4.5^{+3.7}_{1.7}\times10^
{-15}$ erg 
cm$^{-2}$ s$^{-1}$, and 
$\textrm{N}_\textrm{S(H)}=5300^{+2850}_{-1400}$ are 
the best-fit 
parameters with a 68\% confidence level. 
We estimated the flux limit for unresolved point 
sources to be 
$6.61\pm3.68\times10^{-12}$ erg cm$^{-2}$ s$^{-1}$ 
deg$^{-2}$. This flux is consistent with the flux
obtained from spectral fitting.

To constrain the temperature and surface brightness of
Local Hot Bubble ({\tt\string 
apec}$_\textrm{LHB}$) and Milky Way  
({\tt apec}$_\textrm{MW}$) foregrounds, we used the ROSAT 
All-Sky Survey (RASS) data in an annulus 
region of 0.9$^\circ$ - 1.2$^\circ$ from the group 
center where no group emission 
was expected.  
We extracted spectra using the HEASARC X-ray 
background tool\footnote[3]{\url{http://heasarc.gsfc.nasa.gov/cgi-bin/Tools/xraybg/xraybg.pl}}. 
The RASS spectrum was fitted simultaneously 
with the $\suzaku$ data. The background 
fitting results are shown in Figure 
\ref{fig:xspec} and listed
in Table \ref{tab:norm}.

\begin{table}
\caption{X-ray background components of outer regions$^\textrm{a}$}
\begin{tabular}{llllclll}
\hline
Name \ \ \ \ \ \ \ \ & CXB$^\textrm{b}$ \ \ \ \ \ \ \ \ & LHB$^\textrm{c}$ \ \ \ \ \ \ \ \ & MW$^\textrm{d}$\\
\hline
$$\suzaku$$ East & 11.25$^{+0.12}_{-0.13}$ & 13.9$^{+1.1}_{-1.3}$ & 6.21$^{+0.32}_{-0.35}$ \\\\ 
$$\suzaku$$ North & 9.51$^{+0.12}_{-0.12}$ & \ \ \ \ $-$ & \ \ \ \ $-$\\\\
$$\suzaku$$ North-East & 8.63$^{+0.12}_{-0.12}$ & \ \ \ \ $-$ & \ \ \ \ $-$\\\\
\hline
\end{tabular}
\\
$^\textrm{a}$ Results for the normalizations
of various components (non-instrumental) of 
X-ray background for a circular region with 20\arcmin\ 
radii. The LHB and MW components for $\suzaku$ North and 
$\suzaku$ North-East directions are linked to the $\suzaku$ East.\\
$^\textrm{b}$ Normalization of power-law component 
($\Gamma$=1.41) for resolved + unresolved cosmic 
X-ray background in the units of 10$^{-4}$ photons 
s$^{-1}$ cm$^{-2}$ 
keV$^{-1}$ at 1 keV.\\
$^\textrm{c}$ Normalization of the unabsorbed {\tt apec} 
thermal component (kT = 0.08 keV, abun= 
1Z$_{\odot}$) integrated over line of sight, 
$1/4\pi[\textrm{D}_\textrm{A}(1+\textrm{z})]^{2}\int{\textrm{n}_\textrm{e}\textrm{n}_\textrm{H}\ \textrm{dV}}$ 
in the units of 10$^{-18}$ cm$^{-5}$.\\
$^\textrm{d}$ Normalization of the absorbed {\tt apec} 
thermal component (kT = 0.2 keV, abun= 1Z$_{\odot}$)
integrated over line of sight, 
$1/4\pi[\textrm{D}_\textrm{A}(1+\textrm{z})]^{2}\int{\textrm{n}_\textrm{e}\textrm{n}_\textrm{H}\ \textrm{dV}}$ in 
the units of 10$^{-18}$ cm$^{-5}$.\\
\label{tab:norm}

\end{table}

\subsection{Chandra}

MKW4 was observed using $\chandra$ with 
one ACIS-S pointing at the group center and 
three ACIS-I pointings overlapping with the 
outer regions observed with $\suzaku$ . The observation 
logs are listed in Table 
\ref{tab:obs_log}. 

\subsubsection{Data reduction}
The $\chandra$ data reduction was performed 
using HEAsoft 6.25, CIAO-4.11, and a $\chandra$ 
calibration database (CALDB 4.8.3).
We followed a standard data reduction thread 
\footnote[4]{\url{http://cxc.harvard.edu/ciao/threads/index.html}}. 
All data were reprocessed from the level 1 
events using {\tt\string chandra$\_$repro},  
which applied the latest gain, charge transfer 
inefficiency correction, and filtering 
from bad grades. 
The light curves were filtered using 
{\tt\string lc$\_$clean} script 
to remove periods affected by flares. 
The resulting 
filtered exposure times are listed in 
Table \ref{tab:obs_log}. Point 
sources were identified with {\tt\string wavdetect} 
using a range of wavelet radii between 1 and 16 
pixels to maximize the number of detected point sources.
The detection 
threshold was set to $10^{-6}$, 
which guaranteed detection of
$\lesssim$ 1 spurious source per CCD. Detected point
sources were confirmed visually. Figure 
\ref{fig:suzaku_mo} shows the adaptively smoothed 
ACIS-S3 image of MKW4 in the 0.5 - 2.0 keV energy 
band.
 
We extracted spectra from 6 adjoining, 
concentric annuli centered at the X-ray 
centroid in the $\chandra$ ACIS-S3 chip.
The widths of the annular regions were chosen to 
contain approximately the same number of background-subtracted 
counts of 2000 for better spectral 
analysis. 
Count-weighted spectral response matrices 
(ARFs and RMFs) were generated using 
{\tt\string mkwarf} and {\tt\string mkacisrmf} 
tasks for each annulus. The blanksky background
was produced by the {\tt blanksky} tool for 
background subtraction. The blanksky background 
was tailored based on the count rate 
in the 9.5 - 12.0 keV energy band relative to the 
observation.

\subsubsection{Spectral analysis}

All spectra were fitted simultaneously to model 
the ICM properties of different regions. We modeled 
the ICM
emission with a single thermal
{\tt\string apec} model 
associated with a photoelectric absorption model- 
{\tt phabs $\times$ 
apec}. Photoionization cross-sections were taken 
from 
\citet{1992ApJ...400..699B}. We 
obtained the galactic hydrogen column density, 
N$_\textrm{H}$ = 1.72 $\times 10^{20}$\ 
cm$^{-2}$ at the direction 
of MKW4, using HEASARC N$_\textrm{H}$ tool 
\footnote[8]{\url{http://heasarc.gsfc.nasa.gov/cgi-bin/Tools/w3nh/w3nh.pl}}. 
All ICM emission components were allowed 
to vary independently. 
\begin{figure*}
\begin{tabular}{ccc}
\hspace{-10pt} 
\includegraphics[width=0.33\textwidth]{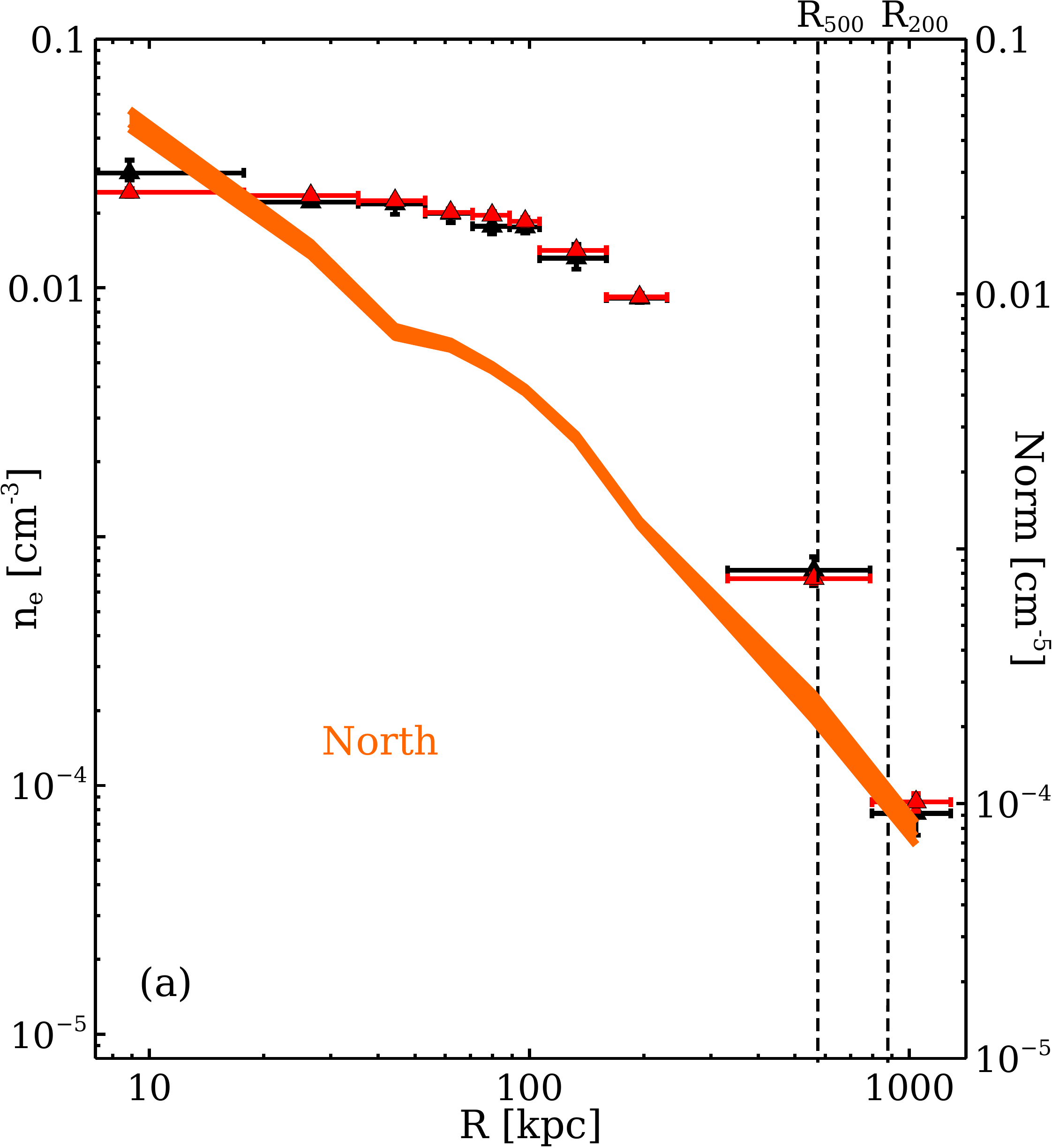}    &  \hspace{-10pt} 
\includegraphics[width=0.33\textwidth]{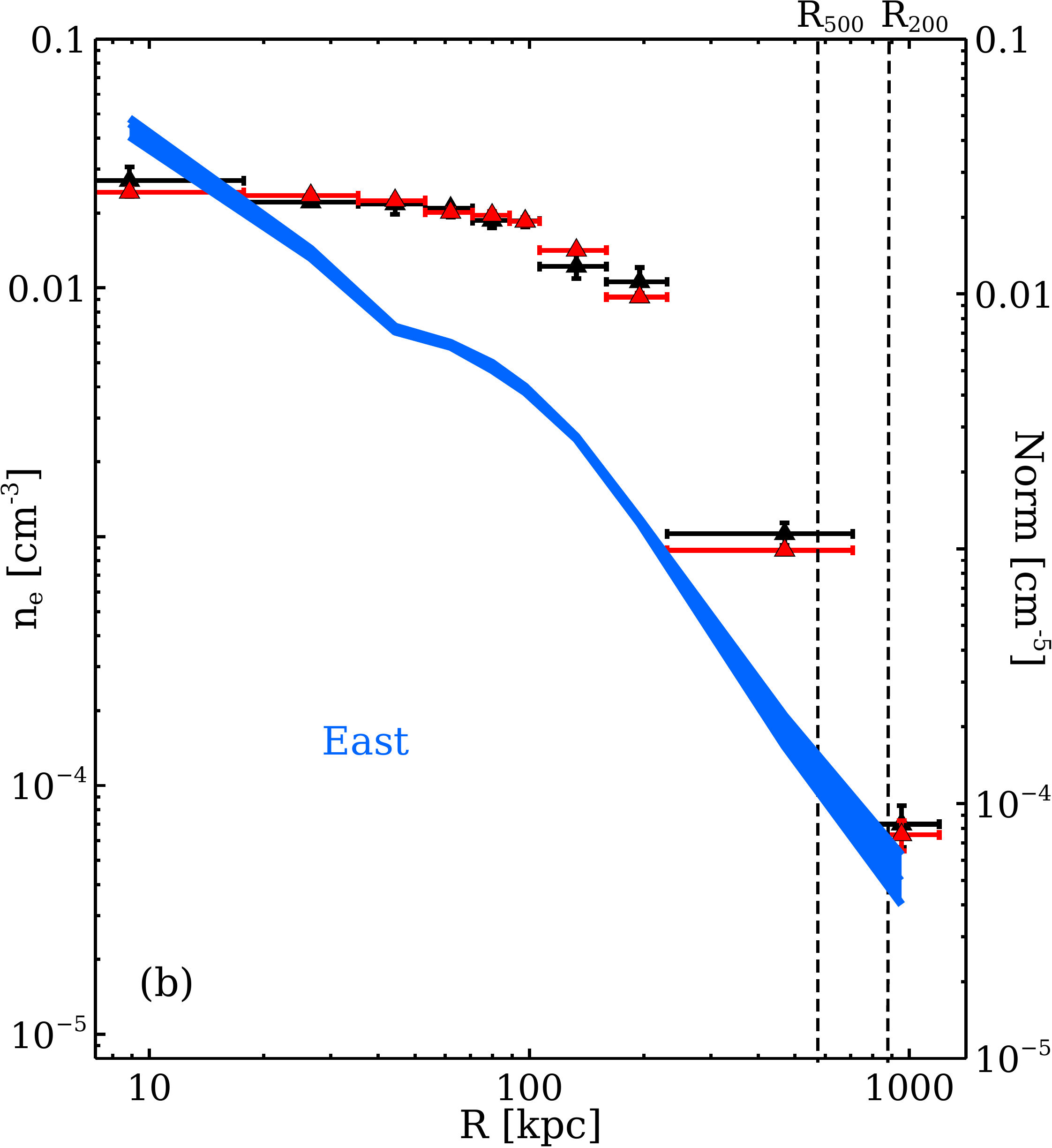} & \hspace{-10pt}
\includegraphics[width=0.33\textwidth]{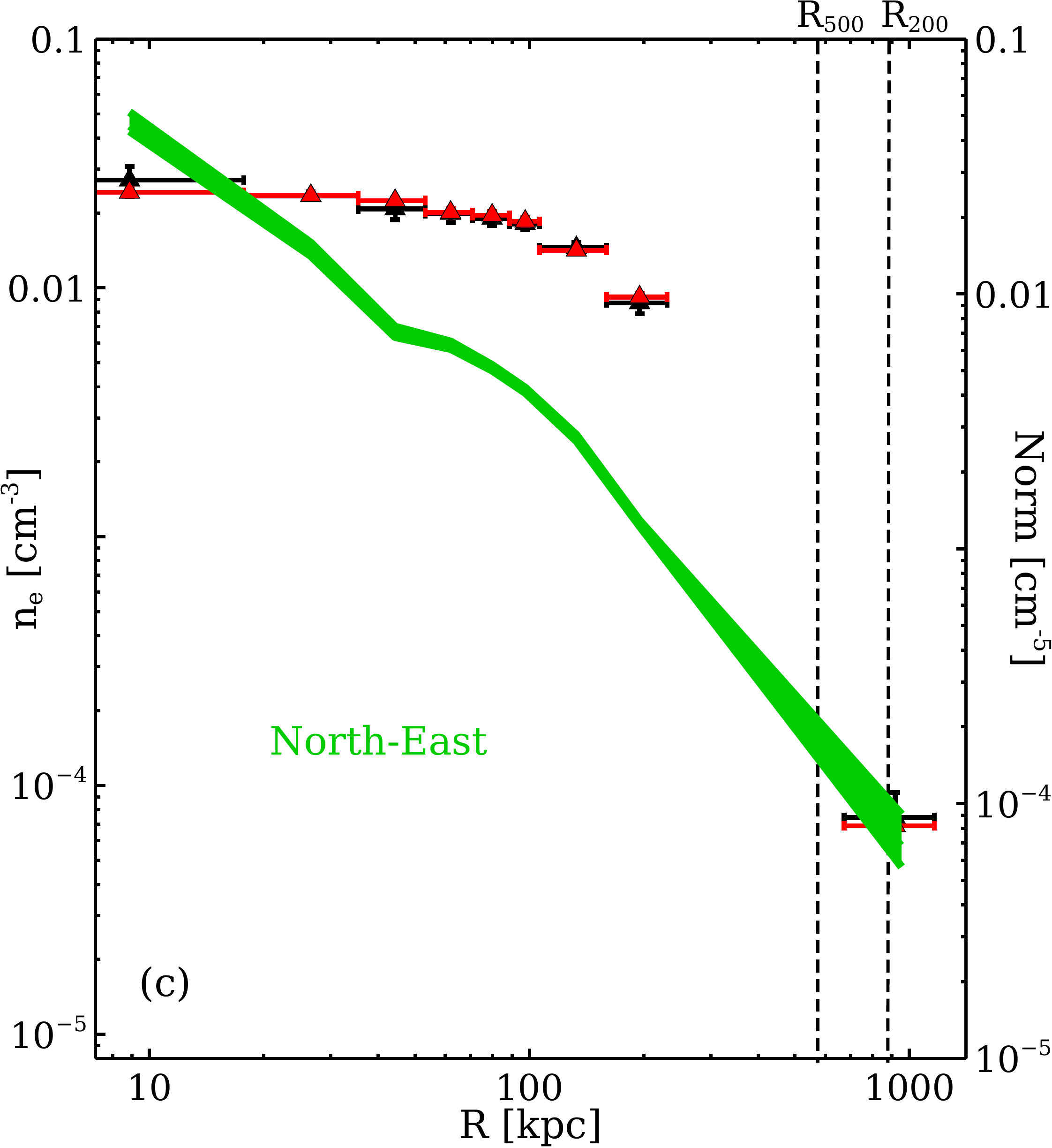}\\
\end{tabular}
\caption{Deprojected density profiles of 
MKW4 in north (orange), east (blue), and 
north-east (green) directions. 
Shaded regions indicate 
1$\sigma$ uncertainties. {Red triangles: the 
best-fit normalizations for the {\tt apec} 
thermal component obtained from 
spectral analysis. Black triangles: the 
normalizations of the {\tt apec} thermal component 
calculated from the resulting 3D density 
profile.}}
\label{fig:den}
\end{figure*}
\begin{figure*}
\begin{tabular}{ccc}
\hspace{-10pt} 
\includegraphics[width=0.33\textwidth]{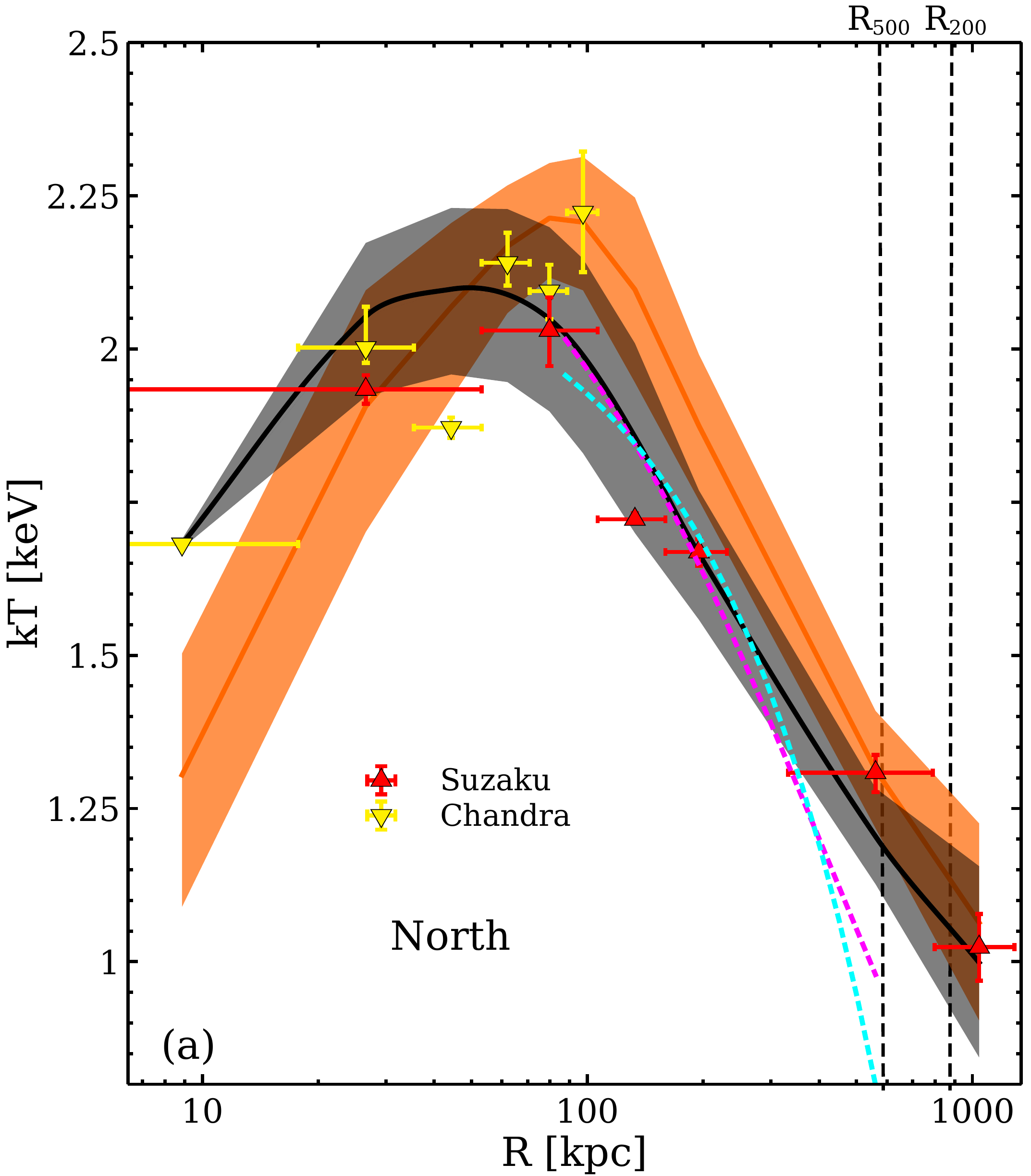}    & \hspace{-10pt}  
\includegraphics[width=0.33\textwidth]{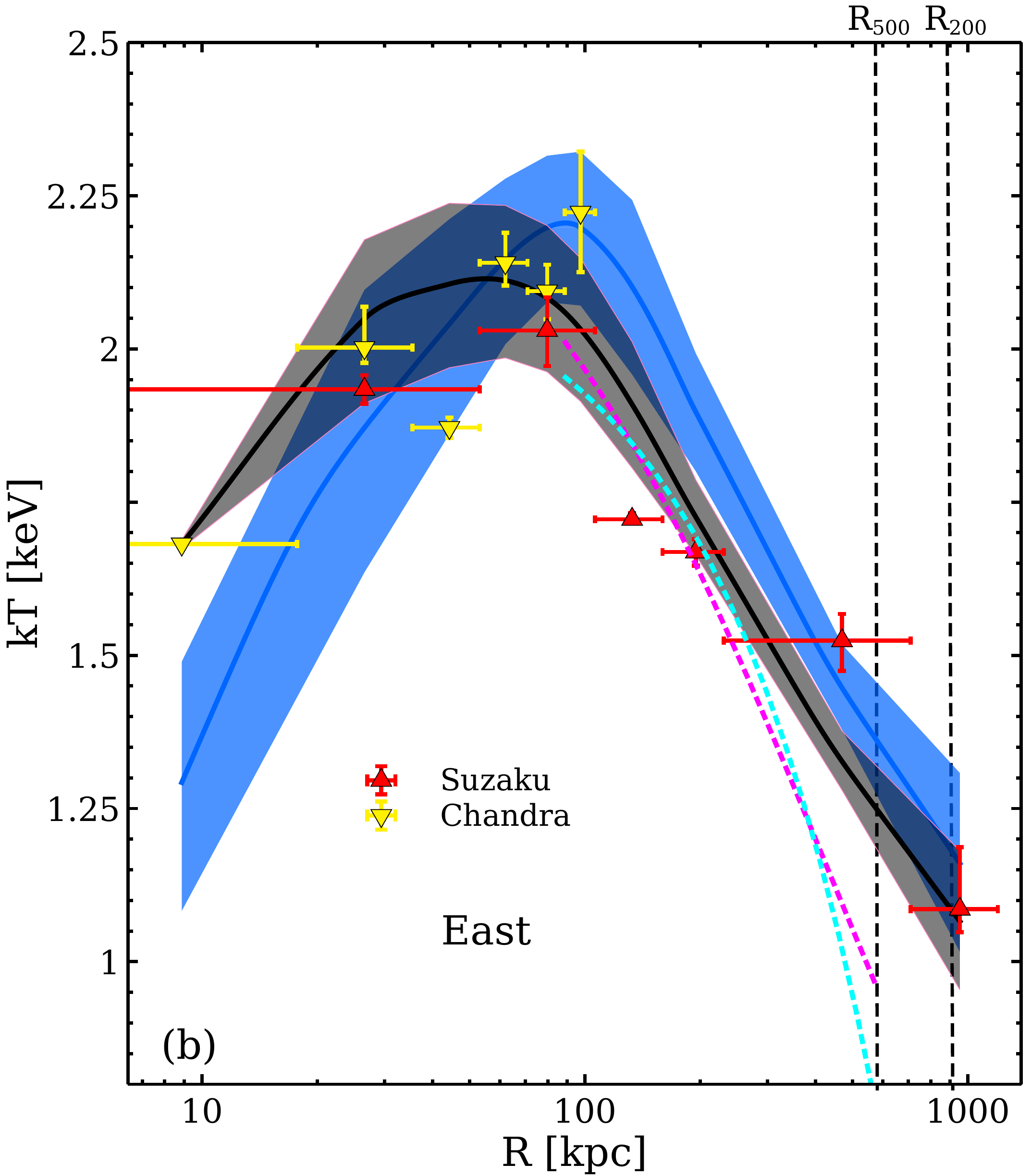} &
\hspace{-10pt} 
\includegraphics[width=0.33\textwidth]{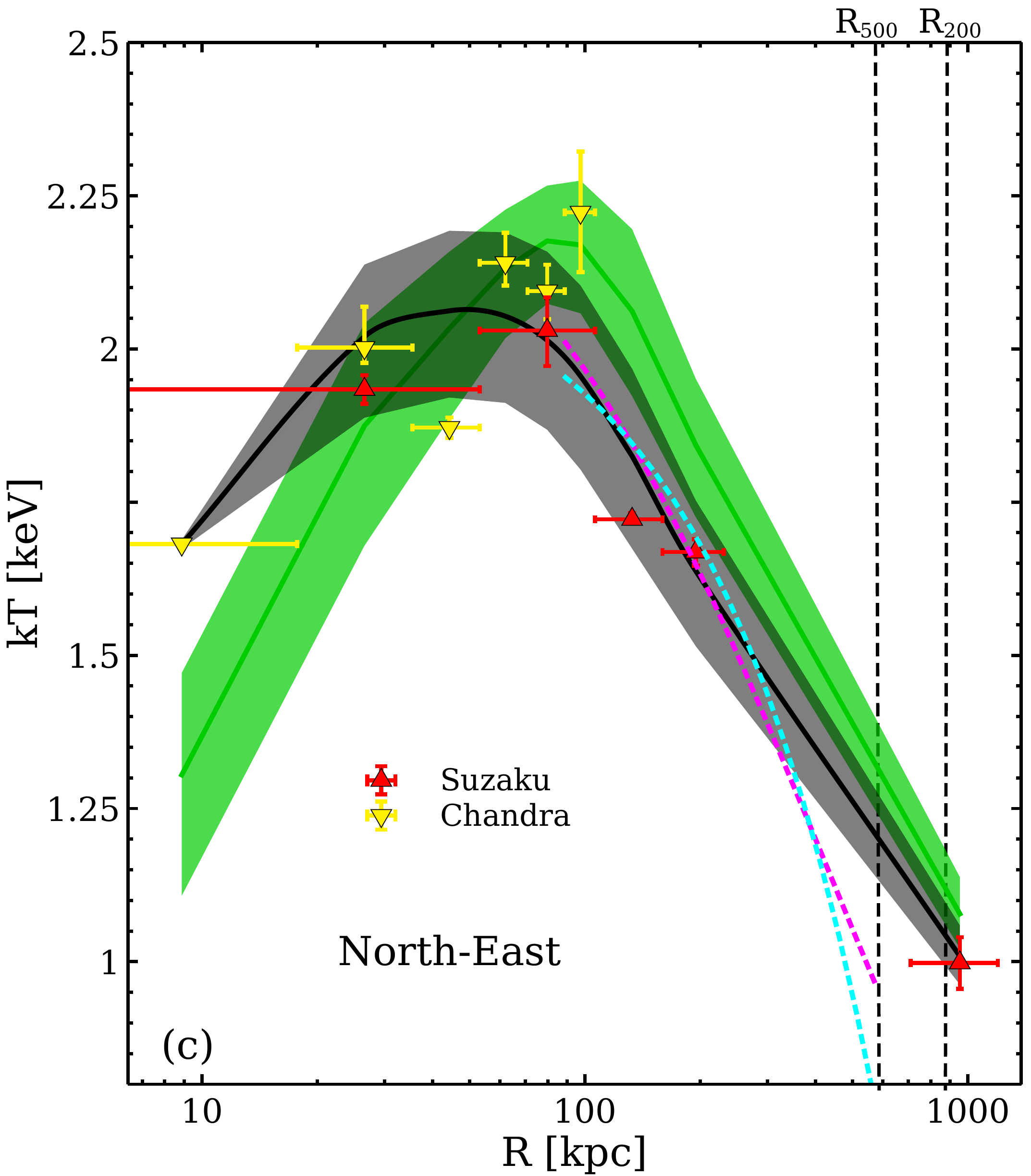}\\
\end{tabular}
\caption{Deprojected temperature 
profiles of MKW4 in 
north (orange), east (blue), and north-east 
(green) directions.  
The 2D temperature profiles of MKW4, 
estimated using Equation \ref{eqn:temp1}, 
are indicated in black. Shaded regions
indicate 1$\sigma$ uncertainties.
Temperatures obtained from the {\tt apec} 
thermal component by fitting the spectra 
extracted from the 
$\chandra$ and $\suzaku$ data are shown 
in yellow and red respectively. 
{ Magenta 
and cyan dashed lines show the average temperature 
profiles derived for groups in
\citet{2002ApJ...579..571L} and 
\citet{Sun_2009} respectively.}
}
\label{fig:temp}
\end{figure*}
\section{Results} \label{sec:result}
\subsection{Density and temperature profile}
We derive the 
deprojected (3D) gas 
density and temperature profiles of MKW4
assuming analytical prescriptions for 
3D gas density and
temperature. { We follow the
analytical expressions described in \citet{Vikhlinin_2006}}. 
The  normalization of the {\tt apec} thermal
component relates to the ICM density as: 
norm = $\int{\textrm{n}_\textrm{e}
\textrm{n}_\textrm{p}\ \textrm{dl}}$, 
where the density profile of electrons 
($\textrm{n}_\textrm{e}$) and 
protons ($\textrm{n}_\textrm{p}$) can analytically
be described as:
\begin{equation}\label{eqn:den}
\begin{split}
{    \textrm{n}_\textrm{e}\textrm{n}_\textrm{p}{\rm(r)}} &{= 
    \frac{\textrm{n}_\textrm{01}^2}{\left(1+\textrm
    {r}^2/\textrm{r}_\textrm{c1}^2\right)^{3\
    \beta_{1}}} 
    +\frac{\textrm{n}_\textrm{02}^2}{\left(1+\textrm
    {r}^2/\textrm{r}_\textrm{c2}^2\right)^{3\
    \beta_{2}}}.
}
    \end{split}
\end{equation}
{ We obtain the 3D density profile of MKW4 by  
projecting the 3D analytic model given in 
Equation \ref{eqn:den} along the line of sight and 
fit it to the measured 2D normalizations 
of thermal {\tt apec} 
component obtained from the spectral 
analysis, assuming n$_{\rm e}$ = 1.2n$_{\rm p}$ 
and a fixed $\gamma$ = 
3. {The uncertainties are
calculated using Monte Carlo simulations with
1000 to 2000 realizations.}}
\begin{figure*}
\begin{tabular}{cc}
\hspace{-10pt} 
\includegraphics[width=0.4\textwidth]{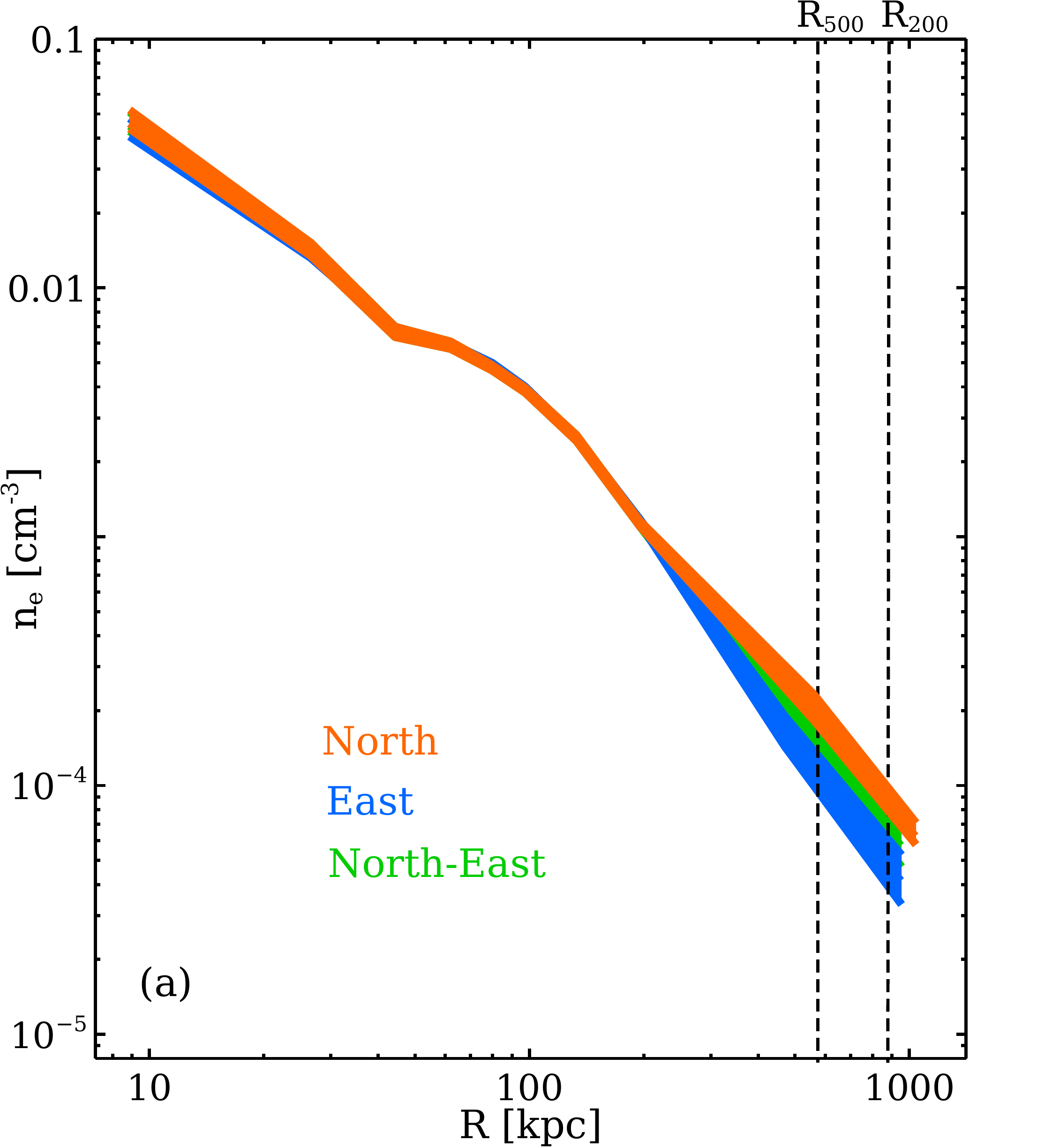}    & \hspace{10pt}  
\includegraphics[width=0.4\textwidth]{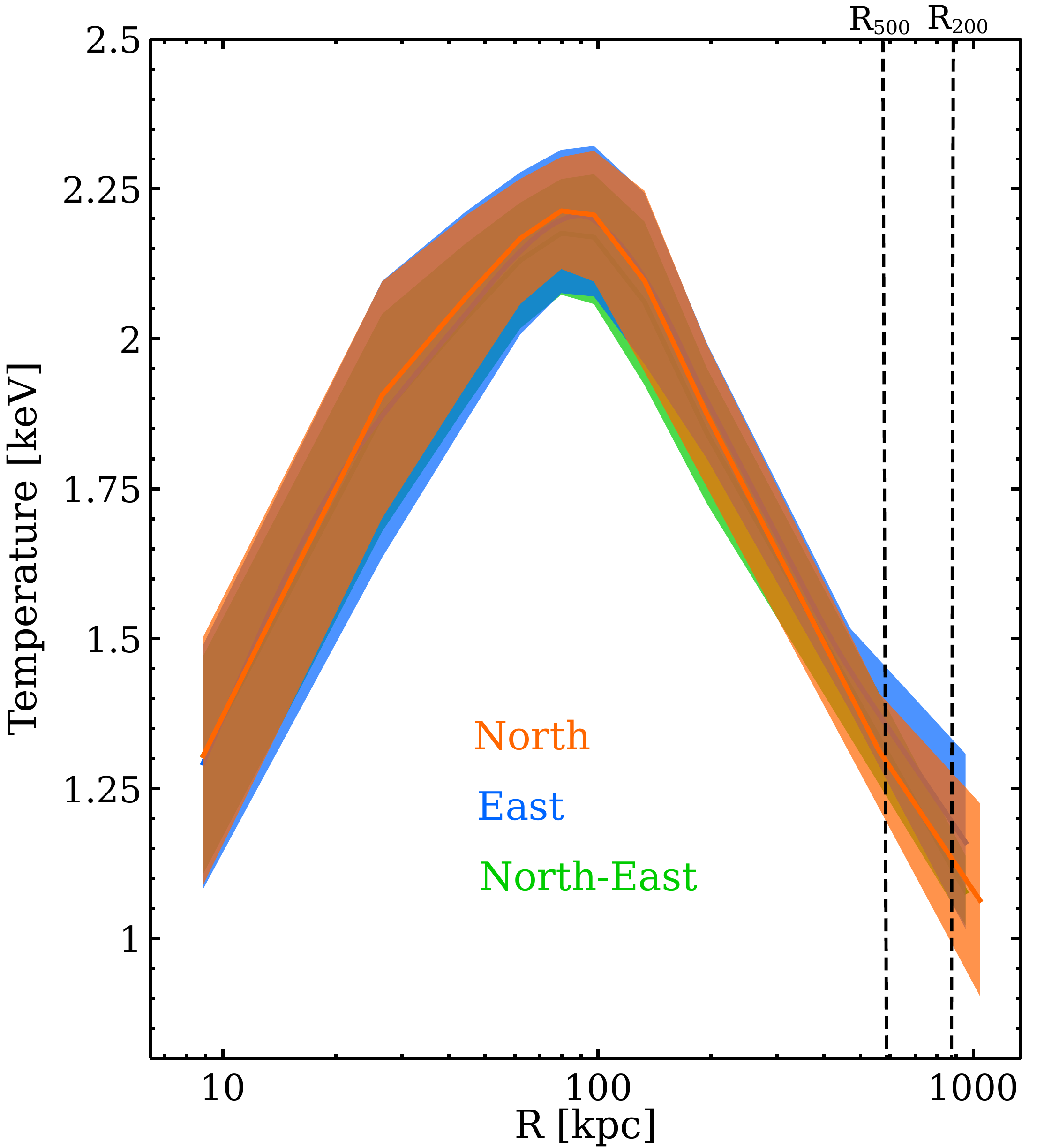}\\
\end{tabular}
\caption{{ 3D density (left) and temperature (right) profiles
of MKW4 in three different directions.}}
\label{fig:compden}
\end{figure*}

The resulting 3D density profiles are shown in 
Figure \ref{fig:den}.
A comparison between the 
normalizations obtained from spectral 
analysis and the normalizations obtained from 
the Equation \ref{eqn:den} are 
also shown in Figure \ref{fig:den}. 
{ Figure \ref{fig:compden} compares the 3D density
profiles of MKW4 in three different directions.}

We derive the 3D 
temperature profile by fitting the  
spectral temperature (2D spectroscopic temperature, 
T$_\textrm{2D}$) obtained from thermal {\tt apec} 
component with the following analytic function \citep{Vikhlinin_2006}:
\begin{equation}\label{eqn:temp1}
    \textrm{T}_\textrm{2D} = 
    \frac{\int{\textrm{n}_\textrm{e}^2\ 
    \textrm{T}_\textrm{3D}^{\frac{1}{4}}\
    \textrm{dV}}}{\int{\textrm{n}_\textrm{e}^2\ 
    \textrm{T}_\textrm{3D}^{-\frac{3}{4}}\
    \textrm{dV}}},
\end{equation}
where
\begin{equation}\label{eqn:temp2}
\begin{split}
    \textrm{T}_\textrm{3D}(\textrm{r}) &= 
    \textrm{T}_0\frac{\left(\textrm{x}+\textrm{T}_\textrm{min}/\textrm{T}_0\right)}{\left(\textrm{x}+1\right)}\ \times\ 
    \frac{\left(\textrm{r}/\textrm{r}_\textrm{t}\right)^{-\textrm{a}}}{\left(1+\textrm{r}^\textrm{b}/\textrm{r}_\textrm{t}^\textrm{b}
    \right)^{\textrm{c}/\textrm{b}}},\\ 
    & \textrm{and}\ \ \textrm{x} = 
    \left(\frac{\textrm{r}}{\textrm{r}_\textrm{cool}}\right)^{\textrm{a}_\textrm{cool}}.
    \end{split}
\end{equation}
Figure \ref{fig:temp} represents the projected (T$_\textrm{2D}$) 
and deprojected (T$_\textrm{3D}$) temperature
profiles of MKW4 from its center out to 
the virial radii in 
north, east, and north-east directions. 

{ 
{The uncertainties in 
temperatures are also
calculated using Monte Carlo simulations.}
The deprojected temperature profiles of MKW4 
decline from 2.20 keV
at 0.1R$_{200}$ 
to 1.14
keV at R$_{200}$. 
A similar temperature drop was found 
in other groups 
\citep[e.g.,][]{2015ApJ...805..104S}. 
The projected temperatures measured with 
$\suzaku$ and $\chandra$ 
are consistent for the overlapping regions.
{ Figure \ref{fig:compden} compares the 3D temperature
profiles of MKW4 in three different directions.}



{Our temperature profile agrees well with 
the $\chandra$ temperature 
profiles of MKW4 out to
R$_{2500}$, as given by \citet{Vikhlinin_2006} 
(V06)
and \citet{Sun_2009} (S09) as well as the 
measurement obtained with $\xmm$ by 
\citet{2007ApJ...669..158G} (G07)
out to $\sim$ 0.6R$_{500}$. We measure a gas temperature
at R$_{500}$ between 
those of V06
and S09 and consistent with the
$\suzaku$ measurement
using the two intermediate pointings 
\citep{2014ApJ...781...36S}.}
We compare the projected temperature profiles of MKW4
to the {empirical} profiles
derived for groups by \citet{Sun_2009}:
\begin{equation}\label{eqn:temp3}
    \frac{\textrm{T}}{\textrm{T}_\textrm{2500}} = (1.22\ \pm\ 0.02) - (0.79\ \pm\ 0.04)\frac{\textrm{R}}{\textrm{R}_{500}}\ ,
\end{equation}
and by \citet{2002ApJ...579..571L}:
\begin{equation}\label{eqn:temp4}
    \frac{\textrm{T}}{\textrm{T}_\textrm{2500}} = (1.37\ \pm\ 0.03) - \left(1 + \frac{\textrm{R}}{\textrm{R}_{500}}\right)^{-(1.34\ \pm\ 0.21)}\ ,
\end{equation} 
{as shown in Figure \ref{fig:temp}.
Both profiles show good agreement with the projected
temperature
profile of MKW4 
out to $\sim$ R$_{2500}$. 
Their values at R$_{500}$ exceed what is measured for 
MKW4 by 30\% and 60\%, 
respectively. 
These empirical profiles are determined with a large 
number
of galaxy groups with a sizable scatter. Many of them
do not have their temperatures actually measured 
out to R$_{500}$.}}

\subsection{Entropy and pressure profile}
\begin{figure*}
\begin{tabular}{ccc}
\hspace{-10pt} 
\includegraphics[width=0.33\textwidth]{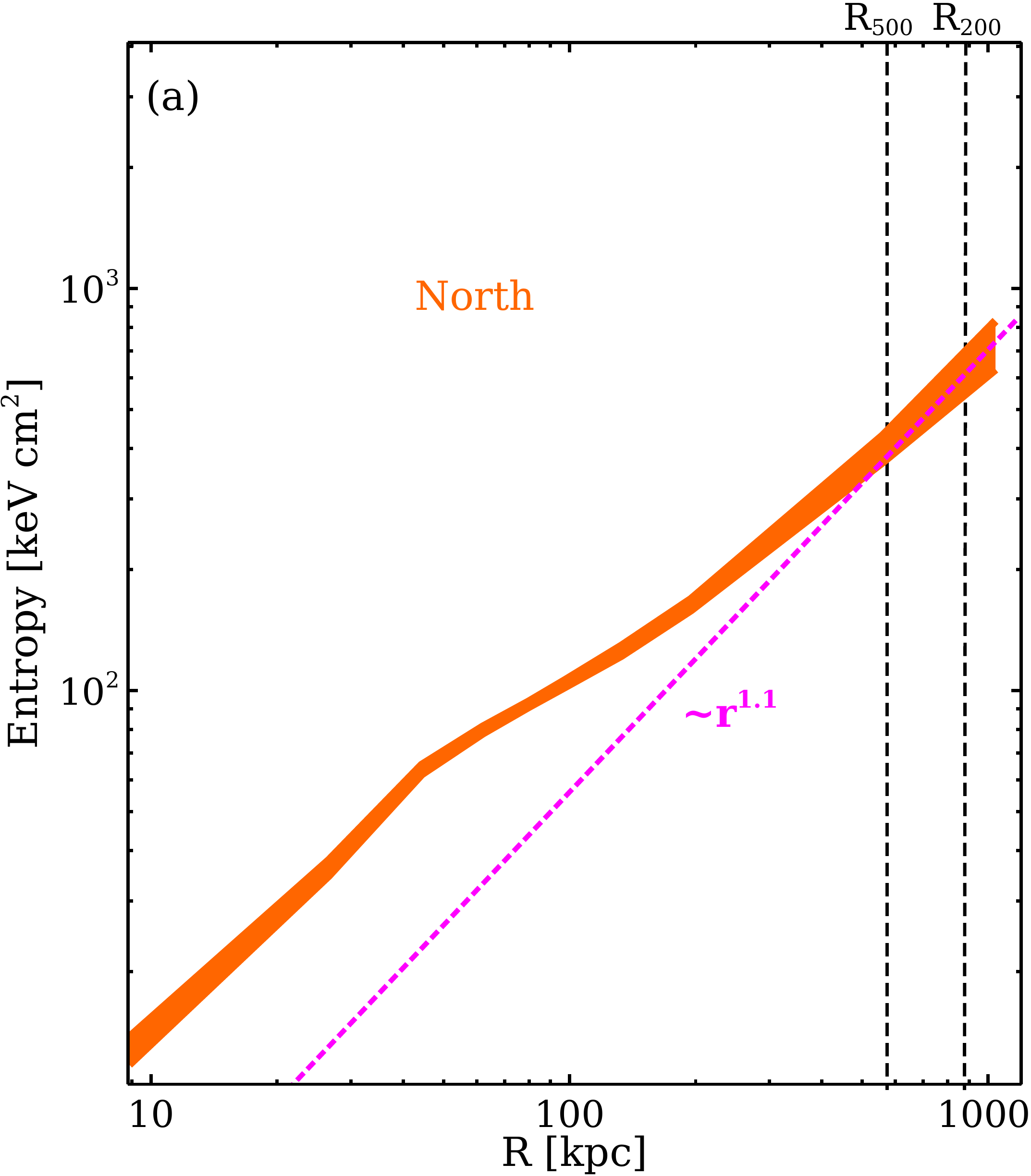}    & \hspace{-10pt}  
\includegraphics[width=0.33\textwidth]{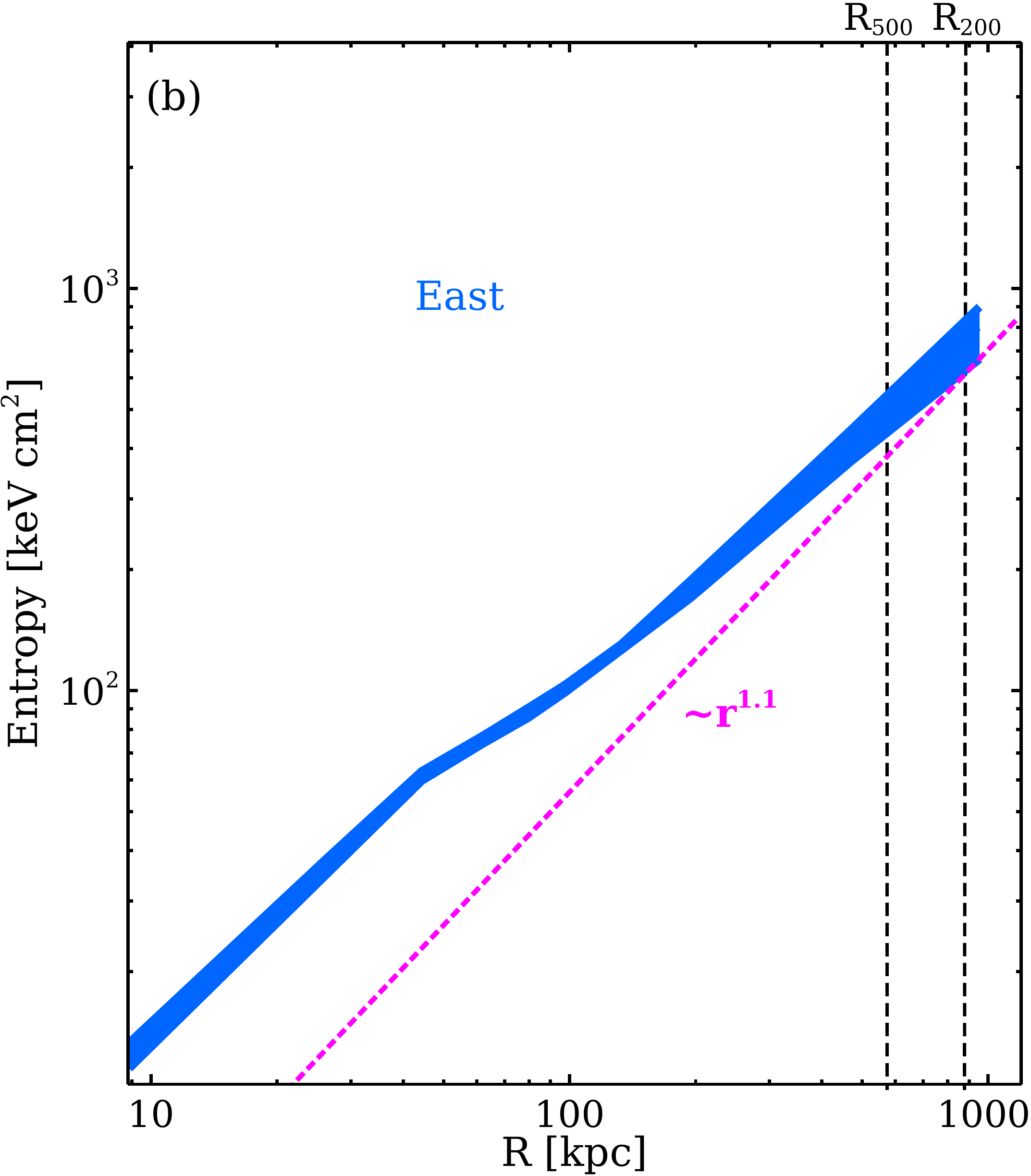} &
\hspace{-10pt} 
\includegraphics[width=0.33\textwidth]{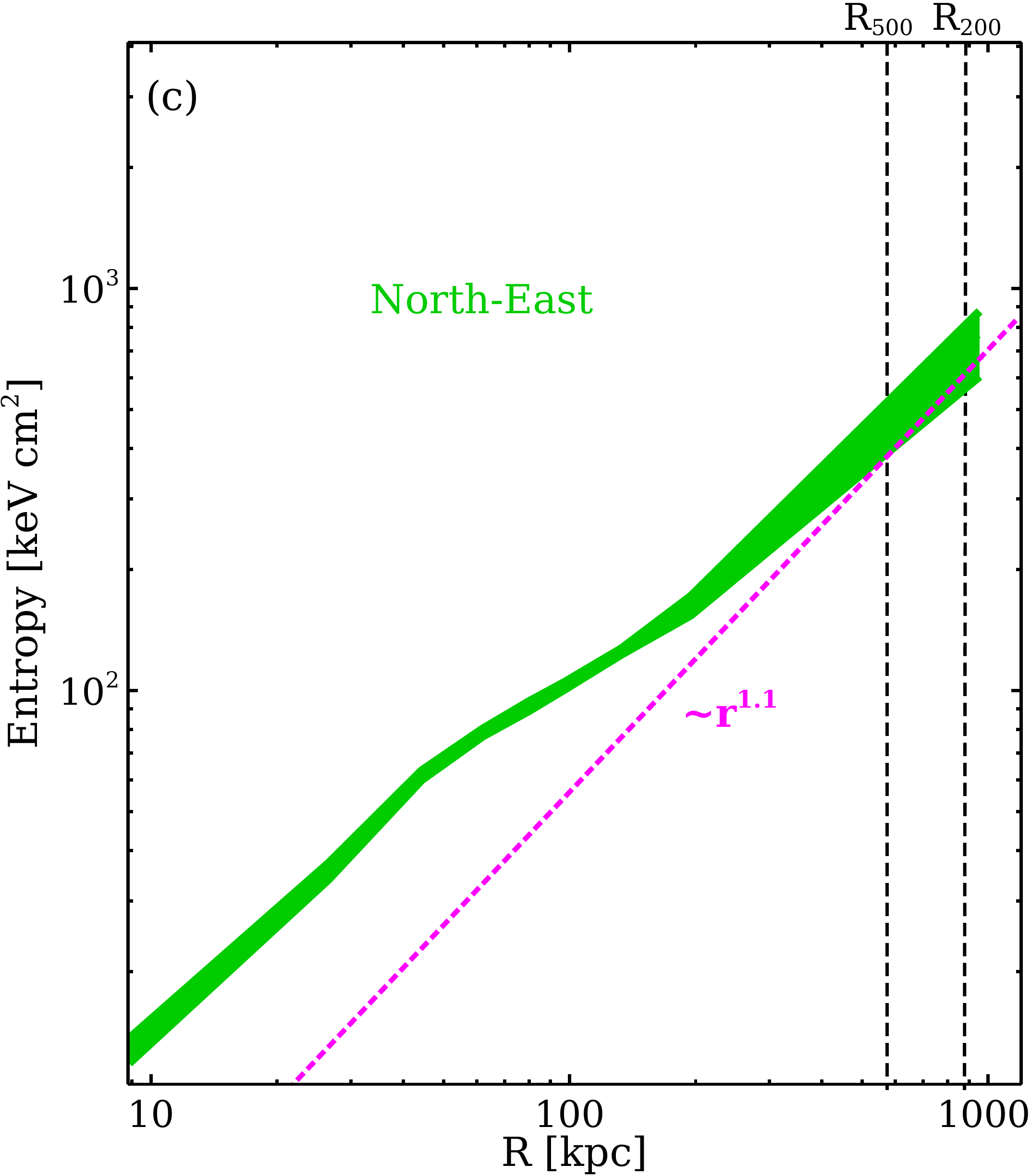}\\
\end{tabular}
\caption{3D entropy profiles of MKW4 in 
north (orange), east (blue), and north-east 
(green) directions. 
Shaded regions 
indicate 1$\sigma$ uncertainties.
Magenta dashed line represents the entropy profile 
derived from 
gravity only cosmological simulation 
\citep{2005MNRAS.364..909V} using Equation
\ref{eqn:entrpy1}.}
\label{fig:entropy}
\end{figure*}

\begin{figure*}
\begin{tabular}{ccc}
\hspace{-10pt} 
\includegraphics[width=0.33\textwidth]{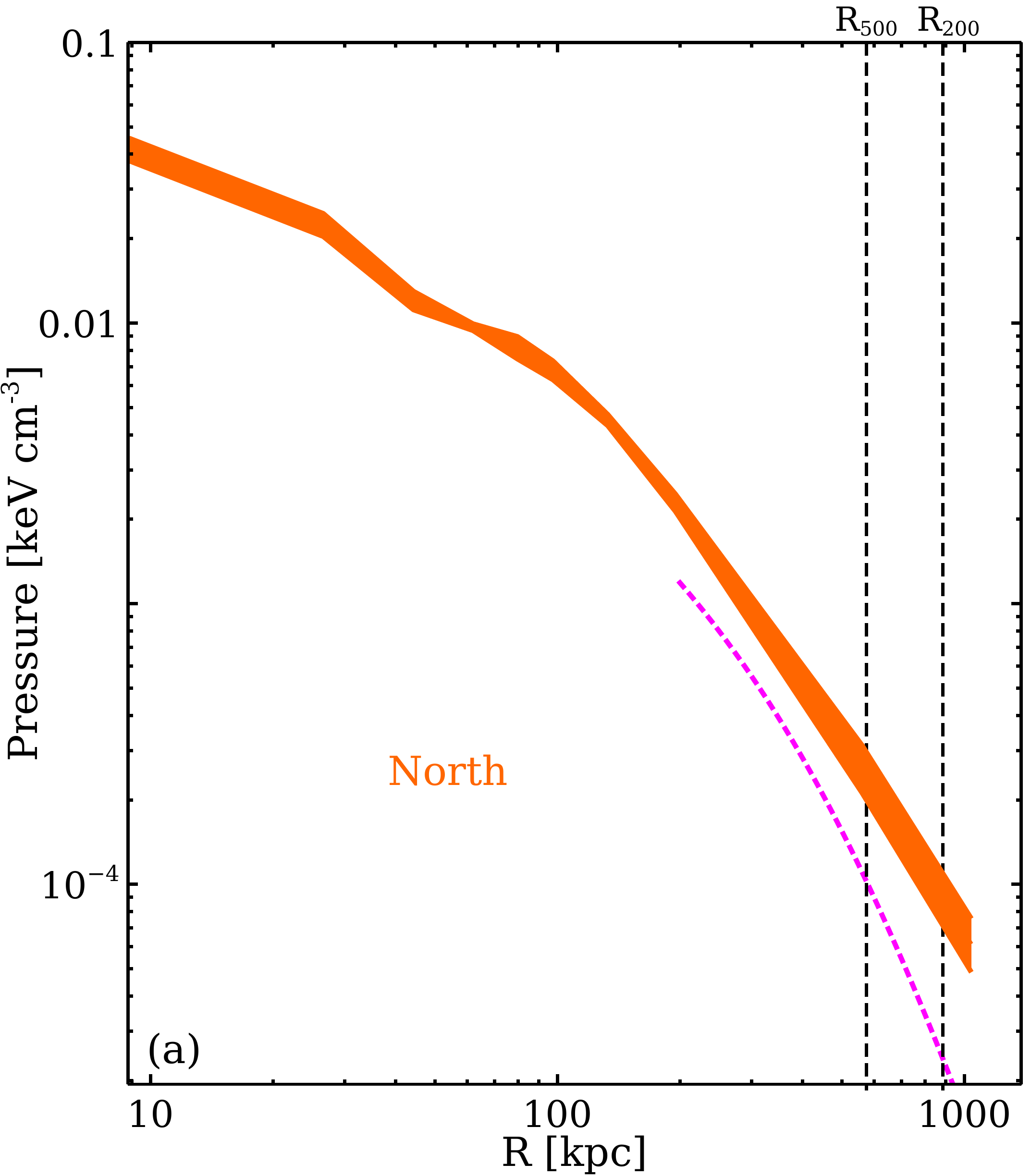}    & \hspace{-10pt}  
\includegraphics[width=0.33\textwidth]{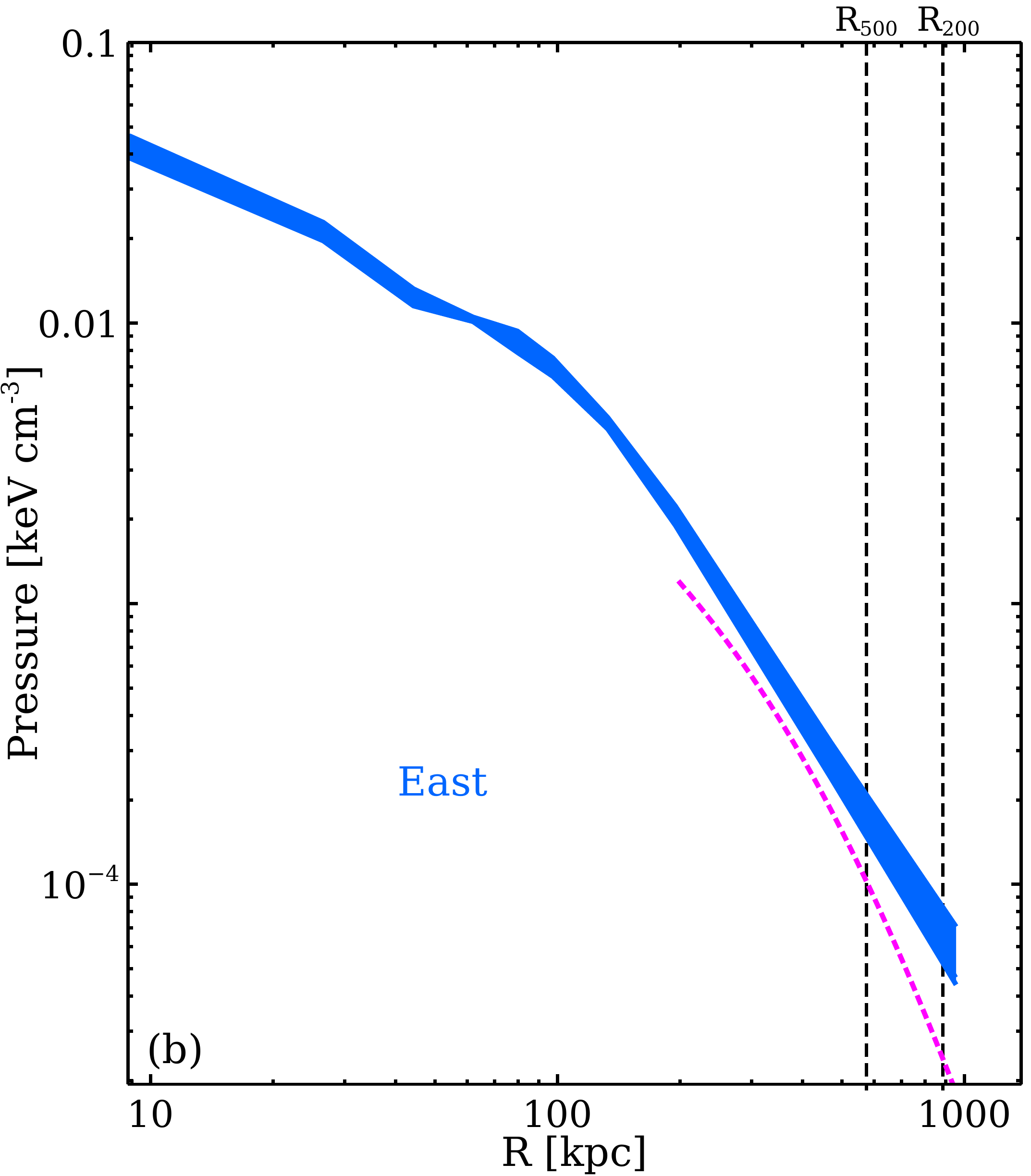} &
\hspace{-10pt} 
\includegraphics[width=0.33\textwidth]{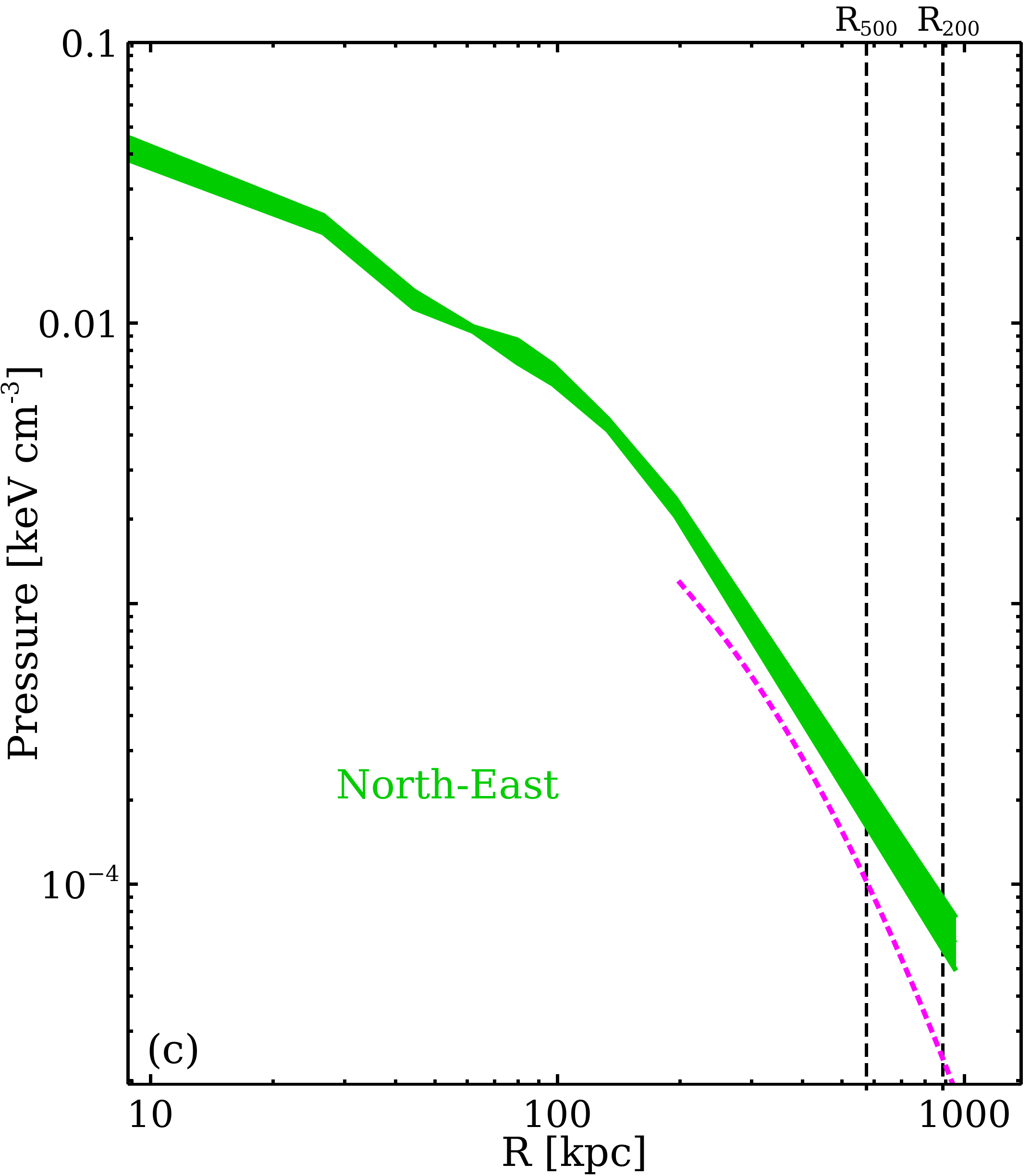}\\
\end{tabular}
\caption{
3D pressure profiles of MKW4 
in north (orange), east (blue), and  
north-east (green) directions. 
Shaded regions indicate 1$\sigma$ 
uncertainties. Magenta dashed line represents the
semi-analytical universal pressure profile
\citep{2010A&A...517A..92A}, derived using 
Equation \ref{eqn:prssr1}. 
}
\label{fig:pressure}
\end{figure*}

We derive the 3D entropy and pressure 
profiles of MKW4 from 3D density and temperature profiles, as shown in Figures   
\ref{fig:entropy} and \ref{fig:pressure}.
We compare our entropy profiles with a 
baseline profile derived 
from purely gravitational structure formation 
\citep{2005MNRAS.364..909V} and 
assuming a hydrostatic 
equilibrium-
\begin{equation}\label{eqn:entrpy1}
    \textrm{K}_\textrm{gra}(\textrm{R}) = 1.32\ 
    \textrm{K}_{200}\ \left ( 
    \frac{\textrm{R}}{\textrm{R}_{200}}\right ) ^{1.1},
\end{equation}
and the normalization K$_{200}$ is defined as- 
\begin{equation}\label{eqn:entrpy2}
    \begin{split}
    \textrm{K}_{200} &= 362\frac{\textrm{G\ M}_{200}\ \mu\  \textrm{m}_\textrm{p}}{2\textrm{R}_{200}} 
    \left ( \frac{1}{\textrm{keV}} \right ) \textrm{E}(\textrm{z})^{-4/3} \\
    & \times \left ( 
    \frac{\Omega_\textrm{m}}{0.3}\right)^{-4/3} \textrm{keV 
    cm}^{-2},
    \end{split}
\end{equation}
\begin{equation}\label{eqn:entrpy3}
    \textrm{where,}\ \textrm{E}(\textrm{z}) = 
    \sqrt{\Omega_\textrm{m}\left(1+\textrm{z} \right)^{3} + 
    \Omega_{\Lambda}}.
\end{equation}
The m$_\textrm{p}$ 
is proton mass and $\mu$ $\sim$ 0.6 is the 
mean molecular weight.  
The 3D entropy profiles of MKW4 increase 
monotonically with radius in 
three directions, 
as seen in 
Figure \ref{fig:entropy}. We observe apparent 
entropy excesses towards 
the central region (up to $\sim$ 0.5R$_{200}$) 
of MKW4 relative to the baseline profile. 
At larger radii, the entropy profiles of 
MKW4 are consistent with the baseline profile, 
i.e., its slope attains a value of $\sim$ 1.1 
between  
R$_{500}$ and R$_{200}$ in all observed directions.

The 3D pressure profiles of MKW4 
are shown in Figure \ref{fig:pressure}. 
We compare the measured pressure profiles 
to a semi-analytical universal pressure profile
\citep[][]{2010A&A...517A..92A} defined as -
\begin{equation}\label{eqn:prssr1}
    \begin{split}
    \textrm{P}(\textrm{r}) & = \textrm{P}_{500} 
    \left [ 
    \frac{\textrm{M}_{500}}{3\times10^{14} 
    \textrm{h}_{70}^{-1}\textrm{M}_{\odot}} 
    \right]^{\alpha_\textrm{p} + 
    \alpha^{'}_\textrm{p}(\textrm{x})} \\
        & \ \ \ \ \ \ \ \ \ \ \ \ \ \ \ \ \ \ \times
        \frac{\textrm{P}_{0}}{\left ( 
        \textrm{c}_{500}x\right)^{\gamma} \left[1 + 
        \left ( 
        \textrm{c}_{500}x\right)^{\alpha} 
        \right]^{\frac{\left (\beta - \alpha 
        \right)}{\alpha}}}
    \end{split}
\end{equation}
where x = $\frac{\textrm{R}}{\textrm{R}_{500}}$, 
$\alpha^{'}_{\textrm{p}}(\textrm{x})$ = $0.10 - 
(\alpha_{\textrm{p}} 
+ 0.10)\frac{(\textrm{x}/0.5)^3}{[1 + 
(\textrm{x}/0.5)^3]}$. 
The P$_{500}$ is the pressure at R$_{500}$ and 
M$_{500}$ (see Table \ref{tab:prop}) is the total 
hydrostatic 
mass within R$_{500}$. 
We adopt  
\begin{align*}
&[\textrm{P}_0,\ \textrm{c}_{500},\ \alpha,\ \beta,\ \gamma,\ 
\alpha_{\textrm{p}}] = \\
& [8.403\textrm{h}_{70}^{-3/2},\ 1.177,\ 0.3081,\ 1.0510,\ 
5.4905,\ 0.12 \pm 0.10]
\end{align*}
from \citet{2010A&A...517A..92A}.
{ {\citet{2011ApJ...727L..49S} adopted
similar parameters and found that the 
\citet{2010A&A...517A..92A} pressure profile 
is also representative for galaxy groups.}}
The 3D pressure profiles of MKW4 in all directions 
from 0.35R$_{500}$ out to $\sim$ 
0.65R$_{500}$ show good agreement 
with this universal pressure profile but exceed it
by more than { 50\%} at R$_{500}$. 


\subsection{Mass and gas fraction}
\begin{figure}
\centering
\hspace{-20pt}
\includegraphics[width=1.1\columnwidth]{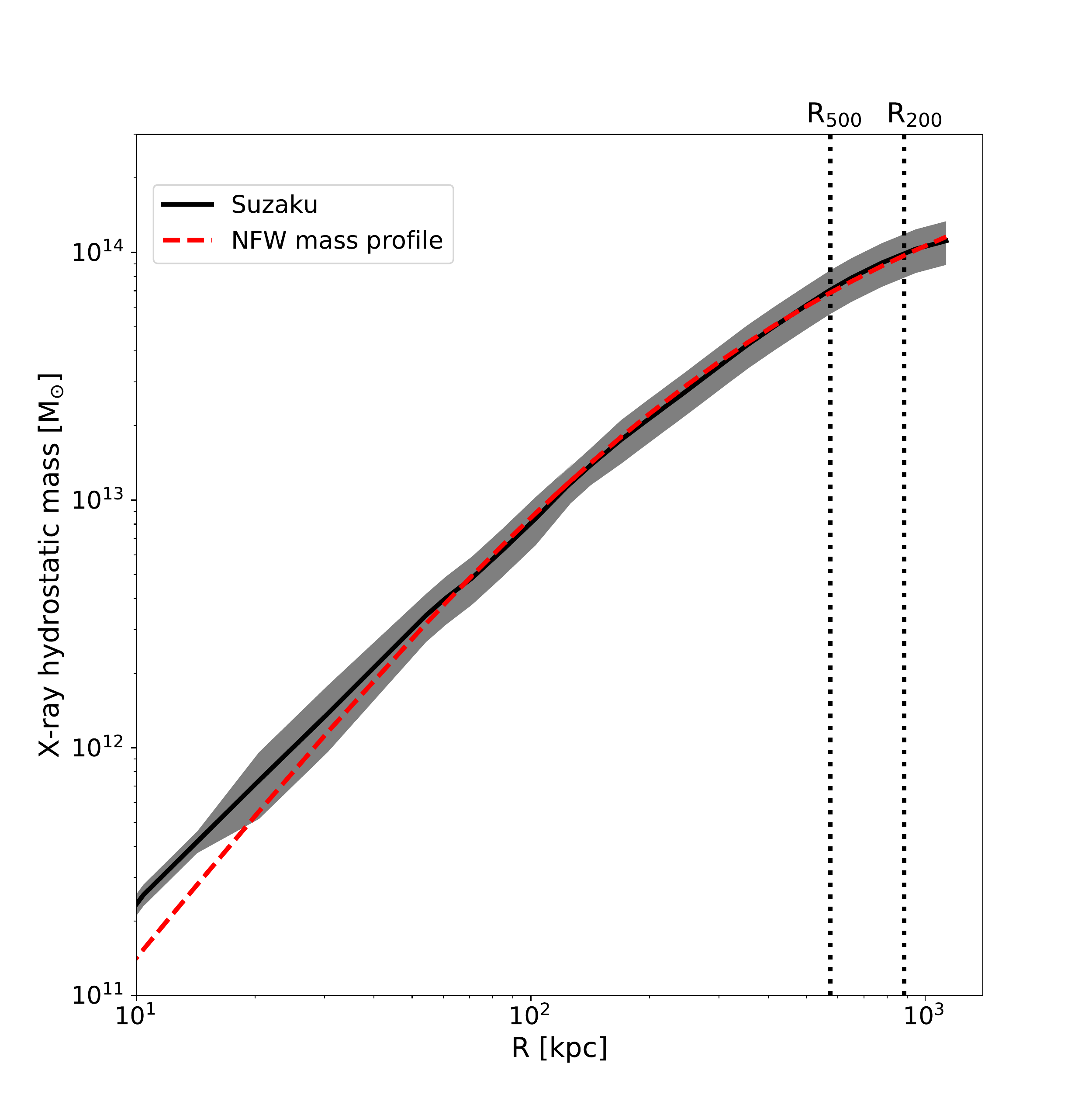}
\caption{The X-ray hydrostatic 
mass profile 
(M$_\textrm{total}$) of MKW4 derived using
Equation \ref{eqn:mass1}. Red dashed line
shows best-fit NFW mass profile obtained by 
integrating Equation \ref{eqn:NFW}.
Shaded region indicates 
1$\sigma$ error.}
\label{fig:mass}
\end{figure}

We derive the X-ray hydrostatic mass of MKW4 
and its gas mass within a specific radius (R) 
from the group center, incorporating
the above 3D 
density and temperature profiles in the following 
equations- 
\begin{equation}\label{eqn:mass1}
    \textrm{M}_\textrm{tot}(<\textrm{R}) = 
    -\frac{\textrm{kT}_\textrm{3D}\textrm{R}}{\textrm{G}\mu \textrm{m}_\textrm{p}} 
    \left( \frac{\textrm{d\ 
    ln}\rho_\textrm{gas}}{\textrm{d\ ln}\textrm{R}} 
    + \frac{\textrm{d\ 
    ln}\textrm{T}_\textrm{3D}}{\textrm{d\ 
    ln}\textrm{R}} \right)
\end{equation}
\begin{equation}\label{eqn:mass2}
    \textrm{M}_\textrm{gas}(<\textrm{R}) = 4\pi\ 
    \int_{0}^\textrm{R}\ 
    \rho_\textrm{g}(\textrm{r}^{\arcmin})\textrm{r}^
    {\arcmin 2} \textrm{dr}^{\arcmin},
    \end{equation}
{ where 
$\rho_\textrm{gas}$ = 1.92$\mu$m$_{\rm H}$n$_{\rm e}$ 
is the gas density, and m$_{\rm H}$ is the proton mass.}
The resulting 
hydrostatic mass
profile of MKW4 is shown in Figure
\ref{fig:mass}.
We obtain $
\textrm{M}_\textrm{tot}(<\textrm{R}_{500})$ 
$\approx$ 6.5 $\pm$ 1.0 
$\times$ 10$^{13}$ $\textrm{M}_{\odot}$ and 
$\textrm{M}_\textrm{tot}(<\textrm{R}_{200})$ 
$\approx$ 9.7 $\pm$ 1.5 
$\times$ 10$^{13}$ $\textrm{M}_{\odot}$. 
{ {Our measured hydrostatic mass within R$_{500}$ is 
between that given by 
\citet{Vikhlinin_2006} (  
7.7 $\pm$ 1.0 
$\times$ 10$^{13}$ $\textrm{M}_{\odot}$) and 
those of \citet{2007ApJ...669..158G} (4.3 $\pm$ 0.2 
$\times$ 10$^{13}$ $\textrm{M}_{\odot}$) and
\citet{Sun_2009} (4.8 $\pm$ 0.7 
$\times$ 10$^{13}$ $\textrm{M}_{\odot}$). }

{We fit our hydrostatic mass profile to the 
NFW mass density profile}
\citep{1997ApJ...490..493N}:
\begin{equation} \label{eqn:NFW}
    \rho({\rm r}) = \frac{\rho_{\rm s}}{({\rm r}/{\rm r}_{\rm s}) (1+{\rm r}/{\rm r}_{\rm s})^{2}},
\end{equation}
as shown in Figure \ref{fig:mass}. {We 
obtain a best-fit r$_{\rm s}$ of 186 $\pm$ 23 kpc,
which gives the central mass concentrations of 
c$_{500}$ = 3.09$^{+0.52}_{-0.39}$ 
at R$_{500}$ and 
c$_{200}$ = 4.75$^{+0.80}_{-0.60}$
at R$_{200}$. }
{We also obtain the sparsity,
the ratio of masses at two 
overdensities \citep{2018ApJ...862...40C}, 
S$_{200, 500}$ = 1.49 $\pm$ 0.31. Our measurement of 
S$_{200, 500}$ for MKW4 is closely align with the 
typical 
S$_{200, 500}$ of 1.5
found for galaxy clusters
in N-body simulation \citep{2018ApJ...862...40C}.
}}

\begin{table*}
\caption{Properties of MKW4}
\resizebox{1\textwidth}{!}{%
\begin{tabular*}{1.15\textwidth}{@{\extracolsep{\fill}\quad}lllllllllllll}
\hline
T$_{2500}^{\textrm{a}}$ \ \ \ \ & T$_{500}$ \ \ \ \  & T$_{200}$ \ \ \ \ & R$_{2500}$ \ \ \ \ & R$_{500}$ \ \ \ & R$_{200}$ \ \ \  & 
M$_{500}^{\textrm{b}}$ \ \
\ & M$_{200}$ \ \ \ & f$_\textrm{gas, 2500}^{\textrm{c}}$ \ \ \ & f$_\textrm{gas, 500}$ \ \ \ & f$_\textrm{gas, 200}$  \ \ \ & c$_{500}$ \ \ \ & c$_{200}$\\
(keV) & (keV)  &  (keV) &  (kpc) &  (kpc) & (kpc) & (10$^{13}$ M$_{\odot}$) & (10$^{13}$ M$_{\odot}$) & & & \\
\hline
1.71$^{+0.09}_{-0.08}$ & 1.36 $\pm$ 0.09 & 
1.14$^{+0.16}_{-0.13}$ & 274 $\pm$ 10 & 574 $\pm$ 20
&
884 $\pm$ 17 & 6.5 $\pm$ 1.0 & 9.7 $\pm$ 1.5 & 0.05
$\pm$ 0.01 & 0.07 $\pm$ 0.01 & 0.090 $\pm$ 0.011 & 3.09$^{+0.52}_{-0.39}$ & 4.75$^{+0.80}_{-0.60}$\\
\hline
\end{tabular*}
}
\flushleft{
$^{\textrm{a}}$T$_{\Delta}$ = deprojected temperature at R$_{\Delta}$.\\
$^{\textrm{b}}$M$_{\Delta}$ = total X-ray hydrostatic mass within R$_{\Delta}$.\\
$^{\textrm{c}}$f$_{\textrm{gas}, \Delta}$ = enclosed gas mass fraction at R$_{\Delta}$.\\
}
\label{tab:prop}
\end{table*}

\begin{figure*}
\begin{tabular}{ccc}
\hspace{-10pt} 
\includegraphics[width=0.33\textwidth]{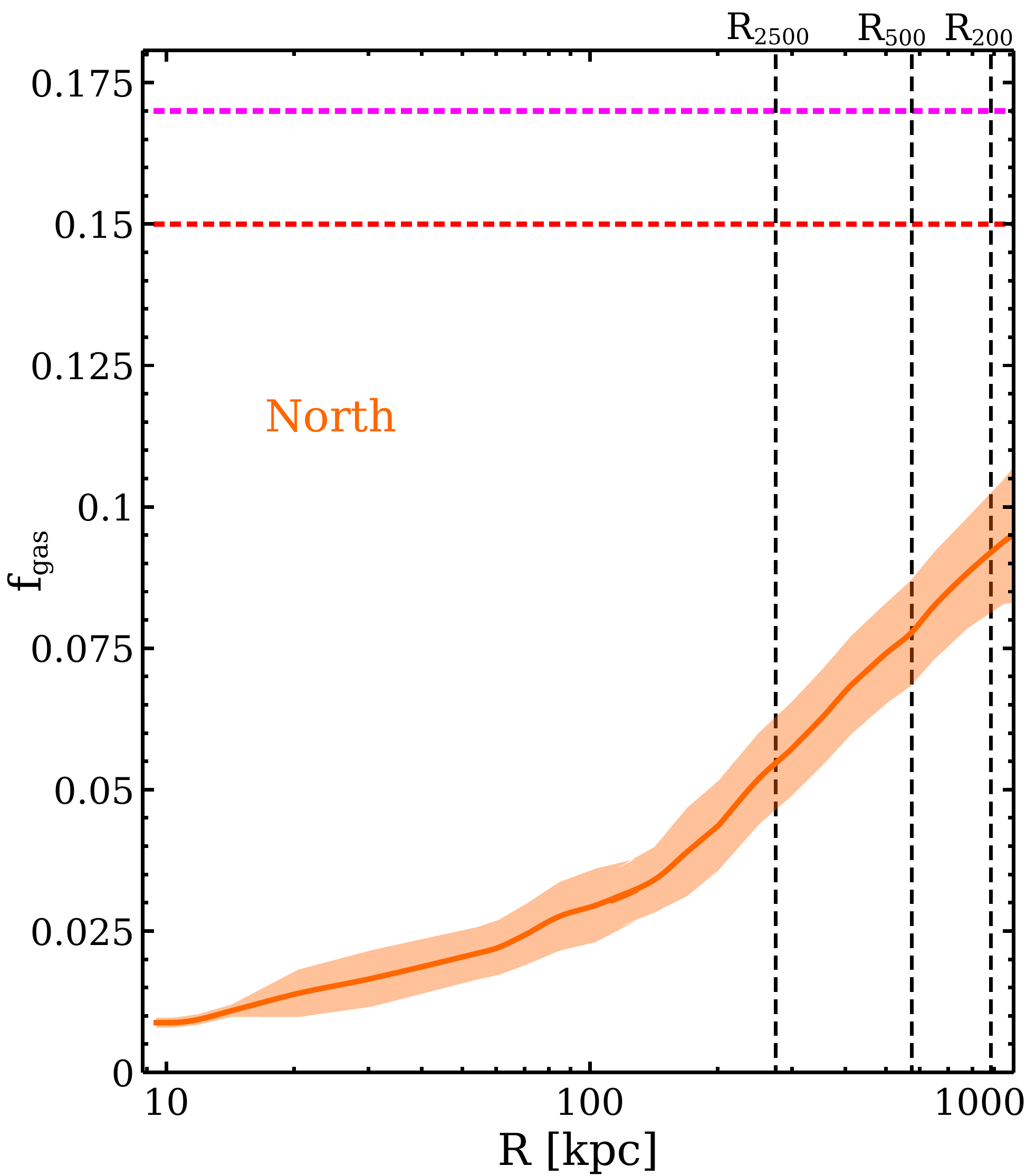}    & \hspace{-10pt}  
\includegraphics[width=0.33\textwidth]{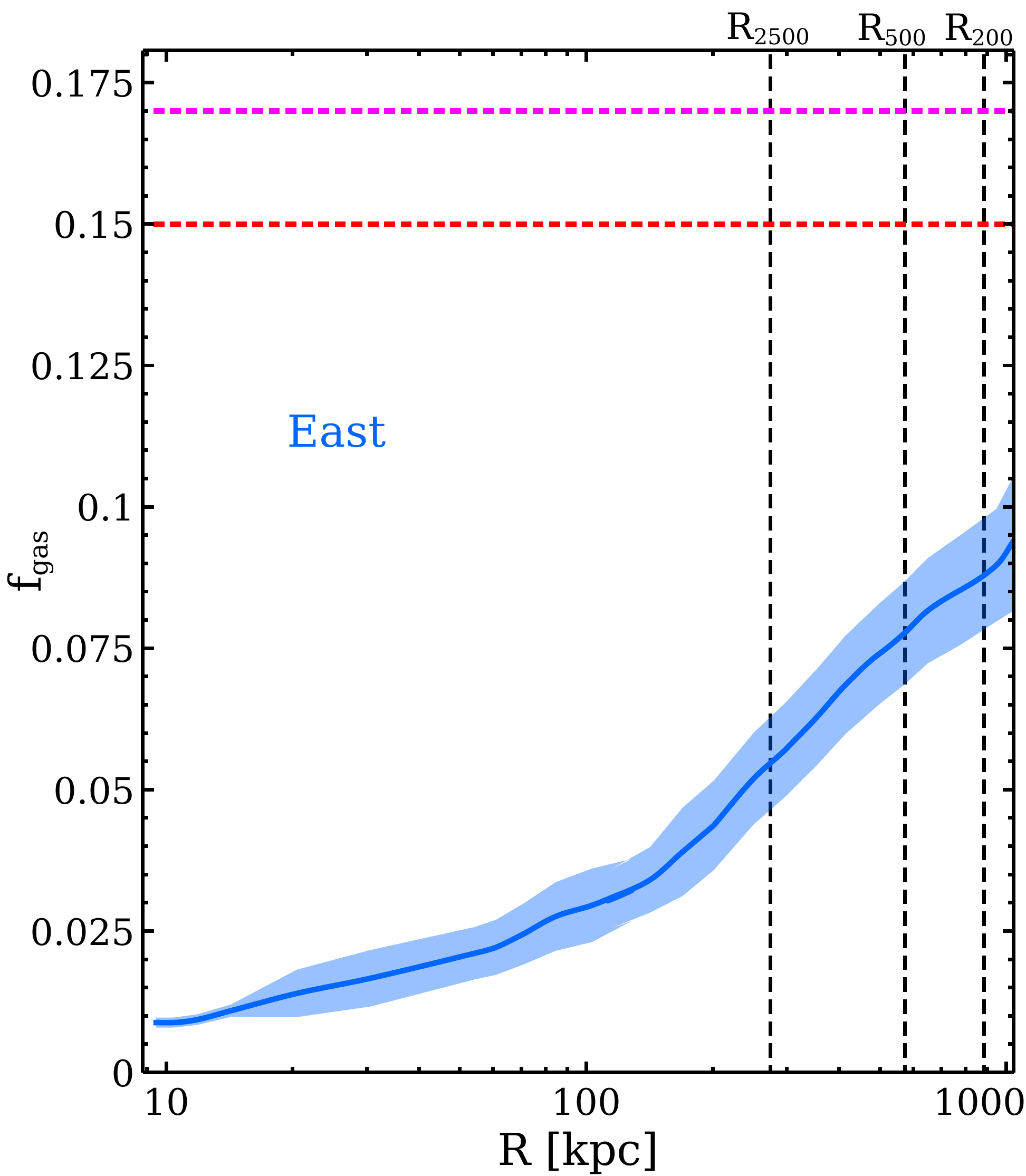} &
\hspace{-10pt} 
\includegraphics[width=0.33\textwidth]{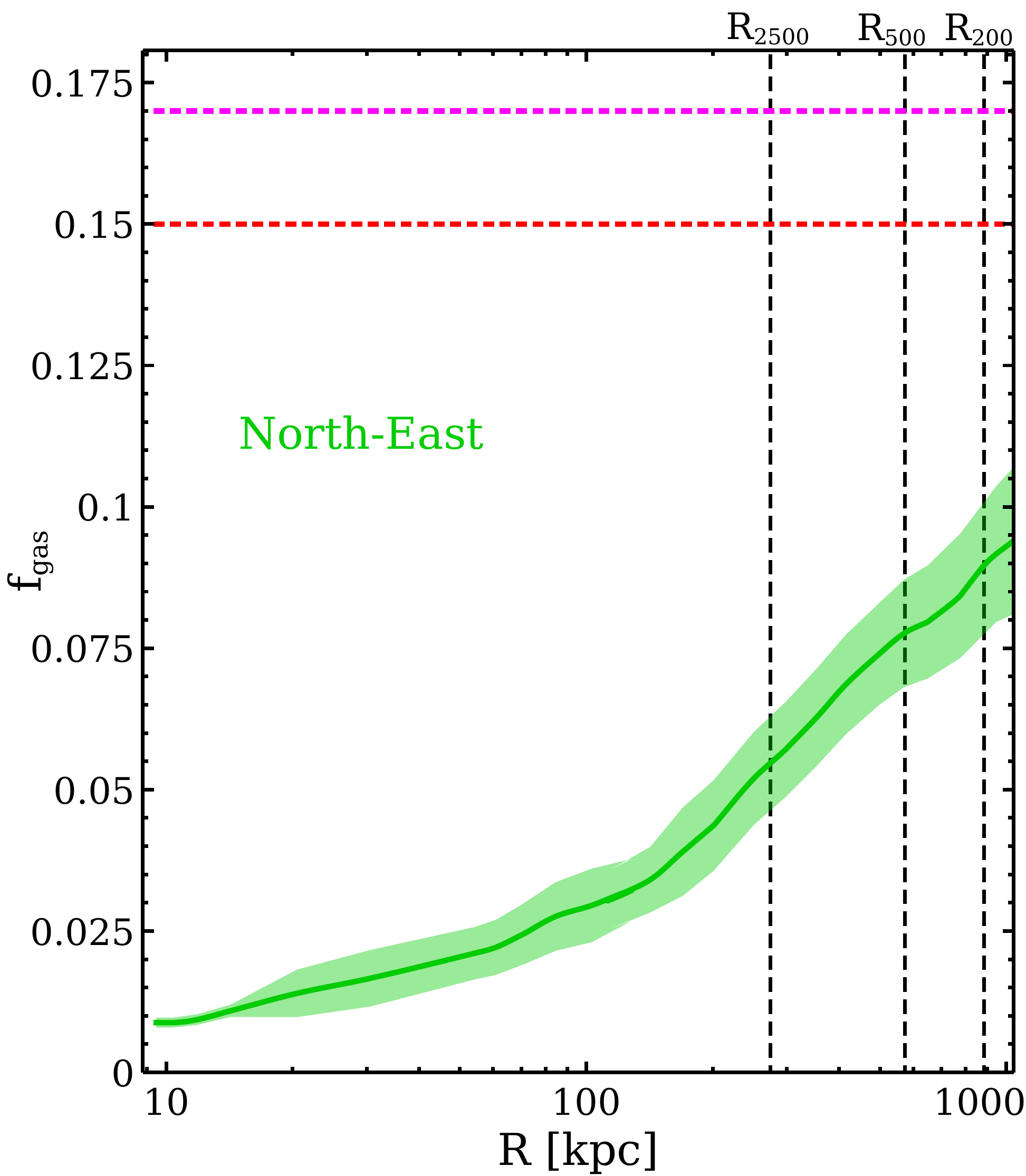}\\
\end{tabular}
\caption{The gas mass fraction of MKW4 in 
north (orange), east (blue), and north-east 
(green) directions.
Red dashed line: cosmic baryon fraction 
estimated by {\sl Planck} 
\citep{2014A&A...571A..16P}. 
Magenta dashed line: cosmic baryon 
fraction estimated by {\sl WMAP} 
\citep{Komatsu_2011}.}
\label{fig:gas_frac}
\end{figure*}

We obtain its enclosed gas mass 
and estimate the gas mass fraction   
f$_\textrm{gas}$ = 
$\frac{\textrm{M}_\textrm{gas}}{\textrm{M}_\textrm{tot}}$ for each 
direction, as shown in Figure \ref{fig:gas_frac}. 
At $\sim$ R$_{500}$, the measured gas mass 
fraction of MKW4 is below 10\% in all directions, 
similar to the other galaxy groups 
\citep[e.g.,][]{Vikhlinin_2006,Sun_2009,2012ApJ...748...11H,2016A&A...592A..37T}. 
At a radii larger than R$_{500}$, 
f$_\textrm{gas}$ grows slowly and 
attains a value of { 0.092 $\pm$ 0.009 in the north, 
0.087 $\pm$ 0.012 in the east, and 0.090 $\pm$ 
0.011} in the north-east at R$_{200}$. 
These gas mass 
fractions are surprisingly low compared
to the cosmic baryon fraction 
of 0.15 
\citep{2014A&A...571A..16P} and 0.17 
\citep{Komatsu_2011}. { A brief comparison of 
the mass and f$_{\rm gas}$ of MKW4 
between our work and previous studies is listed
in Table \ref{tab:com}.}

\begin{table}
\caption{Comparison of our results with other works}
\resizebox{1\columnwidth}{!}{%
\begin{tabular}{cccccc}
\hline
Name \ \  & 
R$_{500}$ \ 
\ & M$_{500}^{\ddagger}$ \ \  &  f$_\textrm{gas, 2500}^{\dagger}$ \ \  &  f$_\textrm{gas, 500}^{\dagger}$ \ \  & c$_\textrm{500}$\\
 & (kpc) & $\times$ 10$^{13}$ (M$_{\odot}$) & & &\\
\hline
This work & 574 $\pm$ 20 & 6.5 $\pm$ 1.0 & 0.05 $\pm$ 
0.01 & 0.07 $\pm$ 0.01 & 3.09$^{+0.52}_{-0.39}$ \\ \\

S09$^{\$}$ & 538$^{+24}_{-29}$ & 4.85$^{+0.71}_{-0.68}$ & 0.047$^{+0.002}_{-0.003}$ & 0.086 $\pm$ 
0.009 & 3.93$^{+1.16}_{-0.78}$ \\ \\

G07 & 527 $\pm$ 8 & 4.27 $\pm$ 0.18 & \ \ $-$ & \ \ $-$ & 6.4 $\pm$ 0.5 \\ \\

V06 & 634 $\pm$ 28 & 7.7 $\pm$ 1.0 & 0.045 $\pm$ 
0.002 & 0.062 $\pm$ 0.006 & 2.54 $\pm$ 0.15 \\ \\

\hline
\end{tabular}
}
\flushleft{
$^{\ddagger}$M$_{\Delta}$ = total X-ray hydrostatic mass within R$_{\Delta}$.\\
$^{\dagger}$f$_{\textrm{gas}, \Delta}$ = enclosed gas mass fraction at R$_{\Delta}$.\\
$^{\$}$ S09, G07, and V06 are referred to \citet{Sun_2009},
\citet{2007ApJ...669..158G}, and \citet{Vikhlinin_2006}
respectively.
}
\label{tab:com}
\end{table}

\section{Systematic uncertainties} \label{sec:sys}
We investigate changes of the best-fit gas density, 
temperature, entropy, pressure, and enclosed
gas mass fraction at R$_{200}$ 
introduced by a variety of systematic effects.
Below we focus on the systematic uncertainties
related to the entropy. All other gas properties
are listed in Table \ref{tab:sys_err}.


\begin{table*}
\caption{Systematic error of gas properties 
at R$_\textrm{200}$}
\label{tab:sys_err}
\begin{tabular*}{\textwidth}{@{\extracolsep{\fill}\quad}lllllll}
\hline
&  Test & Temperature & Density &  Entropy &  
Pressure &  f$_{\textrm{gas}}$ \\
&   &  keV &  10$^{-5}$ cm$^{-3}$ &  keV cm$^2$ &  
10$^{-5}$ keV cm$^{-3}$ &  \\
\hline
& Best fit & 1.13 $\pm$ 0.14 & { 8.85} $\pm$ 1.5 & { 630}
$\pm$ { 82} & { 9.2} $\pm$ { 2.8} & { 0.092} $\pm$ { 0.009} \\
\hline
North & $\Delta$CXB & +0.06, $-$0.04 & +0.08, 
$-$0.46 & $-$4, +176 & +0.17, $-$0.36 & $\pm$0.004\\
    & $\Delta$CXB-$\Gamma$ & +0.14, $-$0.04 & 
    $-$0.24, $-$0.13 & $-$26, +116 & $-$0.8, +0.003 
    & $\pm$0.001\\
    & $\Delta$abun & +0.06, $-$0.04 & +0.22, $-$0.46
    & $-$22, +177 & +0.03, $-$0.36 & +0.0007, 
    $-$0.008\\
    & $\Delta$N$_\textrm{H}$ & $-$0.04, +0.06 & 
    $-$0.22, 
    $-$0.15 & +31, +30 & $-$0.09, +0.4 & $-$0.002, 
    +0.009\\
    & $\Delta$distance & $\pm$0.03 & $\pm$0.2 & 
    $\pm$66 & $\pm$0.34 & $\pm$0.008\\
    & $\Delta$MW & +0.084, +0.024 & +0.001, $-$0.19 
    & $-$51, +72 & +0.20, +0.16 & +0.001, +0.001\\
    & $\Delta$LHB & $\pm$0.09 & $\pm$0.05 & $\pm$165
    & $\pm$0.16 & $\pm$0.01\\
    & $\Delta$solar & +0.02, +0.14 & $-$0.66, 
    $-$0.23 & +159, $-$28 & $\pm$0.78 & +0.01, 
    +0.02\\
\hline
& Best fit & 1.17 $\pm$ 0.15 & { 5.6} $\pm$ { 1.2} & { 712}
$\pm$ { 100} & { 6.55} $\pm$ { 1.3} & { 0.087} $\pm$ { 0.012} \\
\hline
East & $\Delta$CXB & +0.28, $-$0.1 & +1.52, +0.84 & 
+71, $-$141 & +2.4, $-$0.03 & +0.006, $-$0.001\\
& $\Delta$CXB-$\Gamma$ & $-$0.2, +0.28 & +1.06, 
+1.27 & $-$249, +107 & $-$0.27, +2.08 & $-$0.006, 
+0.005\\
& $\Delta$abun & +0.28, $-$0.10 & +0.85, +1.55 & 
$-$68, +143 & $-$0.02, +2.44 & +0.005, $-$0.001\\
& $\Delta$N$_\textrm{H}$ & $-$0.01, +0.05 & +0.12, 
+0.13 & 
$-$63, $-$55 & +0.13, +0.88 & +0.003, +0.02\\
& $\Delta$distance & $\pm$0.02 & $\pm$0.34 & $\pm$39
& $\pm$0.06 & $\pm$0.002\\
& $\Delta$MW & $-$0.06, +0.16 & +0.9, +0.43 & 
$-$186, +167 & +0.23, +0.49 & -0.01, +0.03\\
& $\Delta$LHB & $\pm$0.01 & $\pm$0.62 & $\pm$82 &
$\pm$0.17 & $\pm$ 0.001\\
& $\Delta$solar & +0.06, $-$0.1 & +0.16, +1.27 &
+111, $-$191 & $-$0.11, +0.36 & $-$0.003, +0.008\\ 
\hline
& Best fit & 1.12 $\pm$ 0.07 & { 6.7} $\pm$ { 1.8} & { 682}
$\pm$ { 134} & { 7.54} $\pm$ { 1.7} & { 0.090} $\pm$ { 0.011} \\
\hline
North-East & $\Delta$CXB & +0.09, $-$0.25 & +1.2, 
$-$0.36 & $-$5, +130 & +1.4, $-$0.3 & $-$0.01, 
$-$0.005\\
& $\Delta$CXB-$\Gamma$ & +0.06, $-$0.25 & $-$1.4, 
$-$1.27 & +210, $-$44 & $-$1.7, $-$2.8 & +0.003, 
$-$0.004\\
& $\Delta$abun & $-$0.15, $-$0.07 & $-$0.74, $-$1.8 
& $-$39, +213 & $-$1.8, $-$2.6 & $-$0.003, 
$-$0.004\\
& $\Delta$N$_\textrm{H}$ & $-$0.05, +0.06 & $-$0.14,
$-$0.13 
& $-$34, +40 & $-$0.80, $-$0.70 & $-$0.004, 0.007\\
& $\Delta$distance & $\pm$0.05 & $\pm$1.4 & $\pm$152
& $\pm$2.1 & $\pm$ 0.003\\
& $\Delta$MW & $-$0.15, +0.05 & $-$0.67, $-$0.95 & 
$-$61, +200 & $-$0.93, $-$0.61 & $-$0.004, 
$-$0.008\\
& $\Delta$LHB & $\pm$0.03 & $\pm$0.78 & $\pm$201 & 
$\pm$0.57 & $\pm$0.002\\
& $\Delta$solar & $-$0.07, $-$0.25 & $-$2.0, $-$1.2 
& +235, $-$60 & $-$2.8, +2.7 & $\pm$0.004\\
\hline
\end{tabular*}
\flushleft{
$\Delta$CXB: vary the normalization of the CXB 
component by 20\%, while keeping the slope of the 
power law, $\Gamma$ = 1.4.\\
$\Delta$CXB-$\Gamma$: set the slope of the 
power law for CXB component at $\Gamma$ = 1.3 
and 1.5, respectively.\\
$\Delta$abun: set the metal abundance at 0.1 
Z$_{\odot}$ and 0.3 Z$_{\odot}$, respectively 
for outermost bin. \\
$\Delta$N$_\textrm{H}$: vary the galactic hydrogen 
column density by 20\%. \\
$\Delta$distance: vary the redshift parameter 
by 5\%. \\
$\Delta$MW: vary the normalization for 
{\tt\string apec$_\textrm{MW}$} by 10\%. \\
$\Delta$LHB: vary the normalization for 
{\tt\string apec$_\textrm{LHB}$} by 10\%. \\
$\Delta$solar: set the solar abundance table 
to \citet{ANDERS1989197} and \citet{Lodders_2003}, 
respectively.\\}
\end{table*}

We consider two different sources of systematic 
uncertainties associated with the CXB.
First, we allow the best fit normalization 
of the CXB component in
the background model 
to vary by 20\% with a fixed power-law 
slope of $\Gamma$ = 1.41 
\citep{2004A&A...419..837D}, which leads to a 
maximum change in gas entropy by $\sim$ 18\% at 
R$_{200}$ for the north, east, and north-east 
directions (see $\Delta$CXB in Table 
\ref{tab:sys_err}). 
Second, we fix the CXB power-law slope to 
$\Gamma$ = 1.3 and 1.5, respectively. 
Adopting $\Gamma$ = 1.3 has little effect on our 
results for the north direction, 
while the gas entropy 
varies by $\sim$ 30\% for the other two directions. 
Adopting $\Gamma$ = 1.5, 
the gas entropy is increased by 
$\sim$ 14\% for the north and east directions, 
while no
significant changes are observed for the north-east 
direction. 

We examine the impacts on gas properties due to 
variations in 
the MW 
and LHB foreground 
components by varying the normalization of 
each component 
by 
10\% (see $\Delta$MW and $\Delta$LHB in Table 
\ref{tab:sys_err}). 
We find that the variation in the MW 
component changes the gas 
entropy by 22\% in the east and 27\% 
in the north-east. The 
10\% variation in
the LHB component changes the 
entropy by 20\% in the north and 
27\% in the north-east.

We adopt the solar abundance table
of \citet{ASPLUND20061} for the spectral
analysis, as shown in Figure \ref{fig:xspec}.
Here, we experiment with two different 
solar abundance tables of \citet{ANDERS1989197} and 
\citet{Lodders_2003} to find their impacts
on the  measurement of gas properties.
Results are listed in 
Table 
\ref{tab:sys_err} under $\Delta$solar. 
Using the \citet{ANDERS1989197} solar abundance 
table, the gas entropy 
increases by 20\% in the north, 13\% in the
east, and 32\% in the north-east directions.
In contrast, using the \citet{Lodders_2003} 
abundance 
table, the entropy does not change significantly in 
the north and north-east directions 
but decreases by 
20\% in the east. 

In the spectral analysis, we find it 
necessary to fix the ICM metallicity at 0.2 
$Z_{\odot}$ for regions at R$_{200}$.
Here, we estimate the uncertainties in the 
measurements of gas properties associated
with the possible variation of the ICM
metallicity. We repeat the spectral
analysis by fixing the metallicity
at 0.1 
$Z_{\odot}$ and 0.3 
$Z_{\odot}$ (see $\Delta$abun 
in Table \ref{tab:sys_err}).
Fixing the metal abundance 
at 0.1 $Z_{\odot}$ has little 
impact on gas properties at 
the outskirts of MKW4, while 
adopting a metalicity of 0.3 
$Z_{\odot}$ increases the gas entropy by $\sim$ 20\%
in the north, 15\% in the east,
and 25\% in the north-east directions. 

\section{Discussion}\label{sec:discussion}

Combining the deep $\suzaku$ observations with the  
$\chandra$ ACIS-S and ACIS-I 
observations of MKW4, we measured its gas 
properties from the group center out to the 
virial radii in three directions. 
Its entropy profiles at larger radii
are consistent with the self-similar 
value predicted by 
gravitational collapse alone 
\citep{2005MNRAS.364..909V}.
We estimated the enclosed gas mass fraction of MKW4
as a function of radial distance from the group 
center and obtained a surprisingly low 
value within  
R$_{200}$ compared to clusters.
Below we discuss the implications of these 
results in detail.


\subsection{Entropy profile}
 
The entropy profile describes the thermal 
history of the ICM. Previous works on massive 
clusters (T$_\textrm{X} >$3 keV) have measured 
entropy profiles that flatten between
R$_{500}$ and R$_{200}$ and even tend to fall below 
the self-similar value predicted by 
gravitational collapse alone
\citep[e.g.,][]{Simionescu1576,2013MNRAS.428.2812B,10.1111/j.1365-2966.2012.21282.x}. 
{ The explanations proposed for 
the unexpected entropy profiles include clumpy ICM and 
electron-ion non-equilibrium, either of which can
be induced
as galaxy clusters accrete cold gas from cosmic 
filaments.}
Using numerical 
simulations, \citet{Nagai_2011} predict 
that massive clusters with 
$\textrm{M}_{200}>10^{14}\textrm{M}_{\odot}\textrm{
h}^{-1}$ contain a significant 
fraction of clumpy gas at 
larger radii due to frequent merging events, 
which biases low the entropy by 
overestimating the gas density. 
We do not
observe any flattening in the entropy profiles of 
MKW4 between R$_{500}$ and R$_{200}$. Our results 
suggest that the merger rate is comparably 
lower for MKW4 because of the shallow gravitational 
potential well, which makes the ICM less clumpy at 
its outskirts. This also supports the fact 
that clumping factors are smaller in the 
low mass clusters 
(M$_{200}<10^{14}\textrm{M}_{\odot}\textrm{h}^{-1}$) 
and groups, as suggested by  
\citet{2015ApJ...805..104S}, \citet{2016A&A...592A..37T}, and \citet{2016ApJ...831...55B}. The entropy profiles of 
MKW4 instead follow the baseline profile 
\citep{2005MNRAS.364..909V}, indicating
its gas dynamics at the outskirts may be  
mainly regulated by gravity.  

Another explanation for entropy flattening is 
electron-ion thermal non-equilibrium. A shock wave 
resulting from any recent merger and accretion event
at the outskirts tends to heat the heavy-ions faster
than electrons, causing the ion temperature to
exceed the electron temperature. This thermal 
non-equilibrium could bias low the gas entropy. 
Numerical simulations show that the discrepancy 
between electron and ion temperature is more severe 
in more massive and rapidly growing clusters
\citep[e.g.,][]{2015ApJ...808..176A,2019SSRv..215....7W}. 
The well behaved entropy profiles of MKW4  
suggest that it is a relatively undisturbed system 
and has experienced few recent mergers.

\begin{figure}
\hspace{-10pt} 
\includegraphics[width=\columnwidth]{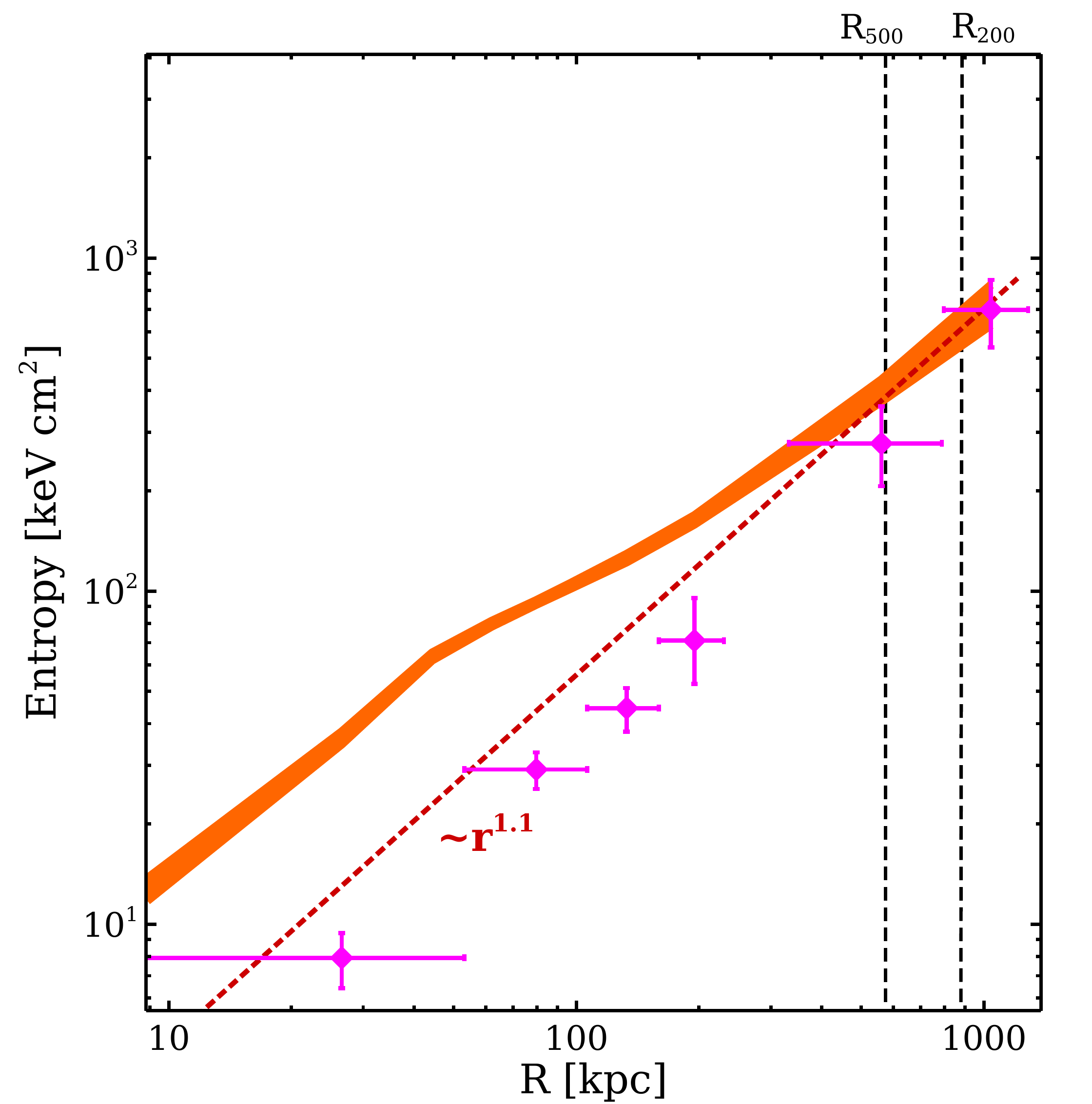}
\caption{Orange: The entropy profile of MKW4 
in north direction with 1$\sigma$ error. 
Red: The baseline entropy 
profile estimated 
using Equation \ref{eqn:entrpy1}. 
Magenta: The entropy profile scaled with 
f$_{\textrm{gas}}$ as shown in Equation \ref{eq:entropy_scaled}.}
\label{fig:scaled_entrpy}
\end{figure}

Despite the agreement between the entropy profiles 
of MKW4 and the baseline profile 
\citep{2005MNRAS.364..909V} at
larger radii, we observe apparent 
entropy excess at the center of MKW4. 
\citet{2010A&A...511A..85P} and \citet{2014MNRAS.441.1270L} advocate that the AGN 
feedback could elevate entropy profiles at the 
center of galaxy clusters by pushing a
substantial fraction of baryons out of R$_{200}$.
\citet{2010A&A...511A..85P} also corrects 
the entropy 
profiles for the redistribution of hot gas due to 
AGN feedback, as follows-
\begin{equation} \label{eq:entropy_scaled}
    \textrm{K}_{\textrm {corrected}} = \textrm{K}_{\textrm{measured}}\ \times\ \left (\frac{\textrm{f}_\textrm{gas}}{\textrm{f}_\textrm{b}} \right )^{2/3}.
\end{equation}
We scale the measured entropy profile of MKW4
in north direction
as Equation \ref{eq:entropy_scaled}, adopting
a cosmic baryon fraction of 
f$_\textrm{b}$ = 0.15 from the  
\citet{2014A&A...571A..16P}, as 
shown in Figure 
\ref{fig:scaled_entrpy}. We find the scaled 
entropy profile follows closely with the baseline 
profile, even out to the virial radii, suggesting that the 
redistribution of the group gas due to
the AGN feedback may have shaped 
its gas entropy \citep{Mathews_2011}.


\subsection{Gas mass fraction}

The enclosed gas mass fraction 
(f$_{\textrm{gas}}$) 
of MKW4 slowly rises from the group center and 
reaches $\sim$ 7\% within R$_{500}$, 
consistent
with the previous $\chandra$ studies of MKW4 
\citep{Vikhlinin_2006} and the 
f$_{\textrm{gas}}$ of 
other groups 
\citep[e.g.,][]{2012ApJ...748...11H,2013ApJ...775...89S,2015A&A...573A.118L,2015ApJ...805..104S,2016A&A...592A..37T}. 
We obtain 
a 
f$_{\textrm{gas}}$
of $\sim$ 9\% within R$_{200}$ for MKW4,
which is remarkably 
small 
compared to the cosmic baryon fraction 
(f$_\textrm{b}$) of 15\%.
\begin{figure}
\hspace{-4pt} \includegraphics[width=\columnwidth]{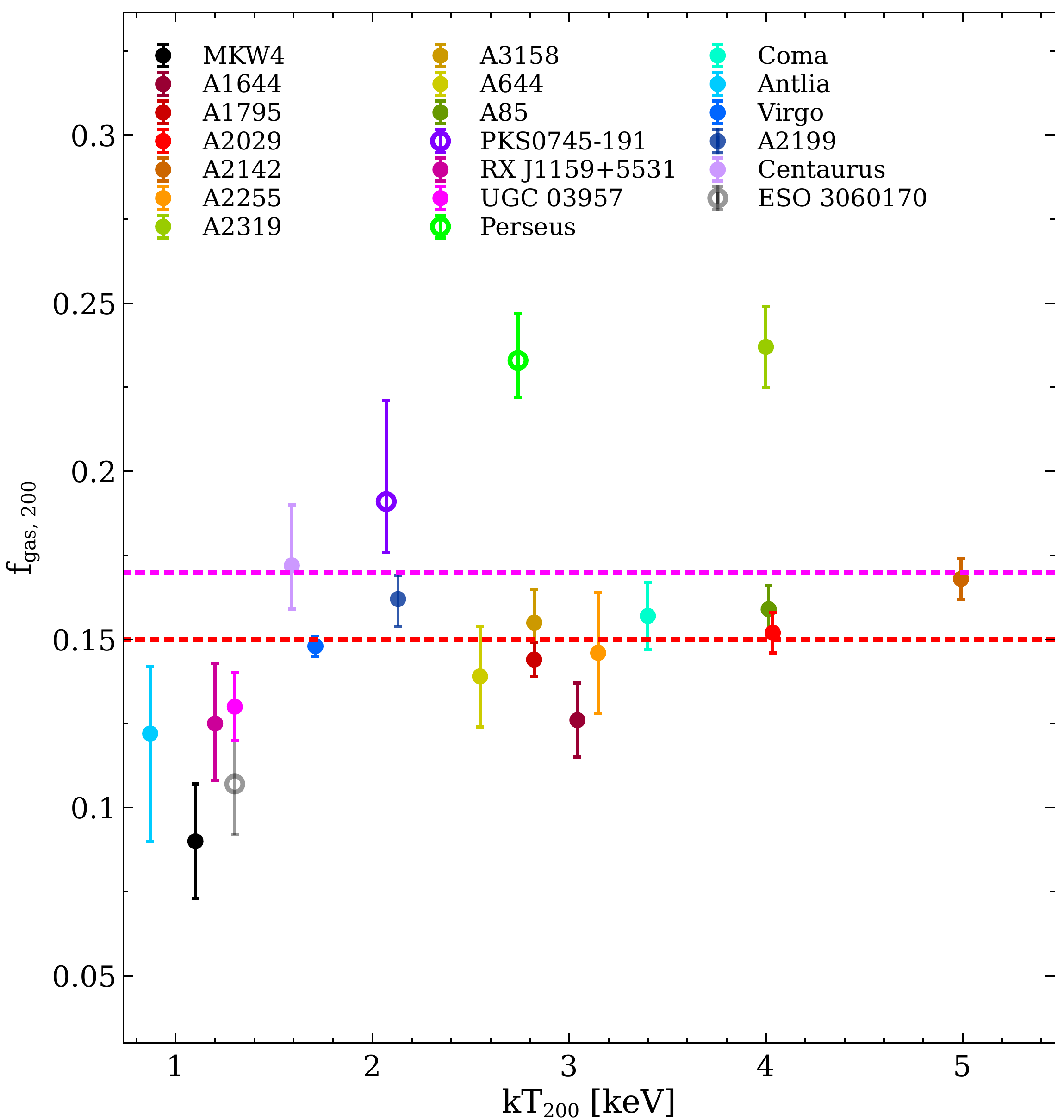}
\caption{The f$_{\rm gas}$ is plotted with the
cluster temperature at R$_{200}$. 
Open circles represent f$_{\rm gas}$ obtained
using $\suzaku$ only and filled circles represent
f$_{\rm gas}$ obtained using $\xmm$ + $\plnk$ or $\suzaku$ + 
$\chandra$.
Color codes for different
clusters are shown in the plot.
The f$_{\rm gas}$ for A1644, A1795, A2029, A2142, A2255, A2319,
A3158, A644, and A85 are taken from \citet{2019A&A...621A..40E}.
We obtain f$_{\rm gas}$ of PKS0745-191 from \citet{2012MNRAS.424.1826W},
RX J1159+5531 from \citet{2015ApJ...805..104S}, UGC 03957 from
\citet{2016A&A...592A..37T}, Coma from 
\citet{2020MNRAS.497.3204M}, Perseus from \citet{Simionescu1576}, Antlia from
\citet{2016ApJ...829...49W}, Virgo from 
\citet{2017MNRAS.469.1476S}, A2199 from 
\citet{2020MNRAS.497.3943M}, Centaurus from
\citet{2013MNRAS.432..554W}, and ESO 3060170 from
\citet{2013ApJ...775...89S}.
Red and magenta dashed lines 
shows cosmic baryon fraction 
estimated using {\sl Planck} 
\citep{2014A&A...571A..16P} and {\sl WMAP} 
\citep{Komatsu_2011} respectively.
}
\label{fig:fgas_t}
\end{figure}
We estimate the stellar mass of MKW4
using Two Micron All-Sky Survey 
(2MASS\footnote[4]{\url{https://irsa.ipac.caltech.edu/applications/2MASS/IM/interactive.html}}) 
data in K$_\textrm{s}$ band, assuming
a stellar mass-to-light ratio of $\sim$ 1 
\citep{2003ApJS..149..289B}. We find that
the stellar 
mass contributes nearly 2\% to the total 
hydrostatic mass, which leads to an
enclosed baryon 
fraction (f$_\textrm{b}$) of $\sim$ 
11\% within 
R$_{200}$ of MKW4, still significantly
lower than the cosmic f$_\textrm{b}$.

Similar discrepancies 
have been found in galaxies 
\citep[e.g.,][]{1998ApJ...503..518F,2005ApJ...635...73H,2010ApJ...719..119D}, 
the so called “missing baryon problem”, which 
remains a grand challenge problem in understanding 
the galaxy evolution 
\citep[e.g.,][]{1998ApJ...503..518F,2010ApJ...719..119D}. 
In the case of MKW4, internal heating caused by AGN 
feedback may have pushed the hot gas towards larger 
radii or even expelled it from the system 
\citep[e.g.,][]{1994ApJ...437..564M,2008MNRAS.390.1399B},
which reduces the amount of hot gas, hence the 
baryon fraction.
Alternatively, a significant amount of hot gas 
associated with the individual galaxies may have 
condensed and cooled out of the X-ray emitting ICM 
and reside in the Circumgalactic Medium (CGM) in the
form of cold gas 
\citep[e.g.,][]{2017MNRAS.466.3810F}
that is undetectable in X-ray and near-infrared
(K$_\textrm{s}$ band).
Massive systems such as clusters of galaxies 
typically have a f$_\textrm{b}$ consistent with the 
cosmic f$_\textrm{b}$ at the virial radii
\citep[e.g.,][]{10.1111/j.1365-2966.2012.21282.x,2013MNRAS.428.2812B}, while the 
f$_\textrm{b}$ of galaxies is approximately 7\% 
\citep{2005ApJ...635...73H} at their
virial radii. 

We note that MKW4, as a
galaxy group, has a 
f$_\textrm{b}$ within R$_{200}$ between those of 
galaxies and clusters. 
{ We compare our estimated
f$_{\textrm{gas}}$ at R$_{200}$ 
(f$_{\textrm{gas, 200}}$) with those of 
other clusters and groups as a function of cluster
temperature at R$_{200}$, 
as shown in Figure \ref{fig:fgas_t}.
The 
f$_{\rm gas, 200}$
of lower-mass clusters tend to
stay below the cosmic baryon fraction.
In contrast, the intermediate and higher-mass
clusters have f$_{\rm gas, 200}$ consistent 
with the cosmic baryon fraction. 
Gravitational potential may have played a critical role 
in retaining hot baryons inside a galaxy cluster.

The measured f$_{\rm gas, 200}$ of 
PKS 0745, Perseus, and A2319 exceed the 
cosmic baryon 
fraction. The f$_{\rm gas, 200}$ of 
PKS 0745 and Perseus were measured with $\suzaku$ alone,
which may introduce positive bias
in the gas mass measurement due to 
unresolved cool gas clumps at their outskirts.
A2319 is a merging cluster, for which
\citet{2018A&A...614A...7G} report
a substantial non-thermal pressure support at its 
outskirts, which may have biased the hydrostatic mass
low.
Our estimated f$_{\rm gas, 200}$ in MKW4 is consistent
with other groups of similar masses,
as seen in Figure 
\ref{fig:fgas_t}.
The 
low f$_{\rm gas, 200}$ found in MKW4 implies that 
its X-ray hydrostatic mass is unlikely to be 
underestimated at R$_{200}$.
}

\subsection{Azimuthal Scatter}
\begin{figure}
\hspace{-4pt} \includegraphics[width=\columnwidth]{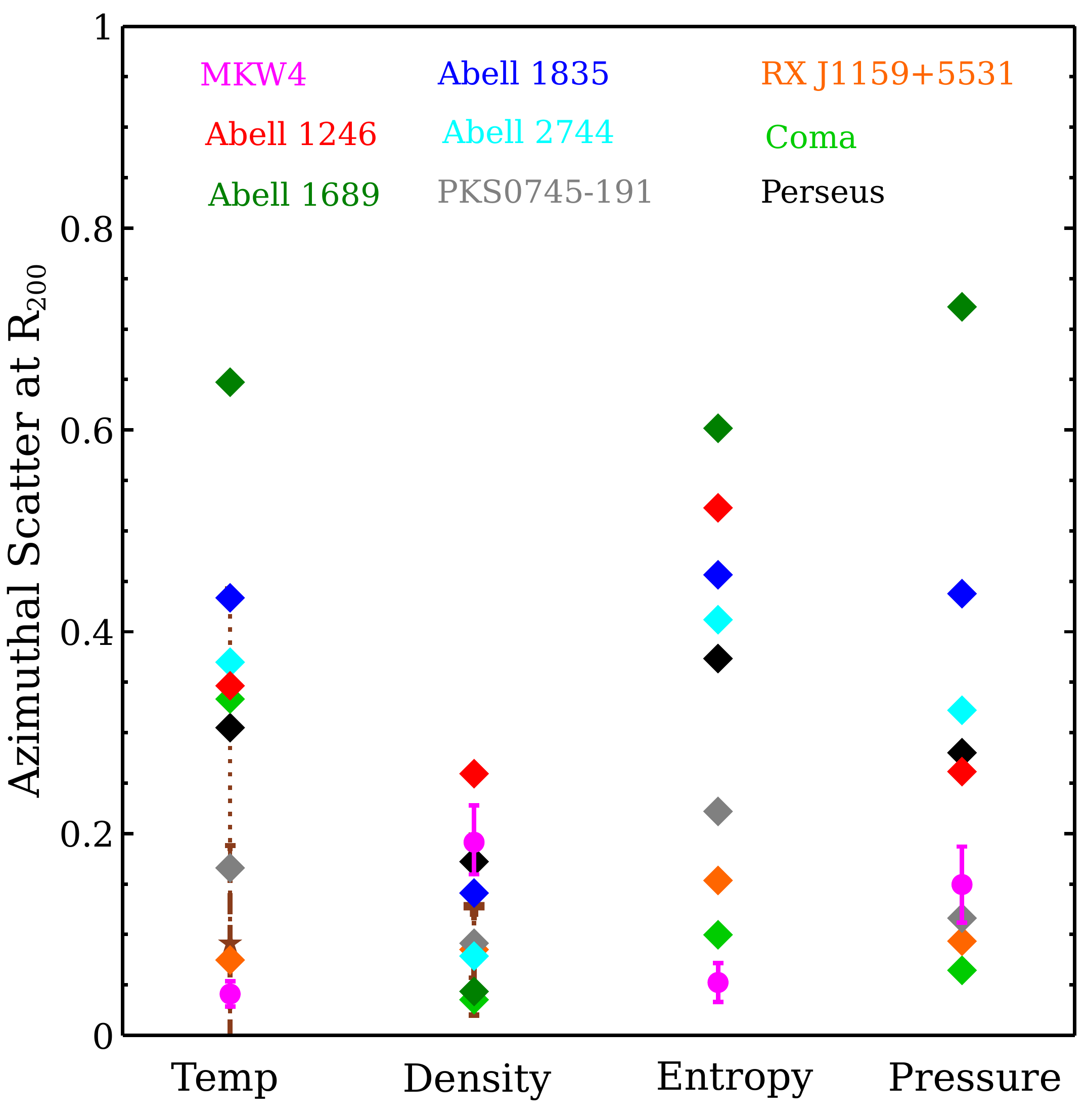}
\caption{
Azimuthal scatter of gas properties at R$_{200}$ of 
MKW4. We compare our results with azimuthal 
variations in other clusters, e.g., A1246: 
\citet{10.1093/pasj/psu061}; A1689: 
\citet{Kawaharada_2010}; A1835: 
\citet{Ichikawa_2013}; A2744: 
\citet{2014A&A...562A..11I}; PKS0745-191: 
\citet{10.1111/j.1365-2966.2009.14547.x}; 
RJX1159+5531: \citet{2015ApJ...805..104S}; Coma: 
\citet{2013ApJ...775....4S}; Perseus: 
\citet{10.1093/mnras/stt2209}. Dashed brown 
line: the azimuthal variation in relaxed simulated 
clusters from \citet{10.1111/j.1365-2966.2010.18120.x}. Dotted brown line: the azimuthal variation in 
perturbed simulated clusters 
from \citet{10.1111/j.1365-2966.2010.18120.x}.}
\label{fig:azi}
\end{figure}

Azimuthal variations in the ICM properties may 
arise for different reasons, including unresolved 
substructures, mergers, and gas clumping  
\citep[e.g.,][]{2012AIPC.1427...13M,2015ApJ...805..104S}.
We adopt the formula given by 
\citet{10.1111/j.1365-2966.2010.18120.x} to 
evaluate the azimuthal scatter of 
different gas properties 
at the outskirts
of MKW4 and compare our results with the other clusters and groups-
\begin{equation}
    \textrm{S}_\textrm{c}(\textrm{r}) = \sqrt{\frac{1}{\textrm{N}} \sum_\textrm{i} \frac{\left [\textrm{y}_\textrm{i}(\textrm{r}) - \textrm{Y}(\textrm{r}) \right ]^2}{\left [ \textrm{Y}(\textrm{r}) \right ]^2}} ,
\end{equation}
where y$_\textrm{i}(\textrm{r})$ is the radial profile of a 
quantity for the i$^{\textrm{th}}$-section,  
Y(r) is the azimuthal 
average of that quantity, and N
is the total 
number of sections. 
We obtain azimuthal scatters of gas properties
at R$_{200}$ of MKW4, as shown in 
Figure \ref{fig:azi}. { We find 
S$_\textrm{c}(\textrm{R}_{200})$ = 0.041 
$\pm$ 0.012 in temperature, 0.191 $\pm$ 0.036 
in density, 0.052 $\pm$ 0.019 in entropy, 
0.149 $\pm$ 0.037 in pressure, and 
0.020 $\pm$ 
0.009 in the gas mass fraction. }

Numerical simulations show that the
azimuthal variation in gas density and 
temperature rises to $\sim$ 10\% at R$_{200}$ 
for relaxed clusters, while it 
increases to $\sim$ 50-80\%
for perturbed clusters 
\citep{10.1111/j.1365-2966.2010.18120.x}, 
as shown in Figure \ref{fig:azi}. The 
azimuthal scatters of the gas properties 
at the outskirts of MKW4 are consistent 
with the expectation for relaxed clusters. 
We also compare our results with a number of 
galaxy clusters observed with $\suzaku$ out 
to R$_{200}$ with $N$ = 3 or more 
azimuthal coverages, as shown in Figure 
\ref{fig:azi}. The azimuthal variations in 
temperature and entropy of MKW4 are 
significantly smaller among those clusters. 
The similar ICM properties at R$_{200}$ in the 
north, east, and north-east directions confirm 
that MKW4 is a spherically symmetric relaxed 
system, and the hydrostatic equilibrium 
is likely to be a good approximation even out to its
virial radii.

\section{Summary}
We have analysed joint $\suzaku$ and $\chandra$ 
observations of the galaxy group MKW4 and measured 
its ICM properties from its center out to the 
virial radius. We have derived the radial 
profiles of gas density, temperature, entropy, 
pressure, and gas mass fraction in three 
different directions. Our findings are 
summarized below.
\begin{itemize}
	\item Using $\chandra$ observations of MKW4, 
we have resolved much of the CXB at its 
center and three outskirt regions. 
{ We are able to model the 
contamination from point sources in $\suzaku$ data
and greatly
reduce the uncertainties in the measurement of 
ICM properties.}
\end{itemize}

\begin{itemize}
	\item The 3D gas densities of MKW4
decline to $\sim$ 10$^{-5}$ cm$^{-3}$ at its 
outskirts and are consistent among 
the three directions.
The 3D temperature profiles decline
from 
2.2 keV at 0.1R$_{200}$ to $\sim$ 1.14 keV 
at the virial radius. { The 
temperature 
profiles of MKW4 in the three directions follow the 
universal profiles are derived 
by \citet{2002ApJ...579..571L} and \citet{Sun_2009}.}
\end{itemize}

\begin{itemize}
	\item The entropy profiles of 
MKW4 follow the 
baseline profile \citep{2005MNRAS.364..909V} 
beyond R$_{500}$ in the north, 
east, and north-east directions, 
which indicates that the gas dynamics 
at the group outskirts is mainly 
regulated by the gravitational collapse. 
Our results also show entropy 
excess towards the group center compared 
to the baseline profile, suggesting that the 
central AGN of MKW4 may have redistributed
the hot gas.
\end{itemize}

\begin{itemize}
	\item We estimated the total X-ray 
hydrostatic mass of MKW4, 
$\textrm{M}_\textrm{tot}(<\textrm{R}_{500})$ 
$\approx$ 6.5 $\pm$ 1.0 
$\times$ 10$^{13}$ $\textrm{M}_{\odot}$ and 
$\textrm{M}_\textrm{tot}(<\textrm{R}_{200})$ 
$\approx$ 9.7 $\pm$ 1.5 
$\times$ 10$^{13}$ $\textrm{M}_{\odot}$. 
Our measurement shows that its baryon fraction
is only $\sim$ 11\% at the virial 
radii, which is lower than the 
cosmic baryon fraction 
\citep{Komatsu_2011,2014A&A...571A..16P}. 
The lower baryon fraction implies 
that the central AGN feedback 
or galactic winds may have
expelled much of its hot gas at its early epoch,
or/and the hot gas 
associated with the individual galaxies may have 
condensed and cooled out of the X-ray emitting ICM 
and reside in the CGM in the
form of cold gas.
\end{itemize}

\begin{itemize}
	\item The azimuthal 
scatter in gas properties  
at the outskirts of MKW4 is 
small, 
suggesting that it is a remarkably 
relaxed system and the bulk of its
ICM is likely to be in hydrostatic equilibrium. 
\end{itemize}

\section*{Acknowledgements}
{
 {We thank the anonymous referee for his or her
helpful suggestions.}
A. S. and Y. S. were partially supported by the
Smithsonian Astrophysical Observatory grants 
AR8-19020A and GO6-17125A. 

}

\section*{Data Availability}
{ The data underlying this article will be shared
on reasonable request to the corresponding author.}




\bibliographystyle{mnras}
\bibliography{sample} 



\bsp	
\label{lastpage}
\end{document}

%% file: macros.tex
\font\manual=manfnt at 7pt \def\dbend{\hbox{\raise0.9ex\hbox{\manual\char127\hspace{0.6em}}}}

\newcounter{INTERNALionstage}

\def\gtsim{\mathrel{\hbox{\rlap{\hbox{\lower4pt\hbox{$\sim$}}}\hbox{$>$}}}}
\def\lesssim{\mathrel{\hbox{\rlap{\hbox{\lower4pt\hbox{$\sim$}}}\hbox{$<$}}}}

\def\ga{{\rm\thinspace gauss}}

\def\la{\mbox{{\rm L}$\alpha$}}

\def\h0{\mbox{{\rm H}$^0$}}
\DeclareMathAlphabet{\vib}{OML}{cmm}{m}{it}
\newcommand*{\satellite}[1]{\textit{#1}}
\newcommand*{\xmm}{\satellite{XMM-Newton}}
\newcommand*{\plnk}{\satellite{Planck}}
\newcommand*{\chandra}{\satellite{Chandra}}
\newcommand*{\suzaku}{\satellite{Suzaku}}



%% file: sample.bbl
\begin{thebibliography}{}
\makeatletter
\relax
\def\mn@urlcharsother{\let\do\@makeother \do\$\do\&\do\#\do\^\do\_\do\%\do\~}
\def\mn@doi{\begingroup\mn@urlcharsother \@ifnextchar [ {\mn@doi@}
  {\mn@doi@[]}}
\def\mn@doi@[#1]#2{\def\@tempa{#1}\ifx\@tempa\@empty \href
  {http://dx.doi.org/#2} {doi:#2}\else \href {http://dx.doi.org/#2} {#1}\fi
  \endgroup}
\def\mn@eprint#1#2{\mn@eprint@#1:#2::\@nil}
\def\mn@eprint@arXiv#1{\href {http://arxiv.org/abs/#1} {{\tt arXiv:#1}}}
\def\mn@eprint@dblp#1{\href {http://dblp.uni-trier.de/rec/bibtex/#1.xml}
  {dblp:#1}}
\def\mn@eprint@#1:#2:#3:#4\@nil{\def\@tempa {#1}\def\@tempb {#2}\def\@tempc
  {#3}\ifx \@tempc \@empty \let \@tempc \@tempb \let \@tempb \@tempa \fi \ifx
  \@tempb \@empty \def\@tempb {arXiv}\fi \@ifundefined
  {mn@eprint@\@tempb}{\@tempb:\@tempc}{\expandafter \expandafter \csname
  mn@eprint@\@tempb\endcsname \expandafter{\@tempc}}}

\bibitem[\protect\citeauthoryear{Akamatsu, Hoshino, Ishisaki, Ohashi, Sato,
  Takei  \& Ota}{Akamatsu et~al.}{2011}]{10.1093/pasj/63.sp3.S1019}
Akamatsu H.,  Hoshino A.,  Ishisaki Y.,  Ohashi T.,  Sato K.,  Takei Y.,   Ota
  N.,  2011, \mn@doi [Publications of the Astronomical Society of Japan]
  {10.1093/pasj/63.sp3.S1019}, 63, S1019

\bibitem[\protect\citeauthoryear{Anders \& Grevesse}{Anders \&
  Grevesse}{1989}]{ANDERS1989197}
Anders E.,  Grevesse N.,  1989, \mn@doi [Geochimica et Cosmochimica Acta]
  {https://doi.org/10.1016/0016-7037(89)90286-X}, 53, 197

\bibitem[\protect\citeauthoryear{{Arnaud}}{{Arnaud}}{2009}]{2009A&A...500..103A}
{Arnaud} M.,  2009, \mn@doi [\aap] {10.1051/0004-6361/200912150}, \href
  {https://ui.adsabs.harvard.edu/abs/2009A&A...500..103A} {500, 103}

\bibitem[\protect\citeauthoryear{{Arnaud}, {Pratt}, {Piffaretti},
  {B{\"o}hringer}, {Croston}  \& {Pointecouteau}}{{Arnaud}
  et~al.}{2010}]{2010A&A...517A..92A}
{Arnaud} M.,  {Pratt} G.~W.,  {Piffaretti} R.,  {B{\"o}hringer} H.,  {Croston}
  J.~H.,   {Pointecouteau} E.,  2010, \mn@doi [\aap]
  {10.1051/0004-6361/200913416}, \href
  {https://ui.adsabs.harvard.edu/abs/2010A&A...517A..92A} {517, A92}

\bibitem[\protect\citeauthoryear{Asplund, Grevesse  \& Sauval]}{Asplund
  et~al.}{2006}]{ASPLUND20061}
Asplund M.,  Grevesse N.,   Sauval] A.~J.,  2006, \mn@doi [Nuclear Physics A]
  {https://doi.org/10.1016/j.nuclphysa.2005.06.010}, 777, 1

\bibitem[\protect\citeauthoryear{{Avestruz}, {Nagai}, {Lau}  \&
  {Nelson}}{{Avestruz} et~al.}{2015}]{2015ApJ...808..176A}
{Avestruz} C.,  {Nagai} D.,  {Lau} E.~T.,   {Nelson} K.,  2015, \mn@doi [\apj]
  {10.1088/0004-637X/808/2/176}, \href
  {https://ui.adsabs.harvard.edu/abs/2015ApJ...808..176A} {808, 176}

\bibitem[\protect\citeauthoryear{{Balucinska-Church} \&
  {McCammon}}{{Balucinska-Church} \& {McCammon}}{1992}]{1992ApJ...400..699B}
{Balucinska-Church} M.,  {McCammon} D.,  1992, \mn@doi [\apj] {10.1086/172032},
  \href {https://ui.adsabs.harvard.edu/abs/1992ApJ...400..699B} {400, 699}

\bibitem[\protect\citeauthoryear{{Bell}, {McIntosh}, {Katz}  \&
  {Weinberg}}{{Bell} et~al.}{2003}]{2003ApJS..149..289B}
{Bell} E.~F.,  {McIntosh} D.~H.,  {Katz} N.,   {Weinberg} M.~D.,  2003, \mn@doi
  [\apjs] {10.1086/378847}, \href
  {https://ui.adsabs.harvard.edu/abs/2003ApJS..149..289B} {149, 289}

\bibitem[\protect\citeauthoryear{{Bonamente}, {Landry}, {Maughan}, {Giles},
  {Joy}  \& {Nevalainen}}{{Bonamente} et~al.}{2013}]{2013MNRAS.428.2812B}
{Bonamente} M.,  {Landry} D.,  {Maughan} B.,  {Giles} P.,  {Joy} M.,
  {Nevalainen} J.,  2013, \mn@doi [\mnras] {10.1093/mnras/sts202}, \href
  {https://ui.adsabs.harvard.edu/abs/2013MNRAS.428.2812B} {428, 2812}

\bibitem[\protect\citeauthoryear{Borgani, Finoguenov, Kay, Ponman, Springel,
  Tozzi  \& Voit}{Borgani et~al.}{2005}]{10.1111/j.1365-2966.2005.09158.x}
Borgani S.,  Finoguenov A.,  Kay S.~T.,  Ponman T.~J.,  Springel V.,  Tozzi P.,
    Voit G.~M.,  2005, \mn@doi [Monthly Notices of the Royal Astronomical
  Society] {10.1111/j.1365-2966.2005.09158.x}, 361, 233

\bibitem[\protect\citeauthoryear{{Bower}, {McCarthy}  \& {Benson}}{{Bower}
  et~al.}{2008}]{2008MNRAS.390.1399B}
{Bower} R.~G.,  {McCarthy} I.~G.,   {Benson} A.~J.,  2008, \mn@doi [\mnras]
  {10.1111/j.1365-2966.2008.13869.x}, \href
  {https://ui.adsabs.harvard.edu/abs/2008MNRAS.390.1399B} {390, 1399}

\bibitem[\protect\citeauthoryear{{Bulbul}, {Markevitch}, {Foster}, {Miller},
  {Bautz}, {Loewenstein}, {Rand all}  \& {Smith}}{{Bulbul}
  et~al.}{2016}]{2016ApJ...831...55B}
{Bulbul} E.,  {Markevitch} M.,  {Foster} A.,  {Miller} E.,  {Bautz} M.,
  {Loewenstein} M.,  {Rand all} S.~W.,   {Smith} R.~K.,  2016, \mn@doi [\apj]
  {10.3847/0004-637X/831/1/55}, \href
  {https://ui.adsabs.harvard.edu/abs/2016ApJ...831...55B} {831, 55}

\bibitem[\protect\citeauthoryear{{Corasaniti}, {Ettori}, {Rasera}, {Sereno},
  {Amodeo}, {Breton}, {Ghirardini}  \& {Eckert}}{{Corasaniti}
  et~al.}{2018}]{2018ApJ...862...40C}
{Corasaniti} P.~S.,  {Ettori} S.,  {Rasera} Y.,  {Sereno} M.,  {Amodeo} S.,
  {Breton} M.~A.,  {Ghirardini} V.,   {Eckert} D.,  2018, \mn@doi [\apj]
  {10.3847/1538-4357/aaccdf}, \href
  {https://ui.adsabs.harvard.edu/abs/2018ApJ...862...40C} {862, 40}

\bibitem[\protect\citeauthoryear{{Dai}, {Bregman}, {Kochanek}  \&
  {Rasia}}{{Dai} et~al.}{2010}]{2010ApJ...719..119D}
{Dai} X.,  {Bregman} J.~N.,  {Kochanek} C.~S.,   {Rasia} E.,  2010, \mn@doi
  [\apj] {10.1088/0004-637X/719/1/119}, \href
  {https://ui.adsabs.harvard.edu/abs/2010ApJ...719..119D} {719, 119}

\bibitem[\protect\citeauthoryear{{De Luca} \& {Molendi}}{{De Luca} \&
  {Molendi}}{2004}]{2004A&A...419..837D}
{De Luca} A.,  {Molendi} S.,  2004, \mn@doi [\aap]
  {10.1051/0004-6361:20034421}, \href
  {https://ui.adsabs.harvard.edu/abs/2004A&A...419..837D} {419, 837}

\bibitem[\protect\citeauthoryear{{Eckert} et~al.,}{{Eckert}
  et~al.}{2019}]{2019A&A...621A..40E}
{Eckert} D.,  et~al., 2019, \mn@doi [\aap] {10.1051/0004-6361/201833324}, \href
  {https://ui.adsabs.harvard.edu/abs/2019A&A...621A..40E} {621, A40}

\bibitem[\protect\citeauthoryear{Eke et~al.,}{Eke
  et~al.}{2004}]{10.1111/j.1365-2966.2004.07408.x}
Eke V.~R.,  et~al., 2004, \mn@doi [Monthly Notices of the Royal Astronomical
  Society] {10.1111/j.1365-2966.2004.07408.x}, 348, 866

\bibitem[\protect\citeauthoryear{{Fielding}, {Quataert}, {McCourt}  \&
  {Thompson}}{{Fielding} et~al.}{2017}]{2017MNRAS.466.3810F}
{Fielding} D.,  {Quataert} E.,  {McCourt} M.,   {Thompson} T.~A.,  2017,
  \mn@doi [\mnras] {10.1093/mnras/stw3326}, \href
  {https://ui.adsabs.harvard.edu/abs/2017MNRAS.466.3810F} {466, 3810}

\bibitem[\protect\citeauthoryear{{Fukugita}, {Hogan}  \& {Peebles}}{{Fukugita}
  et~al.}{1998}]{1998ApJ...503..518F}
{Fukugita} M.,  {Hogan} C.~J.,   {Peebles} P.~J.~E.,  1998, \mn@doi [\apj]
  {10.1086/306025}, \href
  {https://ui.adsabs.harvard.edu/abs/1998ApJ...503..518F} {503, 518}

\bibitem[\protect\citeauthoryear{{Gastaldello}, {Buote}, {Humphrey},
  {Zappacosta}, {Bullock}, {Brighenti}  \& {Mathews}}{{Gastaldello}
  et~al.}{2007}]{2007ApJ...669..158G}
{Gastaldello} F.,  {Buote} D.~A.,  {Humphrey} P.~J.,  {Zappacosta} L.,
  {Bullock} J.~S.,  {Brighenti} F.,   {Mathews} W.~G.,  2007, \mn@doi [\apj]
  {10.1086/521519}, \href
  {https://ui.adsabs.harvard.edu/abs/2007ApJ...669..158G} {669, 158}

\bibitem[\protect\citeauthoryear{George, Fabian, Sanders, Young  \&
  Russell}{George et~al.}{2009}]{10.1111/j.1365-2966.2009.14547.x}
George M.~R.,  Fabian A.~C.,  Sanders J.~S.,  Young A.~J.,   Russell H.~R.,
  2009, \mn@doi [Monthly Notices of the Royal Astronomical Society]
  {10.1111/j.1365-2966.2009.14547.x}, 395, 657

\bibitem[\protect\citeauthoryear{{Ghirardini}, {Ettori}, {Eckert}, {Molendi},
  {Gastaldello}, {Pointecouteau}, {Hurier}  \& {Bourdin}}{{Ghirardini}
  et~al.}{2018}]{2018A&A...614A...7G}
{Ghirardini} V.,  {Ettori} S.,  {Eckert} D.,  {Molendi} S.,  {Gastaldello} F.,
  {Pointecouteau} E.,  {Hurier} G.,   {Bourdin} H.,  2018, \mn@doi [\aap]
  {10.1051/0004-6361/201731748}, \href
  {https://ui.adsabs.harvard.edu/abs/2018A&A...614A...7G} {614, A7}

\bibitem[\protect\citeauthoryear{{Hoekstra}, {Hsieh}, {Yee}, {Lin}  \&
  {Gladders}}{{Hoekstra} et~al.}{2005}]{2005ApJ...635...73H}
{Hoekstra} H.,  {Hsieh} B.~C.,  {Yee} H.~K.~C.,  {Lin} H.,   {Gladders} M.~D.,
  2005, \mn@doi [\apj] {10.1086/496913}, \href
  {https://ui.adsabs.harvard.edu/abs/2005ApJ...635...73H} {635, 73}

\bibitem[\protect\citeauthoryear{Hoshino et~al.,}{Hoshino
  et~al.}{2010}]{10.1093/pasj/62.2.371}
Hoshino A.,  et~al., 2010, \mn@doi [Publications of the Astronomical Society of
  Japan] {10.1093/pasj/62.2.371}, 62, 371

\bibitem[\protect\citeauthoryear{{Humphrey}, {Buote}, {Brighenti}, {Flohic},
  {Gastaldello}  \& {Mathews}}{{Humphrey} et~al.}{2012}]{2012ApJ...748...11H}
{Humphrey} P.~J.,  {Buote} D.~A.,  {Brighenti} F.,  {Flohic} H. M.~L.~G.,
  {Gastaldello} F.,   {Mathews} W.~G.,  2012, \mn@doi [\apj]
  {10.1088/0004-637X/748/1/11}, \href
  {https://ui.adsabs.harvard.edu/abs/2012ApJ...748...11H} {748, 11}

\bibitem[\protect\citeauthoryear{{Ibaraki}, {Ota}, {Akamatsu}, {Zhang}  \&
  {Finoguenov}}{{Ibaraki} et~al.}{2014}]{2014A&A...562A..11I}
{Ibaraki} Y.,  {Ota} N.,  {Akamatsu} H.,  {Zhang} Y.~Y.,   {Finoguenov} A.,
  2014, \mn@doi [\aap] {10.1051/0004-6361/201322806}, \href
  {https://ui.adsabs.harvard.edu/abs/2014A&A...562A..11I} {562, A11}

\bibitem[\protect\citeauthoryear{Ichikawa et~al.,}{Ichikawa
  et~al.}{2013}]{Ichikawa_2013}
Ichikawa K.,  et~al., 2013, \mn@doi [The Astrophysical Journal]
  {10.1088/0004-637x/766/2/90}, 766, 90

\bibitem[\protect\citeauthoryear{Kawaharada et~al.,}{Kawaharada
  et~al.}{2010}]{Kawaharada_2010}
Kawaharada M.,  et~al., 2010, \mn@doi [The Astrophysical Journal]
  {10.1088/0004-637x/714/1/423}, 714, 423

\bibitem[\protect\citeauthoryear{Komatsu et~al.,}{Komatsu
  et~al.}{2011}]{Komatsu_2011}
Komatsu E.,  et~al., 2011, \mn@doi [The Astrophysical Journal Supplement
  Series] {10.1088/0067-0049/192/2/18}, 192, 18

\bibitem[\protect\citeauthoryear{{Le Brun}, {McCarthy}, {Schaye}  \&
  {Ponman}}{{Le Brun} et~al.}{2014}]{2014MNRAS.441.1270L}
{Le Brun} A. M.~C.,  {McCarthy} I.~G.,  {Schaye} J.,   {Ponman} T.~J.,  2014,
  \mn@doi [\mnras] {10.1093/mnras/stu608}, \href
  {https://ui.adsabs.harvard.edu/abs/2014MNRAS.441.1270L} {441, 1270}

\bibitem[\protect\citeauthoryear{Lodders}{Lodders}{2003}]{Lodders_2003}
Lodders K.,  2003, \mn@doi [The Astrophysical Journal] {10.1086/375492}, 591,
  1220

\bibitem[\protect\citeauthoryear{{Loken}, {Norman}, {Nelson}, {Burns}, {Bryan}
  \& {Motl}}{{Loken} et~al.}{2002}]{2002ApJ...579..571L}
{Loken} C.,  {Norman} M.~L.,  {Nelson} E.,  {Burns} J.,  {Bryan} G.~L.,
  {Motl} P.,  2002, \mn@doi [\apj] {10.1086/342825}, \href
  {https://ui.adsabs.harvard.edu/abs/2002ApJ...579..571L} {579, 571}

\bibitem[\protect\citeauthoryear{{Lovisari}, {Reiprich}  \&
  {Schellenberger}}{{Lovisari} et~al.}{2015}]{2015A&A...573A.118L}
{Lovisari} L.,  {Reiprich} T.~H.,   {Schellenberger} G.,  2015, \mn@doi [\aap]
  {10.1051/0004-6361/201423954}, \href
  {https://ui.adsabs.harvard.edu/abs/2015A&A...573A.118L} {573, A118}

\bibitem[\protect\citeauthoryear{Mathews \& Guo}{Mathews \&
  Guo}{2011}]{Mathews_2011}
Mathews W.~G.,  Guo F.,  2011, \mn@doi [The Astrophysical Journal]
  {10.1088/0004-637x/738/2/155}, 738, 155

\bibitem[\protect\citeauthoryear{{McCarthy} et~al.,}{{McCarthy}
  et~al.}{2010}]{2010MNRAS.406..822M}
{McCarthy} I.~G.,  et~al., 2010, \mn@doi [\mnras]
  {10.1111/j.1365-2966.2010.16750.x}, \href
  {https://ui.adsabs.harvard.edu/abs/2010MNRAS.406..822M} {406, 822}

\bibitem[\protect\citeauthoryear{{Metzler} \& {Evrard}}{{Metzler} \&
  {Evrard}}{1994}]{1994ApJ...437..564M}
{Metzler} C.~A.,  {Evrard} A.~E.,  1994, \mn@doi [\apj] {10.1086/175022}, \href
  {https://ui.adsabs.harvard.edu/abs/1994ApJ...437..564M} {437, 564}

\bibitem[\protect\citeauthoryear{{Miller}, {Bautz}, {George}, {Mushotzky},
  {Davis}  \& {Henry}}{{Miller} et~al.}{2012}]{2012AIPC.1427...13M}
{Miller} E.~D.,  {Bautz} M.,  {George} J.,  {Mushotzky} R.,  {Davis} D.,
  {Henry} J.~P.,  2012, in {Petre} R.,  {Mitsuda} K.,   {Angelini} L.,  eds,
  American Institute of Physics Conference Series Vol. 1427, American Institute
  of Physics Conference Series. pp 13--20 (\mn@eprint {arXiv} {1112.0034}),
  \mn@doi{10.1063/1.3696144}

\bibitem[\protect\citeauthoryear{{Mirakhor} \& {Walker}}{{Mirakhor} \&
  {Walker}}{2020a}]{2020MNRAS.497.3204M}
{Mirakhor} M.~S.,  {Walker} S.~A.,  2020a, \mn@doi [\mnras]
  {10.1093/mnras/staa2203}, \href
  {https://ui.adsabs.harvard.edu/abs/2020MNRAS.497.3204M} {497, 3204}

\bibitem[\protect\citeauthoryear{{Mirakhor} \& {Walker}}{{Mirakhor} \&
  {Walker}}{2020b}]{2020MNRAS.497.3943M}
{Mirakhor} M.~S.,  {Walker} S.~A.,  2020b, \mn@doi [\mnras]
  {10.1093/mnras/staa2204}, \href
  {https://ui.adsabs.harvard.edu/abs/2020MNRAS.497.3943M} {497, 3943}

\bibitem[\protect\citeauthoryear{Mitsuda et~al.,}{Mitsuda
  et~al.}{2007}]{10.1093/pasj/59.sp1.S1}
Mitsuda K.,  et~al., 2007, \mn@doi [Publications of the Astronomical Society of
  Japan] {10.1093/pasj/59.sp1.S1}, 59, S1

\bibitem[\protect\citeauthoryear{{Moretti}, {Campana}, {Lazzati}  \&
  {Tagliaferri}}{{Moretti} et~al.}{2003}]{2003ApJ...588..696M}
{Moretti} A.,  {Campana} S.,  {Lazzati} D.,   {Tagliaferri} G.,  2003, \mn@doi
  [\apj] {10.1086/374335}, \href
  {https://ui.adsabs.harvard.edu/abs/2003ApJ...588..696M} {588, 696}

\bibitem[\protect\citeauthoryear{{Moretti} et~al.,}{{Moretti}
  et~al.}{2009}]{2009A&A...493..501M}
{Moretti} A.,  et~al., 2009, \mn@doi [\aap] {10.1051/0004-6361:200811197},
  \href {https://ui.adsabs.harvard.edu/abs/2009A&A...493..501M} {493, 501}

\bibitem[\protect\citeauthoryear{Nagai \& Lau}{Nagai \& Lau}{2011}]{Nagai_2011}
Nagai D.,  Lau E.~T.,  2011, \mn@doi [The Astrophysical Journal]
  {10.1088/2041-8205/731/1/l10}, 731, L10

\bibitem[\protect\citeauthoryear{{Navarro}, {Frenk}  \& {White}}{{Navarro}
  et~al.}{1997}]{1997ApJ...490..493N}
{Navarro} J.~F.,  {Frenk} C.~S.,   {White} S. D.~M.,  1997, \mn@doi [\apj]
  {10.1086/304888}, \href
  {https://ui.adsabs.harvard.edu/abs/1997ApJ...490..493N} {490, 493}

\bibitem[\protect\citeauthoryear{{O'Sullivan}, {Vrtilek}, {Read}, {David}  \&
  {Ponman}}{{O'Sullivan} et~al.}{2003}]{2003MNRAS.346..525O}
{O'Sullivan} E.,  {Vrtilek} J.~M.,  {Read} A.~M.,  {David} L.~P.,   {Ponman}
  T.~J.,  2003, \mn@doi [\mnras] {10.1046/j.1365-2966.2003.07108.x}, \href
  {https://ui.adsabs.harvard.edu/abs/2003MNRAS.346..525O} {346, 525}

\bibitem[\protect\citeauthoryear{{Paul}, {John}, {Gupta}  \& {Kumar}}{{Paul}
  et~al.}{2017}]{2017MNRAS.471....2P}
{Paul} S.,  {John} R.~S.,  {Gupta} P.,   {Kumar} H.,  2017, \mn@doi [\mnras]
  {10.1093/mnras/stx1488}, \href
  {https://ui.adsabs.harvard.edu/abs/2017MNRAS.471....2P} {471, 2}

\bibitem[\protect\citeauthoryear{{Planck Collaboration} et~al.,}{{Planck
  Collaboration} et~al.}{2014}]{2014A&A...571A..16P}
{Planck Collaboration} et~al., 2014, \mn@doi [\aap]
  {10.1051/0004-6361/201321591}, \href
  {https://ui.adsabs.harvard.edu/abs/2014A&A...571A..16P} {571, A16}

\bibitem[\protect\citeauthoryear{{Pratt} et~al.,}{{Pratt}
  et~al.}{2010}]{2010A&A...511A..85P}
{Pratt} G.~W.,  et~al., 2010, \mn@doi [\aap] {10.1051/0004-6361/200913309},
  \href {https://ui.adsabs.harvard.edu/abs/2010A&A...511A..85P} {511, A85}

\bibitem[\protect\citeauthoryear{{Sasaki}, {Matsushita}  \& {Sato}}{{Sasaki}
  et~al.}{2014}]{2014ApJ...781...36S}
{Sasaki} T.,  {Matsushita} K.,   {Sato} K.,  2014, \mn@doi [\apj]
  {10.1088/0004-637X/781/1/36}, \href
  {https://ui.adsabs.harvard.edu/abs/2014ApJ...781...36S} {781, 36}

\bibitem[\protect\citeauthoryear{Sato, Matsushita, Yamasaki, Sasaki  \&
  Ohashi}{Sato et~al.}{2014}]{10.1093/pasj/psu061}
Sato K.,  Matsushita K.,  Yamasaki N.~Y.,  Sasaki S.,   Ohashi T.,  2014,
  \mn@doi [Publications of the Astronomical Society of Japan]
  {10.1093/pasj/psu061}, 66

\bibitem[\protect\citeauthoryear{Simionescu et~al.,}{Simionescu
  et~al.}{2011}]{Simionescu1576}
Simionescu A.,  et~al., 2011, \mn@doi [Science] {10.1126/science.1200331}, 331,
  1576

\bibitem[\protect\citeauthoryear{{Simionescu} et~al.,}{{Simionescu}
  et~al.}{2013}]{2013ApJ...775....4S}
{Simionescu} A.,  et~al., 2013, \mn@doi [\apj] {10.1088/0004-637X/775/1/4},
  \href {https://ui.adsabs.harvard.edu/abs/2013ApJ...775....4S} {775, 4}

\bibitem[\protect\citeauthoryear{{Simionescu}, {Werner}, {Mantz}, {Allen}  \&
  {Urban}}{{Simionescu} et~al.}{2017}]{2017MNRAS.469.1476S}
{Simionescu} A.,  {Werner} N.,  {Mantz} A.,  {Allen} S.~W.,   {Urban} O.,
  2017, \mn@doi [\mnras] {10.1093/mnras/stx919}, \href
  {https://ui.adsabs.harvard.edu/abs/2017MNRAS.469.1476S} {469, 1476}

\bibitem[\protect\citeauthoryear{Springel \& Hernquist}{Springel \&
  Hernquist}{2003}]{10.1046/j.1365-8711.2003.06207.x}
Springel V.,  Hernquist L.,  2003, \mn@doi [Monthly Notices of the Royal
  Astronomical Society] {10.1046/j.1365-8711.2003.06207.x}, 339, 312

\bibitem[\protect\citeauthoryear{{Su}, {White}  \& {Miller}}{{Su}
  et~al.}{2013}]{2013ApJ...775...89S}
{Su} Y.,  {White} Raymond~E. I.,   {Miller} E.~D.,  2013, \mn@doi [\apj]
  {10.1088/0004-637X/775/2/89}, \href
  {https://ui.adsabs.harvard.edu/abs/2013ApJ...775...89S} {775, 89}

\bibitem[\protect\citeauthoryear{{Su}, {Buote}, {Gastaldello}  \&
  {Brighenti}}{{Su} et~al.}{2015}]{2015ApJ...805..104S}
{Su} Y.,  {Buote} D.,  {Gastaldello} F.,   {Brighenti} F.,  2015, \mn@doi
  [\apj] {10.1088/0004-637X/805/2/104}, \href
  {https://ui.adsabs.harvard.edu/abs/2015ApJ...805..104S} {805, 104}

\bibitem[\protect\citeauthoryear{Sun, Voit, Donahue, Jones, Forman  \&
  Vikhlinin}{Sun et~al.}{2009}]{Sun_2009}
Sun M.,  Voit G.~M.,  Donahue M.,  Jones C.,  Forman W.,   Vikhlinin A.,  2009,
  \mn@doi [The Astrophysical Journal] {10.1088/0004-637x/693/2/1142}, 693, 1142

\bibitem[\protect\citeauthoryear{{Sun}, {Sehgal}, {Voit}, {Donahue}, {Jones},
  {Forman}, {Vikhlinin}  \& {Sarazin}}{{Sun}
  et~al.}{2011}]{2011ApJ...727L..49S}
{Sun} M.,  {Sehgal} N.,  {Voit} G.~M.,  {Donahue} M.,  {Jones} C.,  {Forman}
  W.,  {Vikhlinin} A.,   {Sarazin} C.,  2011, \mn@doi [\apjl]
  {10.1088/2041-8205/727/2/L49}, \href
  {https://ui.adsabs.harvard.edu/abs/2011ApJ...727L..49S} {727, L49}

\bibitem[\protect\citeauthoryear{{Th{\"o}lken}, {Lovisari}, {Reiprich}  \&
  {Hasenbusch}}{{Th{\"o}lken} et~al.}{2016}]{2016A&A...592A..37T}
{Th{\"o}lken} S.,  {Lovisari} L.,  {Reiprich} T.~H.,   {Hasenbusch} J.,  2016,
  \mn@doi [\aap] {10.1051/0004-6361/201527608}, \href
  {https://ui.adsabs.harvard.edu/abs/2016A&A...592A..37T} {592, A37}

\bibitem[\protect\citeauthoryear{{Urban} et~al.,}{{Urban}
  et~al.}{2014a}]{2014MNRAS.437.3939U}
{Urban} O.,  et~al., 2014a, \mn@doi [\mnras] {10.1093/mnras/stt2209}, \href
  {https://ui.adsabs.harvard.edu/abs/2014MNRAS.437.3939U} {437, 3939}

\bibitem[\protect\citeauthoryear{Urban et~al.,}{Urban
  et~al.}{2014b}]{10.1093/mnras/stt2209}
Urban O.,  et~al., 2014b, \mn@doi [Monthly Notices of the Royal Astronomical
  Society] {10.1093/mnras/stt2209}, 437, 3939

\bibitem[\protect\citeauthoryear{Vazza, Roncarelli, Ettori  \& Dolag}{Vazza
  et~al.}{2011}]{10.1111/j.1365-2966.2010.18120.x}
Vazza F.,  Roncarelli M.,  Ettori S.,   Dolag K.,  2011, \mn@doi [Monthly
  Notices of the Royal Astronomical Society]
  {10.1111/j.1365-2966.2010.18120.x}, 413, 2305

\bibitem[\protect\citeauthoryear{Vikhlinin, Kravtsov, Forman, Jones,
  Markevitch, Murray  \& Speybroeck}{Vikhlinin et~al.}{2006}]{Vikhlinin_2006}
Vikhlinin A.,  Kravtsov A.,  Forman W.,  Jones C.,  Markevitch M.,  Murray
  S.~S.,   Speybroeck L.~V.,  2006, \mn@doi [The Astrophysical Journal]
  {10.1086/500288}, 640, 691

\bibitem[\protect\citeauthoryear{{Voit}, {Kay}  \& {Bryan}}{{Voit}
  et~al.}{2005}]{2005MNRAS.364..909V}
{Voit} G.~M.,  {Kay} S.~T.,   {Bryan} G.~L.,  2005, \mn@doi [\mnras]
  {10.1111/j.1365-2966.2005.09621.x}, \href
  {https://ui.adsabs.harvard.edu/abs/2005MNRAS.364..909V} {364, 909}

\bibitem[\protect\citeauthoryear{{Walker}, {Fabian}, {Sanders}  \&
  {George}}{{Walker} et~al.}{2012b}]{2012MNRAS.424.1826W}
{Walker} S.~A.,  {Fabian} A.~C.,  {Sanders} J.~S.,   {George} M.~R.,  2012b,
  \mn@doi [\mnras] {10.1111/j.1365-2966.2012.21282.x}, \href
  {https://ui.adsabs.harvard.edu/abs/2012MNRAS.424.1826W} {424, 1826}

\bibitem[\protect\citeauthoryear{Walker, Fabian, Sanders  \& George}{Walker
  et~al.}{2012a}]{10.1111/j.1365-2966.2012.21282.x}
Walker S.~A.,  Fabian A.~C.,  Sanders J.~S.,   George M.~R.,  2012a, \mn@doi
  [Monthly Notices of the Royal Astronomical Society]
  {10.1111/j.1365-2966.2012.21282.x}, 424, 1826

\bibitem[\protect\citeauthoryear{{Walker}, {Fabian}, {Sanders}, {Simionescu}
  \& {Tawara}}{{Walker} et~al.}{2013}]{2013MNRAS.432..554W}
{Walker} S.~A.,  {Fabian} A.~C.,  {Sanders} J.~S.,  {Simionescu} A.,   {Tawara}
  Y.,  2013, \mn@doi [\mnras] {10.1093/mnras/stt497}, \href
  {https://ui.adsabs.harvard.edu/abs/2013MNRAS.432..554W} {432, 554}

\bibitem[\protect\citeauthoryear{{Walker} et~al.,}{{Walker}
  et~al.}{2019}]{2019SSRv..215....7W}
{Walker} S.,  et~al., 2019, \mn@doi [\ssr] {10.1007/s11214-018-0572-8}, \href
  {https://ui.adsabs.harvard.edu/abs/2019SSRv..215....7W} {215, 7}

\bibitem[\protect\citeauthoryear{{Wong}, {Irwin}, {Wik}, {Sun}, {Sarazin},
  {Fujita}  \& {Reiprich}}{{Wong} et~al.}{2016}]{2016ApJ...829...49W}
{Wong} K.-W.,  {Irwin} J.~A.,  {Wik} D.~R.,  {Sun} M.,  {Sarazin} C.~L.,
  {Fujita} Y.,   {Reiprich} T.~H.,  2016, \mn@doi [\apj]
  {10.3847/0004-637X/829/1/49}, \href
  {https://ui.adsabs.harvard.edu/abs/2016ApJ...829...49W} {829, 49}

\makeatother
\end{thebibliography}
